\documentclass[12pt]{article}
\usepackage{epsfig}
\usepackage{graphicx}
\usepackage{amsmath,amssymb}
\def\thebibliography#1{\leftline{\large\bf References}\list
  {[\arabic{enumi}]}{\settowidth\labelwidth{[#1]}\leftmargin\labelwidth
\advance\leftmargin\labelsep
\usecounter{enumi}}
\def\newblock{\hskip .11em plus .33em minus .07em}
\sloppy\clubpenalty4000\widowpenalty4000}

\newcommand{\Det}{\mbox{Det}}

\newcommand{\Dslash}{D \hskip -0.6em /}
\newcommand{\Aslash}{A \hskip -0.6em /}
\newcommand{\dslash}{\partial \hskip -0.6em /}

\newcommand{\pvalt}{\raise0.15ex\hbox{-}\mkern-11.5mu\int}
\newcommand{\be}{\begin{equation}}
\newcommand{\ee}{\end{equation}}
\newcommand{\bea}{\begin{eqnarray}}
\newcommand{\eea}{\end{eqnarray}}
\newcommand{\ben}{\begin{enumerate}}
\newcommand{\een}{\end{enumerate}}
\newcommand{\bit}{\begin{itemize}}
\newcommand{\eit}{\end{itemize}}

\newcommand{\eq}[1]{eq.~(\ref{#1})}

\newcommand{\half}{\frac{1}{2}} 

\renewcommand{\ln}{\,\mbox{ln}\,}

\newcommand{\ba}{\begin{array}}
\newcommand{\ea}{\end{array}}

\newcommand{\bn}{\begin{enumerate}}
\newcommand{\en}{\end{enumerate}}
\newcommand{\dbar}{\bar{d}}
\newcommand{\kslash}{k\hskip-0.5em/}
\newcommand{\lslash}{l\hskip-0.5em/}
\newcommand{\pslash}{p\hskip-0.5em/}
\newcommand{\qslash}{q\hskip-0.5em/}

\newcommand\hide[1]{}

\newcommand{\ole}{{one-loop energy }}

\newcommand\res[1]{\mathcal{#1}}

\newcommand{\flux}{F}
\newcommand{\gaussSub}{{\rm G}}
\newcommand{\EclSub}{{\rm cl}}
\newcommand{\EvacSub}{{\rm vac}}
\newcommand{\EctSub}{{\rm CT}}
\newcommand{\EdeltaSub}{{\delta}}
\newcommand{\EfdSub}{{\rm FD}}
\newcommand{\EDESub}{{\rm DE}}
\begin{document}
 
\begin{center}
{\large \bf
Quantum $QED$ Flux Tubes in $2+1$ and $3+1$ Dimensions 
}
\end{center}

\centerline{
N.~Graham$^{\rm a}$,
V.~Khemani$^{\rm b,c}$,
M.~Quandt$^{\rm d}$, 
O.~Schr\"oder$^{\rm b}$,
H.~Weigel$^{\rm e}$}

\parbox[t]{15cm}{
\begin{center}
{~}\\$^{\rm a}$Department of Physics,
Middlebury College \\
Middlebury, VT 05753 \\~\\
$^{\rm b}$Center for Theoretical Physics,
Laboratory for Nuclear Science\\ and Department of Physics,
Massachusetts Institute of Technology\\ Cambridge, Massachusetts 02139 \\~\\
$^{\rm c}$Department of Physics\\
University of Connecticut \\
Storrs, Connecticut 06269\\~\\
$^{\rm d}$Institute for Theoretical Physics, T\"ubingen University\\
D-72076 T\"ubingen, Germany\\~\\
$^{\rm e}$Fachbereich Physik, Siegen University\\
D-57068 Siegen, Germany\\
{~} \\
{  \qquad MIT-CTP-3547 \qquad UNITU-HEP-10/2004 \\
hep-th/0410171}
\end{center}
}
 
\smallskip

\centerline{\large\bf Abstract}

\smallskip

{\small 
We compute energies and energy densities of static electromagnetic 
flux tubes in three and four spacetime dimensions.  Our 
calculation uses scattering data from the potential 
induced by the flux tube and imposes standard perturbative
renormalization conditions. The calculation is exact to 
one-loop order, with no additional approximation adopted. 
We embed the flux tube in a configuration with 
zero total flux so that we can fully apply standard results from 
scattering theory.  We find that upon choosing the same
on-shell renormalization conditions, the functional dependence of the
energy and energy density on the parameters of the flux tube is very
similar for three and four spacetime dimensions.  We compare our exact
results to those obtained from the derivative and perturbation
expansion approximations, and find good agreement for appropriate
parameters of the flux tube. This remedies some puzzles 
in the prior literature.}

\bigskip

\leftline{\it \small Keywords:~\parbox[t]{15cm}{
flux tubes, vacuum polarization energy, renormalization, 
derivative expansion}}

\leftline{\it \small PACS:~\parbox[t]{15cm}{11.10.Gh, 11.15.Kc,
11.27.+d, 12.20.Ds.}}
\newpage
\bigskip
\section{Introduction} 
Flux tubes in QED coupled to fermions exhibit a number of interesting
phenomena, such as the Aharonov-Bohm effect \cite{Aharonov:1979}, its
consequences for fermion scattering \cite{Alford89}, parity
anomalies~\cite{Redlich:1984dv}, formation of a 
condensate~\cite{Cangemi:1995by}, and exotic quantum 
numbers~\cite{Blankenbecler:1986ft,Kiskis97,Niemi:1983rq}.  
The same (non-perturbative) features of the theory
that give rise to these unusual phenomena make it more difficult to
analyze this system with standard techniques, especially in calculations
that require renormalization. The analysis in
ref.~\cite{Bordag:1998tg} and the world line formalism in
ref.~\cite{Langfeld:2002vy} have addressed some of these issues. Here we
provide a comprehensive approach drawing on techniques from scattering
theory to analyze the \ole and charge of this system.

Our primary motivation for this analysis is to shed light on
vortices in more complicated field theories, especially the Z-string
in the standard electroweak theory~\cite{Achucarro:1999it}. 
The Z-string is a vortex
configuration carrying magnetic flux in the field of the Z-gauge boson.  
If a network of such objects existed at the electroweak phase transition, 
then it would be one key ingredient in a viable mechanism for electroweak 
baryogenesis without a first-order phase transition.  Since the classical 
Z-string is known to be unstable \cite{Vachaspati:1992fi}, this scenario 
would require stabilization via quantum effects~\cite{Farhi:2000ws}, 
perhaps by trapping heavy quarks 
along the string. We also expect that by extending our results to the 
Abelian Higgs model, they could be applied to Abrikosov flux tubes in 
Type II superconductors \cite{Abrikosov:1956sx} or supersymmetric
Abelian Higgs models~\cite{Goldhaber:2004kn}.

We compare the one-loop energies and energy densities of electromagnetic
vortices in $D=2+1$ and $D=3+1$ spacetime dimensions.  The classical
calculation is of course the same in the two cases.  The quantum
corrections to the energy could possibly be very different
\cite{Langfeld:2002vy} because of the different divergence structure.
In $D=3+1$, the bare one-loop energy is divergent and only after we impose
renormalization conditions do we get a finite result. In $D=2+1$, 
in contrast, the bare energy is finite.  However, a comparison
between the two dimensionalities is sensible only when we use the same
renormalization conditions, which requires a finite counterterm in the
$D=2+1$ case to keep the photon field normalization the same.  Without
this finite renormalization, the $D=2+1$ and $D=3+1$ energies 
are qualitatively different.  But after proper renormalization, we
find that both the energies and energy densities are closely related.  

We also study this problem to get a handle on several technical issues
associated with the computation of the one-loop energy of a flux
tube.  As described in Section~\ref{sec:FunctionalDeterminant}, an
efficient way to compute the energy is to use scattering data of
fermions in the background of the flux tube.  However, vortex
configurations give long-range potentials, which do not satisfy standard
conditions in scattering theory \cite{Newton:1982qc}, which would
guarantee the analytic properties of scattering data. This leads to
subtleties that are discussed in Section~\ref{sec:subtleties}.  We
show that these puzzles arise only because an isolated flux tube is
unphysical, and once a region of return flux is included, the
scattering  problem is well-defined and the puzzles disappear.  In the
limit where the return flux is infinitely spread out, the energy 
density becomes entirely localized at the original flux tube.

The paper is organized as follows: In the following Section we
describe the theory and introduce electromagnetic flux tubes.
We outline the computation of their classical and renormalized 
one--loop vacuum polarization energies in Section~3. We describe puzzles 
in this calculation that arise from an isolated flux tube in Section~4
and explain how these puzzles can be resolved by appropriate embedding
of the flux tube in Section~5. We present numerical results 
for the energies, energy densities and charge densities in Section~6. 
In Section~7 we summarize our results and provide an outlook on related 
studies.  Three Appendices give the technical details needed for the 
computation of quantum contributions to the energy and charge densities.

\section{The Theory} 
We consider QED in spacetime dimensions $D=2+1$ and $D=3+1$, with a
four-component fermion field, $\psi$, in both cases.  The Lagrangian
density is 
\be
\mathcal{L}^{(D)} = - \frac{1}{4} F_{\mu \nu} F^{\mu \nu} + \bar{\psi} (i
\dslash + e \Aslash - m ) \psi + \mathcal{L}^{(D)}_\EctSub \, ,
\ee  
where the indices $\mu$ and $\nu$ run from $0$ to $D-1$ and 
$\mathcal{L}^{(D)}_\EctSub$ is
the counterterm Lagrangian in $D$ spacetime dimensions.  In $D=3+1$, this
Lagrangian describes electromagnetism with a single four-component
fermion of mass $m$ and charge $e$.  In $D=2+1$, we have parity-invariant
electromagnetism with two flavors of two-component fermions of equal
mass $m$ and charge $e$.  We are interested in static magnetic flux tubes.  These are localized,
cylindrically symmetric magnetic fields (pointing in the $z$ direction
in $D=3+1$), with a net flux $\flux$ through the $xy$-plane.  
In the radial gauge, flux tubes arise from vortex configurations of
gauge fields:
\be
A_0 = 0\,, \quad  \vec{A} = \frac{\flux}{2 \pi r} f(r)\hat{e}_{\varphi} \,,
\label{eq:vortex}
\ee  
where $r^2 = x^2 + y^2$ measures the planar distance from the center
of the vortex. The radial function $f(r)$ goes from 0 to 1. In 
$D=2+1$ the magnetic field is the radial function
\be
B ( r ) = \frac{F}{2 \pi r} \frac{d f(r)}{d r} \, ,
\label{eq:vortexB}
\ee
while in $D=3+1$ it is the vector field 
$\vec{B\,}(\vec{r\,})=B(r)\hat{e}_z$.
For small $r$, we must have $f(r)$ going to zero at least 
quadratically for $B(r)$ and $\vec{A}(\vec{r})$ to be non--singular. 

We often find it convenient to specify the flux in units of $2\pi/e$ 
and define the dimensionless quantity
\be
\res{\flux} =\frac{e}{2\pi}\flux\,.
\label{flux1}
\ee

In most cases we will take the example of a Gau{\ss}ian flux tube of 
width $w$,
\be
f_G(r)=1-e^{-r^2/w^2} \hbox{\qquad so that\qquad }
B_\gaussSub ( r ) = B_\gaussSub( 0 ) e^{-r^2/w^2} \, ,
\label{eq:gauss}
\ee
whose flux is
\be
\flux_\gaussSub = \pi w^2 B_\gaussSub( 0 ) \,.
\label{eq:gaussf}
\ee

\section{The Energy}
We would like to to compute the total energy $E^{(3)}$ for $D=2+1$ and
the energy per unit length $E^{(4)}$ for $D=3+1$, to one-loop order.  In
general, we would have to evaluate the appropriate matrix element of 
the energy momentum tensor.  However, as we review in
Appendix~\ref{Feynmanappendix}, for static configurations the
total energy is simply given by the negative one-loop effective action 
per unit time.  Since the theory is Abelian, the photons do not
have self-interactions and the one-loop effective action is obtained by
integrating out the fermion field.  The photon fluctuations start
contributing only at two loops and higher, which we ignore.

To one loop order, the total energy is the sum of the classical 
energy, $E_\EclSub$, and the fermion vacuum polarization energy,
$E^{(D)}_\EvacSub$,
\be
E^{(D)} = E_\EclSub + E^{(D)}_\EvacSub\,.
\ee
For both cases, $D=2+1$ and $D=3+1$, the classical contribution is
\be
E_\EclSub = \frac{1}{2} \int d^2 r B^2(\vec{r\,}) \, .
\label{eq:Ecl}
\ee
We extract the renormalized fermion vacuum polarization energy
from the one--loop effective action 
\bea
E_{\rm vac}^{(3)} &=& \lim_{T \rightarrow \infty} \frac{i}{T}
\left[ \ln \Det (i\dslash + e\Aslash - m) - \ln \Det (i\dslash - m)
  \right] + E^{(3)}_\EctSub\,, \cr
E_{\rm vac}^{(4)} &=& \lim_{T,L_z \rightarrow \infty} \frac{i}{TL_z}
\left[ \ln \Det (i\dslash + e\Aslash - m) - \ln \Det (i\dslash - m)
  \right] + E^{(4)}_\EctSub \, , 
\label{eq:det}
\eea
by dividing out the arbitrarily long time interval in both cases, and
the arbitrary length of the vortex in $D=3+1$ dimensions.
Note that the energy is always defined relative to a background where the
electromagnetic fields are zero everywhere.

\subsection{Renormalization}
\label{sec:Renormalization}

The counterterms introduced in eq.~(\ref{eq:det}) 
originate from the standard counterterm Lagrangian,
${\cal L}_{\rm ct}=-\frac{C^{(D)}}{4} F_{\mu\nu}^2$. Since
we do not consider electric fields, the counterterms are quadratic 
in the magnetic field
\be
E^{(D)}_\EctSub = \frac{C^{(D)}}{2} \, \int d^2 x B^2 \, .
\label{eq:counterterm}
\ee
To fix the counterterm coefficient, $C^{(D)}$, we choose on-shell
renormalization conditions, which require that the residue at the $q^2=0$
pole of the photon propagator is unity to ensure that one-photon states
remain normalized.\footnote{
Although fermions are logarithmically confined in $D=2+1$, asymptotic photon
states can still be observed experimentally through photon-photon scattering.}
We obtain
\be
C^{(D)} = - \frac{4 e^2}{3 (4\pi)^{D/2}} \frac{\Gamma \left(2 -
  \frac{D}{2} \right)}{m^{4-D}}\,
\ee
which becomes 
\be
C^{(3)} = - \frac{e^2}{6 \pi m} \, ,
\ee
in $D=2+1$ dimensions, and the dimensionally regularized quantity
\be
C^{(4-\epsilon)} = - \frac{e^2}{12 \pi^2} \left(
\frac{2}{\epsilon} - \gamma + \ln \frac{4 \pi}{m^2} \right) \, .
\ee
for dimensions approaching $D=3+1$, where $\epsilon=4-D$.

The counterterm, ${\cal L}_{\rm ct}$ renormalizes the bare photon field 
and fermion charge,
\be
A^\mu = (1+C^{(D)})^{-1/2} A_{\rm bare}^\mu 
\quad {\rm and} \quad e = (1+C^{(D)})^{1/2}
e_{\rm bare} \, .   
\ee
In the absence of photon fluctuations, the bare fermion field and mass
do not get renormalized.  In $D=3+1$ the counterterm is divergent and
combines with the divergent functional determinant to give a finite,
renormalized one-loop energy.  In $D=2+1$, the counterterm is finite,
but it still must be taken into account to ensure that the photon field
remains canonically normalized.  Since we want to compare the
energies in $D=2+1$ and $D=3+1$, we must impose identical
renormalization conditions in the two cases.

\subsection{Functional Determinant}
\label{sec:FunctionalDeterminant}
We use the phase shift approach \cite{Graham:1999pp,Graham:2002fi} to
compute the functional determinant in \eq{eq:det} exactly. The fermion
vacuum energy is given by the renormalized sum over the shift in the
zero-point energies of the fermion modes due to the presence of  the
background magnetic field. This calculation comprises a sum over bound state
energies and an integral over the continuum energies weighted by the
change in the density of states.  These quantities are obtained from
the Dirac equation for the fermion fields in the background of the
flux tube. The explicit form of this equation is given in
Appendix~\ref{energyappendix}, eqs.~(\ref{Dirac4})
and~(\ref{Dirac2}). Since the configuration that characterizes the
flux tube is cylindrically symmetric, the fermion scattering matrix
may be decomposed into partial waves labeled by the
$z$-component of the total angular momentum, $M$.  Then, in $D=2+1$,
the change in the density of continuum states is given in terms of the phase
shifts of the fermion scattering wave functions in the flux tube background, 
\be
\Delta \rho(k) = \sum_{M,\sigma} \frac{1}{\pi}\, 
\frac{d}{dk}\left[\delta_{M,\sigma} (k)\right] \, , 
\label{eq:fundamental}
\ee
where $k$ is the magnitude of the momentum and $\sigma$ refers to
all other discrete labels. For a given momentum~$k$ there are 
two energy eigenvalues $\pm\sqrt{k^2+m^2}$ and 
two spin states. This four-channel scattering problem can 
easily be diagonalized, yielding four identical phase shifts (with 
the exception of the threshold at $k=0$, as discussed below). Therefore
summing over the discrete labels merely gives a factor of four.  We
denote the result of this sum as $\delta_M(k)$.  In $D=3+1$, 
we must also integrate over momenta in the $z$-direction
\cite{Graham:2001dy}.  The details of the phase shift calculation are
given in Appendix \ref{energyappendix}.

To renormalize the continuum integral, 
we subtract the first two terms in 
the Born expansion of the phase shift and add back in the
corresponding energy in the form of the two-point Feynman diagram (a
fermion loop with two insertions of the background potential).  The
details of this calculation are described in Appendices
\ref{energyappendix} and~\ref{Feynmanappendix}. In $D=3+1$, the Born
subtraction renders the integral over the continuum energies
convergent.  The ultraviolet divergences are isolated in the Feynman
diagram, which when combined with the counterterm,
eq.~(\ref{eq:counterterm}), gives the properly renormalized finite
result.  In $D=2+1$ the Born subtraction is finite, but implements the
on-shell renormalization condition.

Our final expression for the renormalized fermion vacuum energy is
the sum of the subtracted contribution of the bound
and continuum modes, the Feynman diagram, and the counterterm:
\be
E^{(D)}_\EvacSub=E^{(D)}_\EdeltaSub+E^{(D)}_\EfdSub+E^{(D)}_\EctSub \,.
\label{eq:totalenergy}
\ee
The contribution from the 
Born-subtracted mode sum can be expressed through scattering data,
\bea
E^{(3)}_\EdeltaSub & = & 
\half\sum_j\left(|\omega_j|-m\right)
+ \frac{1}{2\pi} \int_0^\infty dk
\frac{k}{\sqrt{k^2+m^2}} \sum_M \bar{\delta}_M(k) \, , 
\label{eq:phaseshiftD3}\\ 
E^{(4)}_\EdeltaSub & = &  
-\frac{1}{8\pi}\sum_j\left(\omega_j^2\ln \frac{\omega_j^2}{m^2}
+m^2-\omega_j^2\right)
- \frac{1}{4\pi^2} \int_0^\infty dk k \ln
\frac{k^2+m^2}{m^2} \sum_M \bar{\delta}_M (k) \, , 
\label{eq:phaseshiftD4}
\eea
where $\omega_j$ are the energy eigenvalues of the 
discrete bound states.  (Even though the configurations that
we consider here do not have such bound states, we have
included their contribution for completeness.)  We have defined
\be
\bar{\delta}_M(k) = \delta_M (k) - \delta_M^{(1)} (k) -
\delta_M^{(2)}(k) \, ,   
\ee
where $\delta^{(i)}_M$ denotes the $i^{\rm th}$ term in the Born 
series expansion of the exact phase shift, $\delta_M$. Since we have 
subtracted the two leading contributions of the Born series, the remainder, 
$E^{(D)}_\EdeltaSub$ comprises the sum of the third and higher order 
pieces.

Next we add back the two leading Born terms 
in form of Feynman diagrams. The first order diagram vanishes
by Furry's theorem. In dimensional regularization the unrenormalized 
second-order Feynman diagram energy is
\be
E^{(D)}_\EfdSub = \frac{8 \pi \res{\flux}^2}{(4 \pi)^{D/2}}
\int_0^\infty dp \left[\int_0^\infty dr \frac{d f(r)}{d r}
J_0(p r)\right]^2 \int_0^1 dx \frac{x(1-x) p \Gamma( 2-
D/2)}{[m^2+p^2 x(1-x)]^{2-D/2}} \, .
\label{eq:EFD}
\ee
Combining this expression with the counterterm energy, 
eq. (\ref{eq:counterterm}), gives the renormalized Feynman diagram 
energy at the physical spacetime dimension. The result is to impose
perturbative renormalization conditions on this non--perturbative calculation.

\section{Subtleties of Configurations With Net Flux}
\label{sec:subtleties}
We have described the general computation of the renormalized 
one-loop energies of magnetic flux tubes.  Before 
proceeding with the calculation, however, we must address
subtleties in the calculation of the phase shifts at zero momentum.
These questions arise because the background potential does not
satisfy standard asymptotic conditions in scattering theory.
As is well-known from the Aharonov-Bohm effect,
even if the magnetic fields are exponentially localized, the vector
potential falls only as $1/r$, which is too slowly for many of the
standard theorems of scattering theory to hold \cite{Newton:1982qc}.
As a result, the phase shifts are discontinuous at threshold.
Although this discontinuity does not 
cause problems for calculating the vacuum polarization energy,
because the integrands in eqs.~(\ref{eq:phaseshiftD3}) 
and~(\ref{eq:phaseshiftD4}) remain smooth, we can no longer determine
the number of bound states through Levinson's theorem~\cite{Farhi:2000gz}.

As described in Appendix \ref{energyappendix}, we calculate the phase
shifts in the background of a flux tube by deriving second-order 
differential equations from the Dirac equation.  The asymptotic
behavior as $r\to\infty$ of the  coefficient functions in these
differential equations is extracted from eqs.~(\ref{LL}) in 
with $h(r)=\res{\flux}$ and $\ell=M-\half$,
\bea
g^{(1)''}_\ell(r) + \frac{g^{(1)'}_\ell(r)}{r} + \left( k^2 -
\frac{(M-\half-\res{\flux})^2}{r^2} \right) g^{(1)}_\ell(r) &=& 0 \, ,
\nonumber \\
g^{(2)''}_\ell(r) + \frac{g^{(2)'}_\ell(r)}{r} + \left( k^2 -
\frac{(M+\half-\res{\flux})^2}{r^2} \right) g^{(2)}_\ell(r) &=& 0 \, , 
\label{eq:asymptotic}
\eea
where primes denote derivatives with respect to $r$. For large $r$ we do 
not have free Bessel differential equations whose index equals the 
angular momentum; rather the equations describe an 
\emph{ideal} flux tube with profile function $f(r)\equiv 1$. If
extended to the origin, such an ideal flux tube would generate a 
singular background potential.  

Since the centrifugal barrier is shifted by the amount of the flux,
the regular solutions to the asymptotic differential
equations~(\ref{eq:asymptotic}) are Bessel functions with a 
correspondingly shifted index, 
\be
g^{(1)}_\ell(r) = J_{|M-\half-\res{F}|} (k r) \quad  {\rm and} \quad
g^{(2)}_\ell(r) = J_{|M+\half-\res{F}|} (k r) \, . 
\label{eq:Bess1}
\ee
From the asymptotic behavior of the Bessel functions, 
we read off a phase shift relative to the trivial configuration 
$A_\mu=0$~\cite{Ruijsenaars:1983fp}
\be
\Delta \delta_{M} (k) = \pi \left( |M-\half| +
|M+\half| - |M-\half-\res{\flux}|  - |M+\half-\res{\flux}|\right) \, ,      
\label{deltarel}
\ee 
where we have summed the phase shifts associated with the two radial 
functions, which corresponds to summing over the spin degrees of
freedom. For a radially varying profile function $f(r)$ with 
$\lim_{r\to\infty}f(r)=1$, the second-order equations~(\ref{LL}) allow us to
compute phase shifts relative to the ideal flux tube,\footnote{The 
corresponding scattering wavefunctions approach
$H^{(2)}_{M\pm\half-\res{\flux}}(kr)-{\rm e}^{2i\delta_{M,\rm ideal}}
H^{(1)}_{M\pm\half-\res{\flux}}(kr)$ at large radii, where
$H^{(i)}$ are Hankel functions.} $\delta_{M,\rm ideal}$, to 
which $\Delta\delta_{M}$ must be added to
obtain the phase shift relative to the trivial configuration: 
\be
\delta_{M} = \delta_{M, \rm ideal} +\Delta\delta_{M} \, .
\ee  
However, for $k=0$, there is no $r$ large enough where the
asymptotic form of the Bessel functions holds. Hence 
equation~(\ref{deltarel}) is valid only for $k>0$ and 
and it is unclear what $\Delta\delta_{M}(0)$, and hence 
$\delta_{M}(0)$, should be.\footnote{See 
refs.~\cite{Boyanovsky:1985nv,Jaroszewicz:1986ss} for discussions of 
the relation between this discontinuity of the phase shifts and the 
anomaly in $D=2+1$.}

We also note that the second-order differential equations 
are not equivalent to the Dirac equations for $k=0$. 
For $\omega\to+m$ the lower
component of the spinor becomes identically zero, while for
$\omega\to-m$ the upper component is zero,\footnote{We use the
Bjorken--Drell convention for Dirac matrices.} so the elimination 
of one component in favor of the other becomes singular. 
For most applications, this is not a problem 
because the phase shifts have a smooth limit as $k\to0$. 
However, as discussed in Appendix~\ref{subsec:threshold}, 
the long-range potentials associated with flux tube
configurations allow for bound states with zero binding energy, 
which introduce discontinuities in the phase shift at threshold.
This singularity is reflected in our numerical analysis of the
phase shifts.  For $k$ significantly greater than $0$, the phase
shifts that we obtain from the two radial functions $g^{(1)}_\ell(r)$ and
$g^{(2)}_\ell(r)$ are identical, as they should be.  In the limit $k\to 0$,
we find that for $M\ge\half$ only the differential equation for
$g^{(2)}_\ell(r)$ converges to the phase shift, while the equation for
$g^{(1)}_\ell$ gives discontinuities in $\delta_M(k)$ because of the
threshold bound states.  The discontinuity is smeared out from $k=0$
because in practice we integrate the second-order equations starting from a
large-distance cutoff, rather than infinity.  For $M \le -\half$,
situation is the same, with the roles of $g^{(1)}_\ell$ and
$g^{(2)}_\ell$ reversed.  For small but nonzero $k$, the phase shifts
(without singularities) converge to
\be
\lim_{k \rightarrow 0} \delta_{M} (k) = \pi \left\{ \begin{array}{lll}
   |M+\half| - |M+\half-\res{F}| 
&\quad {\rm for}\quad & M \ge \half \\
   |M-\half| - |M-\half-\res{F}| 
&\quad {\rm for}\quad & M \le -\half \\
\end{array} \right. \,.
\label{eq:smallk}
\ee

According to ref.~\cite{Farhi:2000gz} we should be able to
compute the number of bound states from the the phase shifts at $k=0$,
\be
N^{\rm B}_M=\frac{1}{\pi}\left[
\delta_M(\omega\to+m)+\delta_M(\omega\to-m)\right]
=\frac{2}{\pi}\lim_{k \rightarrow 0} \delta_{M} (k)
\label{eq:Levinson}
\ee
since $\lim_{k \rightarrow \infty} \delta_{M} (k)=0$. With the
help of an explicit example we will now show that this result 
does not hold in the present case.  For definiteness we consider
$\res{\flux}=-1.2$.  From eq.~(\ref{eq:fluxcond}) in
Appendix~\ref{subsec:threshold} we conclude that there is a threshold
state at $\omega=-m$ with $M=-1/2$.  However eq.~(\ref{eq:Levinson}) yields
\be
N^{\rm B}_M = \left\{ \begin{array}{ccc}
1.6 &\quad{\rm for}\quad& M = -\half \\
2.4 &\quad{\rm for}\quad& M < -\half \\
-2.4 &\quad{\rm for}\quad& M \ge \half
\end{array} \right.
\ee
and we have a discrepancy between the number of states
that leave the continuum ($N_M^B$) and the number of bound states ($1$).

These problems can be seen more concretely by considering the case
of $D=2+1$ with a single two-component fermion.  Then the net
fermion charge in the background of the flux tube is $-\res{\flux}/2$,
which is connected to the parity anomaly \cite{Niemi:1983rq}.  
Thus charge conservation is violated as the flux tube is turned on.
Furthermore, if we hold the flux fixed but spread the magnetic field
out over a larger and larger region the total energy of the flux tube
approaches zero (as shown in Section~\ref{sec:fixedFlux}), leading to
the implausible conclusion that  there exist charged, massless objects
in the theory and the fundamental fermion cannot be stable. 

The puzzles addressed in this Section do not necessarily prevent us
from calculating the vacuum polarization energy, because the 
small momentum region is suppressed in the integrals in
eqs.~(\ref{eq:phaseshiftD3},~\ref{eq:phaseshiftD4}) and we 
have methods other than eq.~(\ref{eq:Levinson}) available to find the 
number of bound states and compute their binding energies.  However, 
to be certain of the consistency of our results, we turn to a 
method to eliminate these puzzles by extending the background potential 
to satisfy the standard conditions of scattering theory. This approach
will then also allow us to directly generalize the methods of
ref.~\cite{Graham:2002xq} to compute energy densities.

\section{Embedding}
\label{sec:embedding}
All the issues discussed in the previous Section can be resolved
by recalling that it is not possible to create a configuration
carrying non-zero net flux starting from nothing.
In 3+1 dimensions we have the Bianchi identity
\be
\epsilon^{\alpha \beta \mu \nu} \partial_\beta F_{\mu \nu} = 0 \, .
\ee
It is a mathematical identity and not an equation of motion derived
from the action principle.  In terms of electric and magnetic fields,
it gives us two of the Maxwell equations: 
\be
\nabla \cdot \vec{B}  =  0 \, , \qquad
\frac{\partial \vec{B}}{\partial t}  =  - \vec{\nabla} \times \vec{E} \, .
\ee
The fact that the magnetic field is divergenceless requires all
magnetic flux lines to be closed and there can be no net flux
through a plane unless closure of the flux lines 
is arranged at spatial infinity, which may not be appropriate
to compute the vacuum polarization energy since that requires 
integrating over all space.  
In 2+1 dimensions, only the 
second of the two equations remains in the form 
\be
\frac{\partial B}{\partial t} =  - \partial_x E_y + \partial_y E_x \, .
\ee
Integrating this equation over space, it is clear that the magnetic flux
is time independent for localized fields and it is not possible to
create a net flux starting from no flux.

Therefore, we consider flux tube configurations with a \emph{return flux}
such that all flux lines are closed and the total flux is zero. We say
that the flux tube is embedded in a physical no net flux configuration. Any
apparently missing states or charge can be accounted for by
considering the corresponding quantities localized around the return
flux. Once the flux tube is embedded in a no net flux configuration,
the resulting potentials in the second order differential equations
for $g^{(1)}_\ell$ and $g^{(2)}_\ell$  fall into the class of potentials that
are well understood in scattering theory.
(Ref. \cite{Blankenbecler:1986ft} uses similar ideas as part of 
a complementary approach to the scattering problem.)  The separation
between the locations of the flux tube and the return flux should be
large compared to the extent of both the flux tube and the return
flux.  As we will demonstrate, the vacuum polarization energy of the
configuration with zero net flux approaches a well-defined limit as
the return flux is sufficiently separated from the flux tube and is
sufficiently diffuse.  This limit represents the energy of the flux
tube alone, which we can verify by computing the corresponding density
over  the volume of the flux tube.  We find that this result
agrees with the energy obtained by using the phase shifts in the
background  of an isolated flux tube, ignoring the subtleties
discussed in the previous Section.

We now describe the above embedding in more detail.  We 
consider a special subset of return flux configurations with magnetic
field $B_R$ for which the extension, $w_R$, is proportional to the
distance, $R$, from the flux tube.  Since then $B_R$ is
characterized by a single length scale, and the flux is held
fixed, simple scaling arguments based on the perturbative expansion
show (see Section~\ref{sec:fixedFlux}) that the classical 
energy goes to zero like $1/R^2$ and the one-loop energy 
with on-shell renormalization goes like $1/R^4$ 
in both $D=2+1$ and $D=3+1$.  Thus the energy of the no
net flux configuration should approach the energy of the flux tube 
alone as $R\to\infty$. To show that this value corresponds to
computing the energy using the phase shifts in the problem
with no return flux, we consider the Gau{\ss}ian flux tube,
$B_\gaussSub(r)$, defined in \eq{eq:gauss} with flux $F_G$. We take
\be 
B_R(r) = - \frac{16 F_G}{\pi R^2 \left( 1+ 256 \left(
  r^2/R^2 - 1\right)^2 \right) \left( \pi/2 + \arctan(16)
  \right)} \,,
\label{eq:returnB}
\ee
which originates from the profile function
\be
f_R(r)=-\frac{\arctan\left[16(r^2/R^2-1)\right]+\arctan(16)}
{\pi/2+\arctan(16)} \label{eq:profile-return}
\ee
for our return flux.
By construction $f_R(r)\to-1$ as $r\to\infty$ such that
$f_G(r)+f_R(r)\to0$ in the same limit, with $f_G(r)$ defined
in eq.~(\ref{eq:gauss}).  Since this sum also vanishes
as $r\to 0$, it generates a scattering potential satisfying
standard conditions in scattering theory. 
Analogously for the magnitic field, we have the no net flux embedding 
\be
B_0 (r) = B_\gaussSub(r) + B_R(r) \, . 
\label{eq:zeroNF}
\ee

\begin{figure}
\centerline{
\includegraphics[height=8cm]{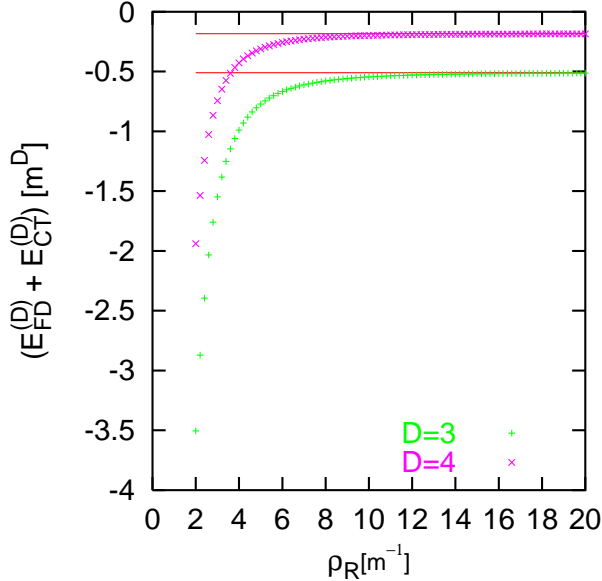}
}
\caption{\label{fig:embedFD}\sl Renormalized $D=2+1$ and $D=3+1$ Feynman
Diagram energies in appropriate units of $m$ as functions of the 
return flux radius for $w=1/m$ and $\res{\flux} = 4.8$.  The solid
lines correspond to the energies without the return flux.} 
\end{figure}

\begin{figure}
\centerline{
\includegraphics[width=6cm]{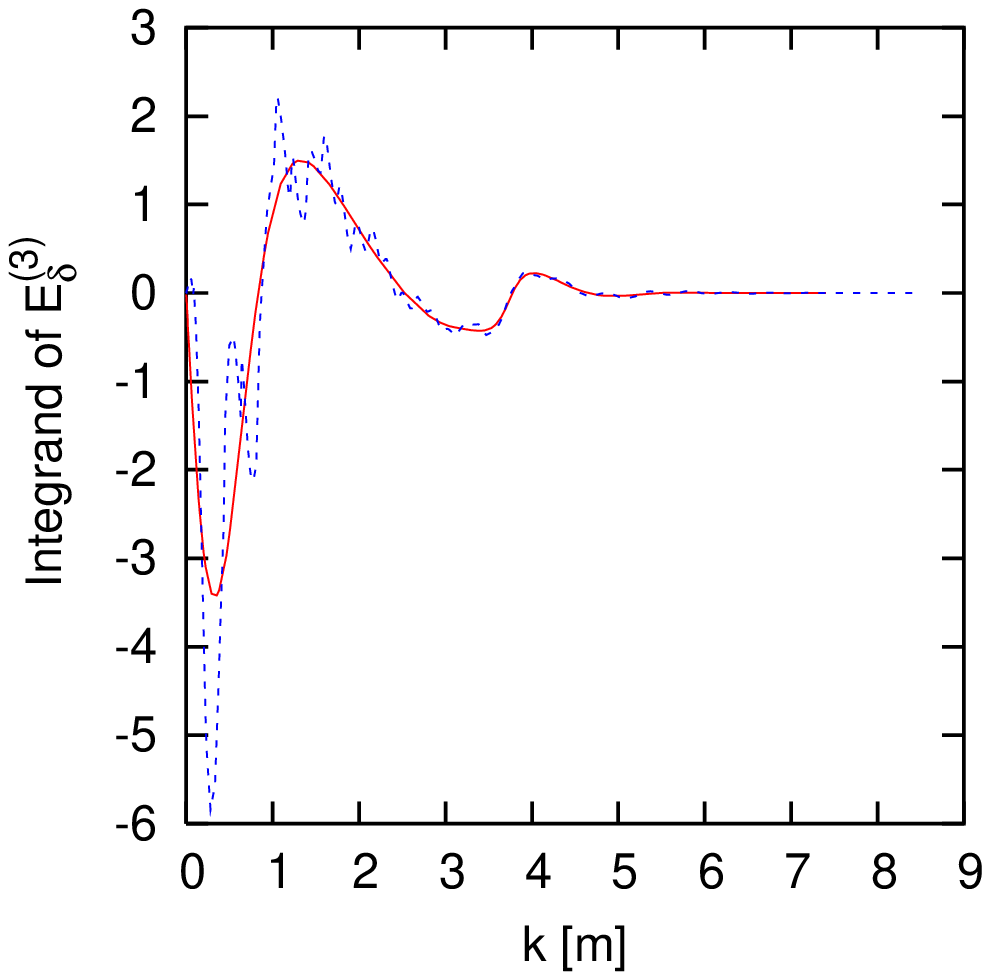}
\hskip 1.0cm
\includegraphics[width=6cm]{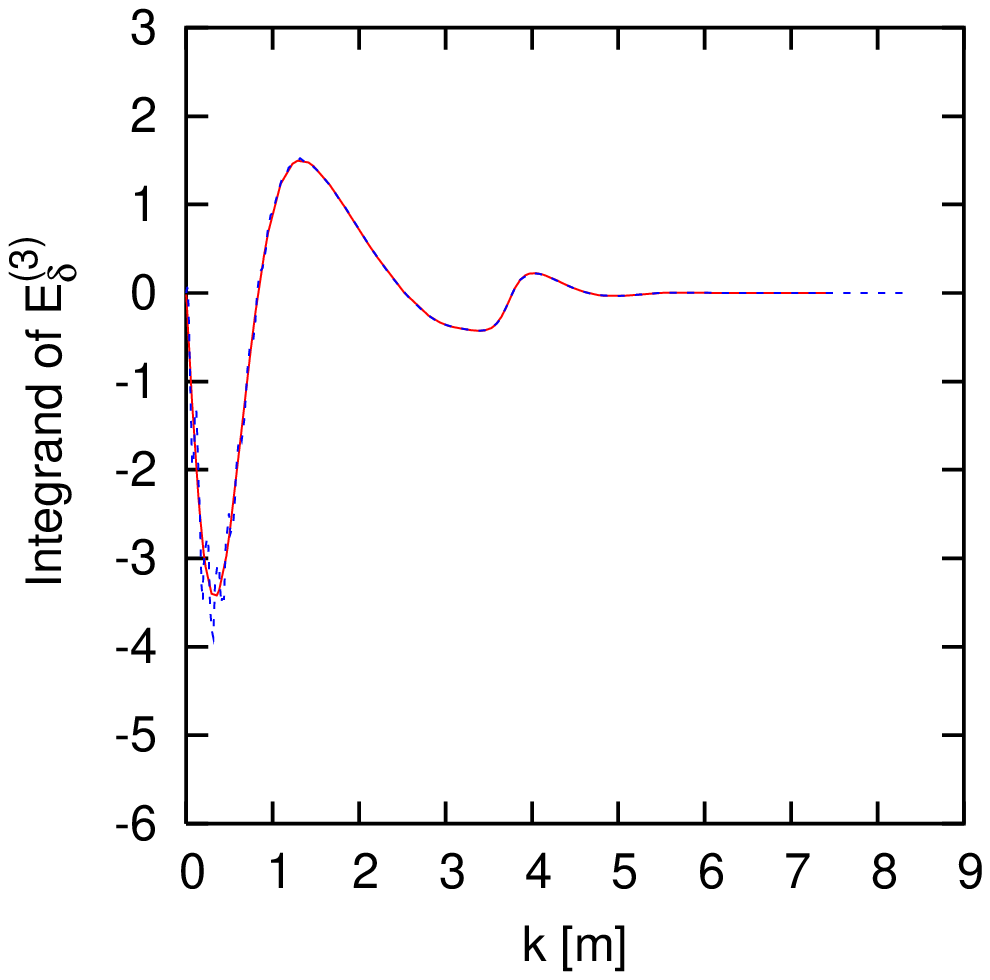}
}
\caption{\label{fig:embedDelta}\sl Integrand of the phase shift
contribution to the energy in $D=2+1$ for $R=6/m$ (left panel) and
$R=26/m$ (right panel).  The solid lines
correspond to the isolated flux tube problem and the dashed lines to
the embedded problem. }
\end{figure}

The total energy consists of three parts: the classical
energy, the renormalized Feynman diagram energy and the phase shift
energy.  We consider these different energy contributions for $B_0$ as
a function of $R$.  We choose $\res{\flux}_G = 4.8$ and $w=1/m$ 
to fix the flux tube, $B_\gaussSub$. Using the expression for the 
classical energy, eq.~(\ref{eq:Ecl}), it is straightforward to verify that 
\be
\lim_{R \rightarrow \infty} E_\EclSub [B_0] =
E_\EclSub [B_\gaussSub] \, .
\ee
In Fig.~\ref{fig:embedFD}, we show the renormalized Feynman diagram 
contribution to the energy , $E^{(D)}_\EfdSub + E_\EctSub$ as a function
of $R$ as computed from eqs.~(\ref{eq:EFD}) and~(\ref{eq:counterterm})
for the no net flux configuration, eq.~(\ref{eq:zeroNF}).
In both cases, $D=2+1$ and $D=3+1$, we find that the limit is well
saturated for $R > 10/m$.
In Fig.~\ref{fig:embedDelta}, we show the integrand of the phase shift
part of the energy, eqs.~(\ref{eq:phaseshiftD3},~\ref{eq:phaseshiftD4}), 
in both the embedded 
configuration and the flux tube configuration, for two fixed values of 
$R$.  The details of the calculation of the phase shifts of the
non--zero flux configuration may be found in Appendix~\ref{subsec:nonzeroNF}.
We observe that the integrands disagree only for small values of 
the momentum, at which the fermion is sensitive to the presence of the
spread-out return flux.  The integrand for the embedded configuration 
oscillates around the integrand of the flux tube configuration with
an amplitude that decreases as $k$ increases. The region of
disagreement gets pushed to smaller values of $k$ as $R$ increases.
In the limit $R\to\infty$, we find that the phase shift 
contribution to the energy is identical for the embedded and the flux 
tube problems.  Analogous results hold in $D=3+1$ because the integrand
differs from that in $D=2+1$ only by a background-independent function of 
$k$, {\it cf.} eqs.~(\ref{eq:phaseshiftD3}) and~(\ref{eq:phaseshiftD4}).

Since the energy in the embedded problem approaches the energy in the
isolated flux tube problem, for computational efficiency we can use
the phase shifts in the background of the isolated flux tube to
compute the energy, with the implicit understanding that there is a
spread-out return flux present that does not affect the energy.

\section{Results}
In this Section we discuss our numerical results in more detail.
The central issue will be the energy of the flux tube.  We will also
consider its energy and charge densities, as obtained from the embedded
problem, and compare our results to the derivative expansion.
\subsection{Energy}
As already argued in the previous Section, we can compute the 
energy of a magnetic flux tube in three and four dimensions
either independently or by embedding it in a
zero--net--flux configuration.  We then use scattering data to 
compute the vacuum polarization energy.  Of course, there are 
other (approximate) methods to compute the vacuum polarization
energy of such a configuration, most notably the derivative 
expansion~\cite{Cangemi:1995by,Gusynin:1998bt,Lee:1989vh}, which 
represents an expansion in derivatives of the magnetic field.
However, this expansion has recently been criticized, 
in particular in the four-dimensional case~\cite{Langfeld:2002vy}.
Since our approach is exact, we can easily test the validity 
of the derivative expansion. In addition, we may compare our results
to a perturbative expansion in the magnitude of the
magnetic field.  For the Gau{\ss}ian flux tube, eq.~(\ref{eq:gaussf}) we 
can formulate both expansions by varying the width, $w$ under
appropriate constraints.

\subsubsection{Derivative Expansion}
\label{sec:derivativeExpansion}
As its width approaches infinity, a Gau{\ss}ian flux tube with a
fixed value of magnetic field at its center approaches a spatially
constant magnetic field, for which the derivative expansion is reliable.
Therefore, to compare with the derivative expansion, we compute the
total energy of configurations with varying $w$ and fixed
$B_\gaussSub(0)$ for the Gau{\ss}ian magnetic field,
eq.~(\ref{eq:gauss}).  The flux, eq.~(\ref{eq:gaussf}), is not held
fixed in this approach.

In $D=2+1$, the derivative expansion of the \emph{unrenormalized} one-loop
vacuum polarization energy is given by \cite{Cangemi:1995by,Gusynin:1998bt} 
\be
E^{(3)}_{\EvacSub,DE} - E^{(3)}_\EctSub = E^{(3)}_{\EDESub, 0} +
E^{(3)}_{\EDESub, 2} + \ldots \, ,  
\label{eq:derexpD3}
\ee
to next-to-leading order in the derivative, where
\bea
 E^{(3)}_{\EDESub, 0} & = & \int d^2 x \frac{|eB|^{3/2}}{4 \pi^{3/2}}
 \int_0^\infty ds e^{-sm^2/|eB|} s^{-3/2} \left( \coth(s) -
 \frac{1}{s} \right) \, , \nonumber \\ 
 E^{(3)}_{\EDESub, 2} & = & \frac{1}{4} \int d^2 x
 |\nabla(eB)|^2 |4 \pi eB|^{-3/2} \int_0^\infty ds
 e^{-sm^2/|eB|} s^{-1/2} \frac{d^3 (s \coth s)}{ds^3} \, . 
\label{eq:derexpD3explicit}
\eea
Note that the above expressions are for four-component
fermions in the loop, which give twice the result for two-component
spinors.  Scaling the spatial integration
variable by $\xi = x/w$ straightforwardly yields 
$E^{(3)}_{\EDESub, 0} \propto w^2$ and $E^{(3)}_{\EDESub,2} 
\propto w^0$.  The counterterm contribution to the energy is also
proportional to $w^2$.  We stress that keeping $B(0)$ fixed is
important to obtain this simple power law behavior.  (Taking the limit
differently, for example with the flux fixed, the proper-time
integrals in eqs.~(\ref{eq:derexpD3explicit}) would induce
more complicated dependences on $w$.)
For the special case of the Gau{\ss}ian flux tube we find   
\be
E^{(3)}_\EctSub = - \frac{e^2 B_\gaussSub^2(0) w^2}{24 m} \, .
\ee
We add the counterterm to get the \emph{renormalized} derivative expansion
\be
E^{(3)}_{\EvacSub,DE} = E^{(3)}_\EctSub + E^{(3)}_{\EDESub, 0} +
E^{(3)}_{\EDESub, 2} +\ldots \, .  
\label{eq:renormderex}
\ee
It is straightforward to verify that this expression satisfies the 
on-shell renormalization condition that we imposed on our scattering 
data result in subsection~\ref{sec:Renormalization}. The first 
two terms on the right hand side of eq.~(\ref{eq:renormderex}) are 
proportional to $w^2$, while the last term is independent of $w$. 
All omitted terms contain more derivatives, and thus they
vanish as $w\to\infty$.

In $D=3+1$, the next-to-leading order derivative expansion of the
renormalized one-loop energy is
\be
E^{(4)}_{\EvacSub,DE} = E^{(4)}_{\EDESub, 0} +  E^{(4)}_{\EDESub, 2} 
\, +\ldots , 
\label{eq:derexpD4}
\ee
where \cite{Lee:1989vh} 
\bea
 E^{(4)}_{\EDESub, 0} & = & \int d^2 x \frac{|eB|^2}{8 \pi^2}
 \int_0^\infty ds e^{-sm^2/|eB|} s^{-2} \left( \coth(s) - \frac{1}{s}
 - \frac{s}{3} \right) \, , \nonumber \\ 
 E^{(4)}_{\EDESub, 2} & = & - \int d^2 x |\nabla(eB)|^2 |32
 \pi^2 eB|^{-1}
\label{eq:derexpD4explicit} \\ &&\hspace{2cm}\times
 \int_0^\infty ds e^{-sm^2/|eB|} \left( 1 - 4 \coth^2 s
 + 3 \coth^4 s + 
   \frac{3 \coth s}{s}(1 - \coth^2 s)  \right) \, .
\nonumber
\eea
Again we have $E^{(4)}_{\EDESub, 0} \propto w^2$ and
$E^{(4)}_{\EDESub, 2} \propto w^0$, with higher orders going as
inverse powers of $w$.

\begin{figure}
\centerline{
\includegraphics[width=6cm]{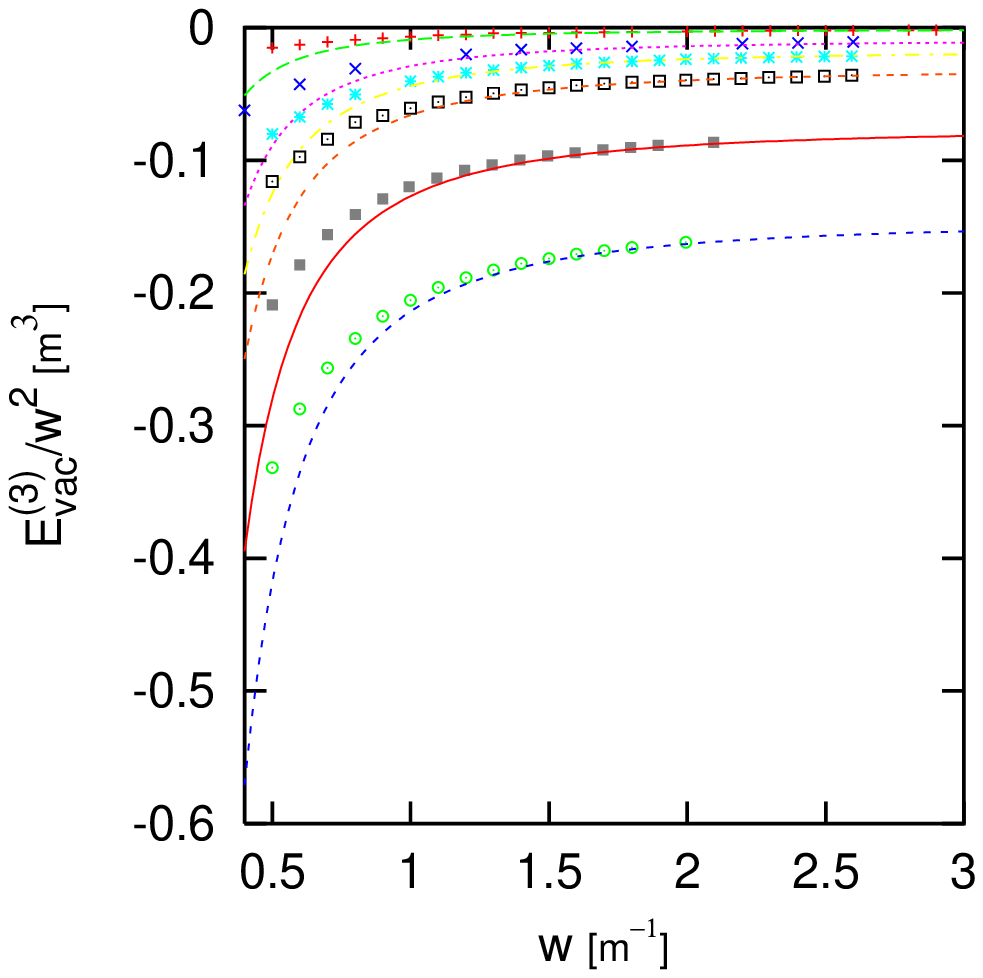}
\hskip 1.0cm
\includegraphics[width=6cm]{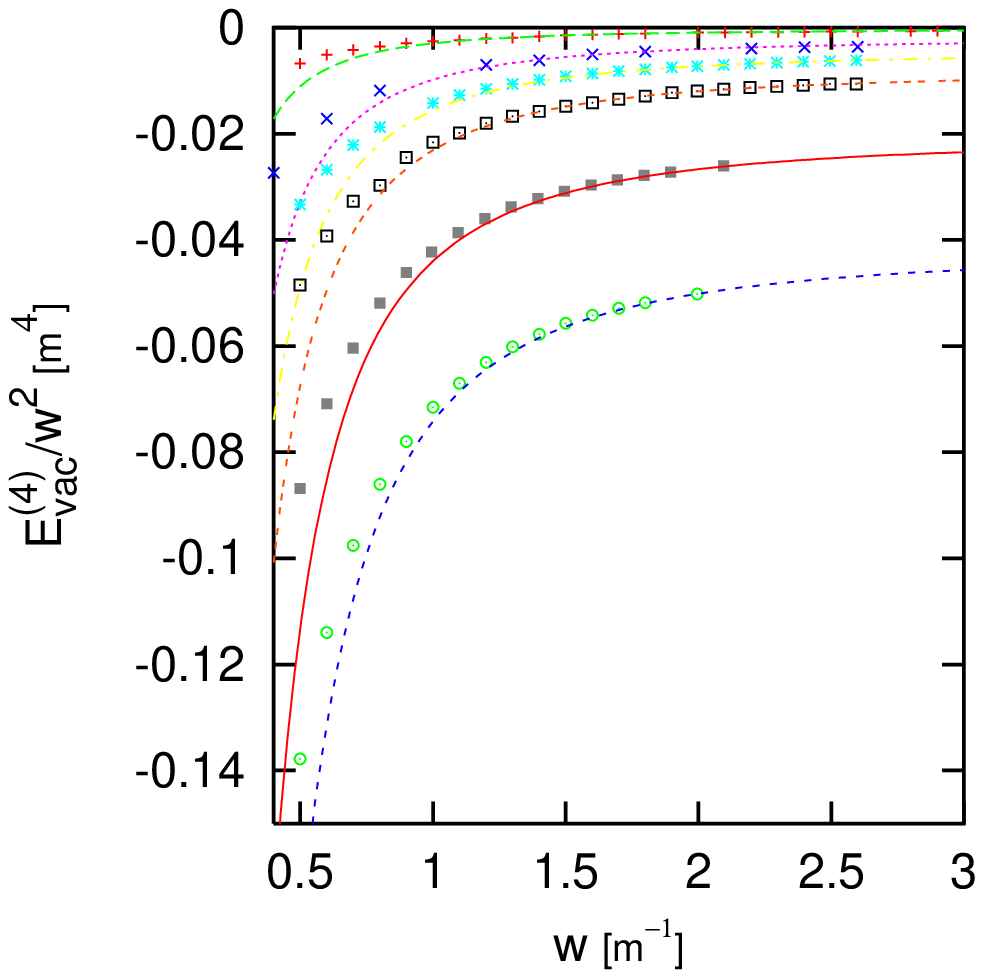}
}
\caption{\label{fig:derivExp}\sl Renormalized one-loop energies in $D=2+1$
(left panel) and $D=3+1$ (right panel), for fixed values of the magnetic
field at the origin, as a function of the width of the Gau{\ss}ian flux
tube.  The lines correspond to the derivative expansion to lowest
nontrivial order in the derivative.  From top to bottom,
$e B_\gaussSub (0)/m^2 = 1.1, 2, 2.5, 3, 4, 5$.}
\end{figure}

In Fig.~\ref{fig:derivExp}, we compare our exact result $E^{(D)}_\EvacSub$
with the derivative expansion approximation $E^{(D)}_{\EvacSub,DE}$
for different values of $eB_\gaussSub(0)$, and find excellent
agreement for widths larger than 1.  There appears to be no
qualitative difference $D=2+1$ and $D=3+1$.  
However, had we not included the counterterm in $D=2+1$, as is done in 
\cite{Langfeld:2002vy}, we would be effectively imposing a
different renormalization condition from the one in $D=3+1$,
and would see a qualitative difference between these two cases.  
As we will see below, the extra contribution from the off-shell scheme
can dominate the underlying result.  In particular, if this scheme is
consistently imposed both exactly and in the derivative expansion,
this dominant contribution can obscure differences between the exact
result and the derivative expansion in cases where the derivative
expansion is not valid.
\subsubsection{Fixed Flux}
\label{sec:fixedFlux}
We next turn to such a limit, which is the case of configurations for
which the flux, rather than the magnetic field at the origin,
is held fixed as the width approaches infinity. 
This limit was considered in ref.~\cite{Langfeld:2002vy}.  
In this case the energy of the flux tube goes to zero as $w$ goes to
infinity, and it is the perturbation expansion, rather than the
derivative expansion, that becomes exact.

First consider the classical energy, eq.~(\ref{eq:Ecl}),
\be
E_\EclSub = \frac{ \pi \res{\flux}^2}{e^2} \int_0^\infty \frac{dr}{r}\,
\left( \frac{d f_G(r)}{d r}  \right)^2 
=\frac{\pi \res{\flux}^2}{e^2 w^2}
\, , 
\ee
for the Gau{\ss}ian flux tube, eq.~(\ref{eq:gauss}). For large widths, 
it goes to zero like $1/w^2$. In this limit, the magnetic 
field becomes weak, as can easily be seen from eq.~(\ref{eq:gaussf}). 
Hence the dominant contribution to the vacuum polarization energy comes from
the two-point function.  In $D=2+1$, the unrenormalized two-point energy
can be expressed as a series in $1/w^2$. The leading term proportional 
to $1/w^2$ turns out to be exactly equal to minus the counterterm energy 
in the on--shell subtraction scheme.  Thus only the subleading term survives  
in the renormalized two-point energy. For a Gau{\ss}ian flux tube, 
\bea
E^{(3)}_\EfdSub & = & \frac{\res{\flux}^2}{6 m w^2} \left( 1 -
\frac{1}{5 m^2 w^2} + \mathcal{O}(1/w^4) \right) \nonumber \\ 
E^{(3)}_\EfdSub + E^{(3)}_\EctSub & = & - \frac{\res{\flux}^2}{30 m^3
  w^4} + \mathcal{O}(1/w^{6})  \, .
\eea
For large widths, renormalization not only changes the sign of the
one-loop energy, but also the rate at which zero is approached. The
renormalized results in $D=3+1$ are similar to the results in $D=2+1$: 
\be
E^{(4)}_\EfdSub + E^{(4)}_\EctSub  =
-\frac{\res{\flux}^2}{30 \pi m^2 w^4}
+ \mathcal{O}(1/w^{6}) 
\ee
for the Gau{\ss}ian flux tube.  In the limit 
$w\rightarrow \infty$, the total energy for both $D=2+1$ and $D=3+1$ 
vanishes.  This result enables us to introduce a return flux in the
embedded problem without changing the energy.

While such fixed flux configurations
are well approximated by the perturbative expansion, they are 
inappropriate to test the derivative expansion, as was
done in ref.~\cite{Langfeld:2002vy}.
Since the magnitude of the magnetic field at the
center of the vortex then depends on the width, so too will
the proper-time integrals in eqs.~(\ref{eq:derexpD3explicit},
\ref{eq:derexpD4explicit}), and there is no longer any reason to
expect that the the two-derivative contribution is less
important than the no-derivative contribution for large widths.

\begin{figure}
\centerline{
\includegraphics[width=5cm]{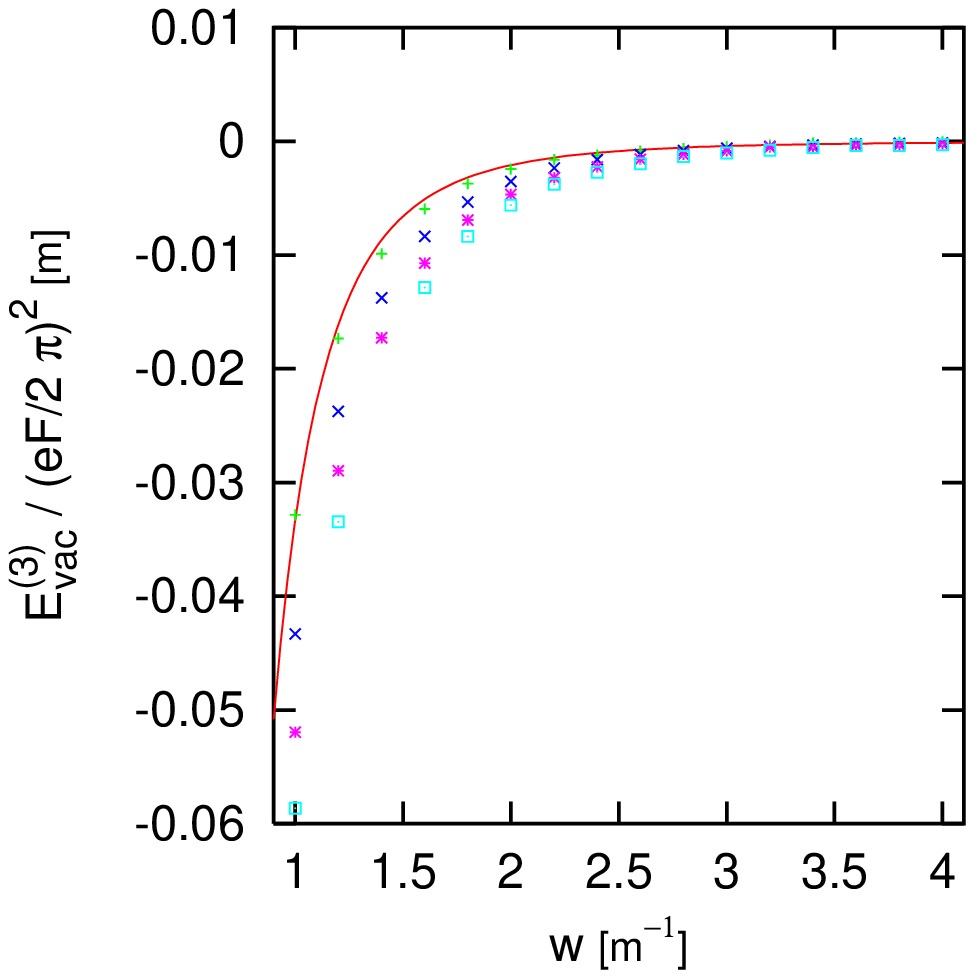}
\hskip 1.0cm
\includegraphics[width=5cm]{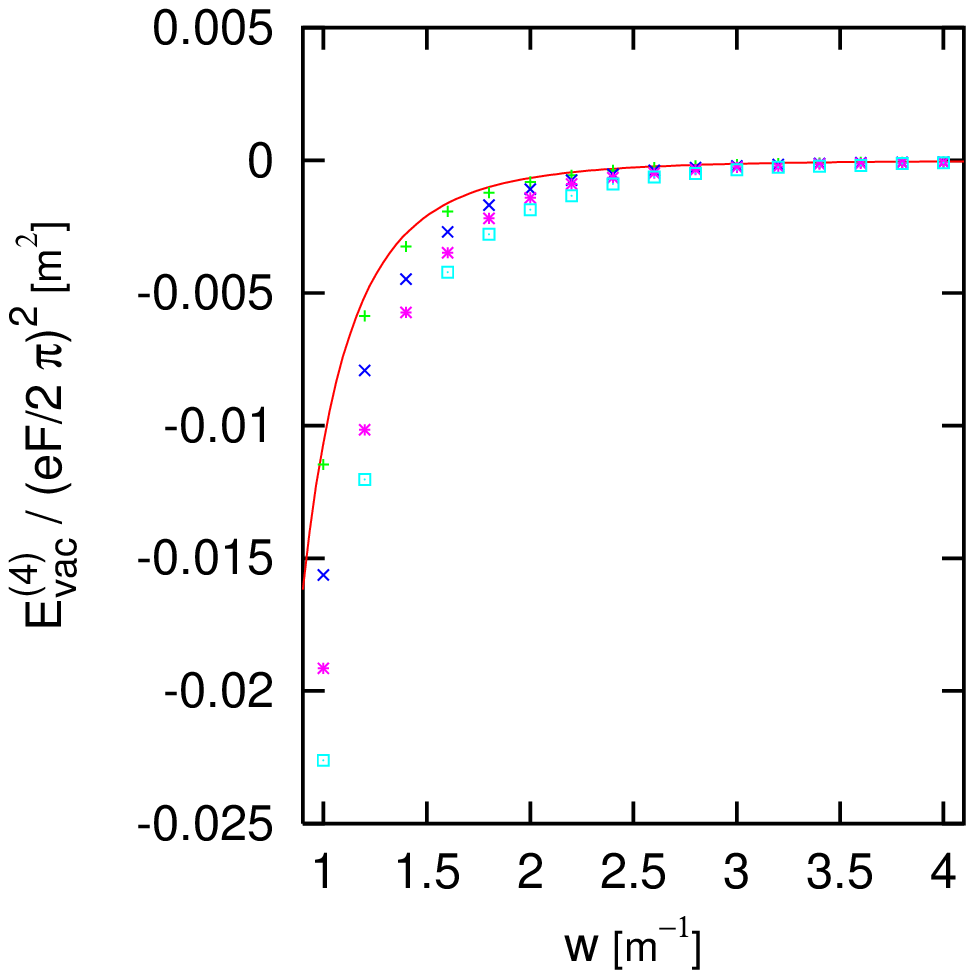}
}
\caption{\label{fig:fixedFlux}\sl Renormalized fermion vacuum 
polarization energy in units of $\res{\flux}^2$ as a function of the width,
for various fixed values of the flux $\res{\flux}$ (2.5, 4.5, 6.5, 8.5
from top to bottom) in the Gau{\ss}ian flux tube. The full line
represents the two-point function contribution. The left panel is for
$D=2+1$ and the right panel for $D=3+1$.}
\end{figure}

In Fig.~\ref{fig:fixedFlux}, we plot the exact one-loop energies for
various values of the flux as a
function of the width.  We normalize 
the energies in units of $\res{\flux}^2$ so that all differences
between the different fluxes are due to three-point and higher
contributions.  We compare the energies with the leading order
two-point energies and find good agreement for large widths.  As the
flux increases, we need to go to larger widths to get a weak magnetic
field everywhere, {\it cf.} eq.~(\ref{eq:gaussf}). Hence the agreement
with the leading order two--point function contribution to the energy
sets in at larger widths when the flux increases.  

\subsubsection{Further Comments}
\label{sec:fcomments}

Methods similar to our phase shift calculation have been 
used in ref.~\cite{Pasipoularides:2000gg}.  However, in this calculation 
infinite quantities are used without proper renormalization, such as
eq.~(26) of \cite{Pasipoularides:2000gg} for $D=3+1$.  Although
convergent results are reported, we believe that they are an artifact
of taking a low, fixed upper limit in the sum over channels.
Instead, this limit should scale with the momentum $k$.  By keeping more
terms in the sum, one should see this divergence.

n Ref.~\cite{Bordag:1998tg} a step function background $B(r)$
was studied in $D=2+1$ dimensions. In order to separate divergent 
contributions to the vacuum polarization energy the authors of 
Ref.~\cite{Bordag:1998tg} extracted the asymptotic behavior of the 
Jost function and identified that with pieces in the heat kernel 
expansion. Instead of using on-shell renormalization 
conditions, the renormalization is defined by requiring
that the renormalized vacuum polarization energy vanishes
in the limit where the fermion becomes infinitely 
heavy.  However, this requirement is not sufficient to uniquely
determine the renormalization scheme, as has been noted in 
\cite{Bordag:2002dg}.  As a result, we cannot make a 
quantitative comparison with our calculation, but qualitatively, our
results seem to agree in the sense that they also find 
a negative vacuum polarization energy. Since different renormalization 
conditions amout to different contributions from the positive definite 
counterterm, eq.~(\ref{eq:counterterm}), this agreement may be 
not be conclusive.

\subsection{Energy Density}

So far we have only discussed the total vacuum polarization energy of
the flux tube.  We have argued that it is preferable to consider a
flux tube embedded in system with no net flux, where the region 
carrying the return flux is well separated from the flux tube core. 
In this Section we make this argument more precise by studying
the radial density of the vacuum polarization energy. 
We will observe that the contribution to the energy density from the
central flux tube and the return flux can clearly be distinguished, giving
a unique definition for the energy of the central flux tube alone.
Numerical evaluation  shows that it equals the energy of the
non-embedded flux tube that we computed in
Section~\ref{sec:embedding}.  We will also take the
opportunity to compare our results for the energy density  with the
derivative expansion approximation, which has recently been discussed
in ref.~\cite{Langfeld:2002vy}.

\begin{figure}
\centerline{
\includegraphics[width=7cm]{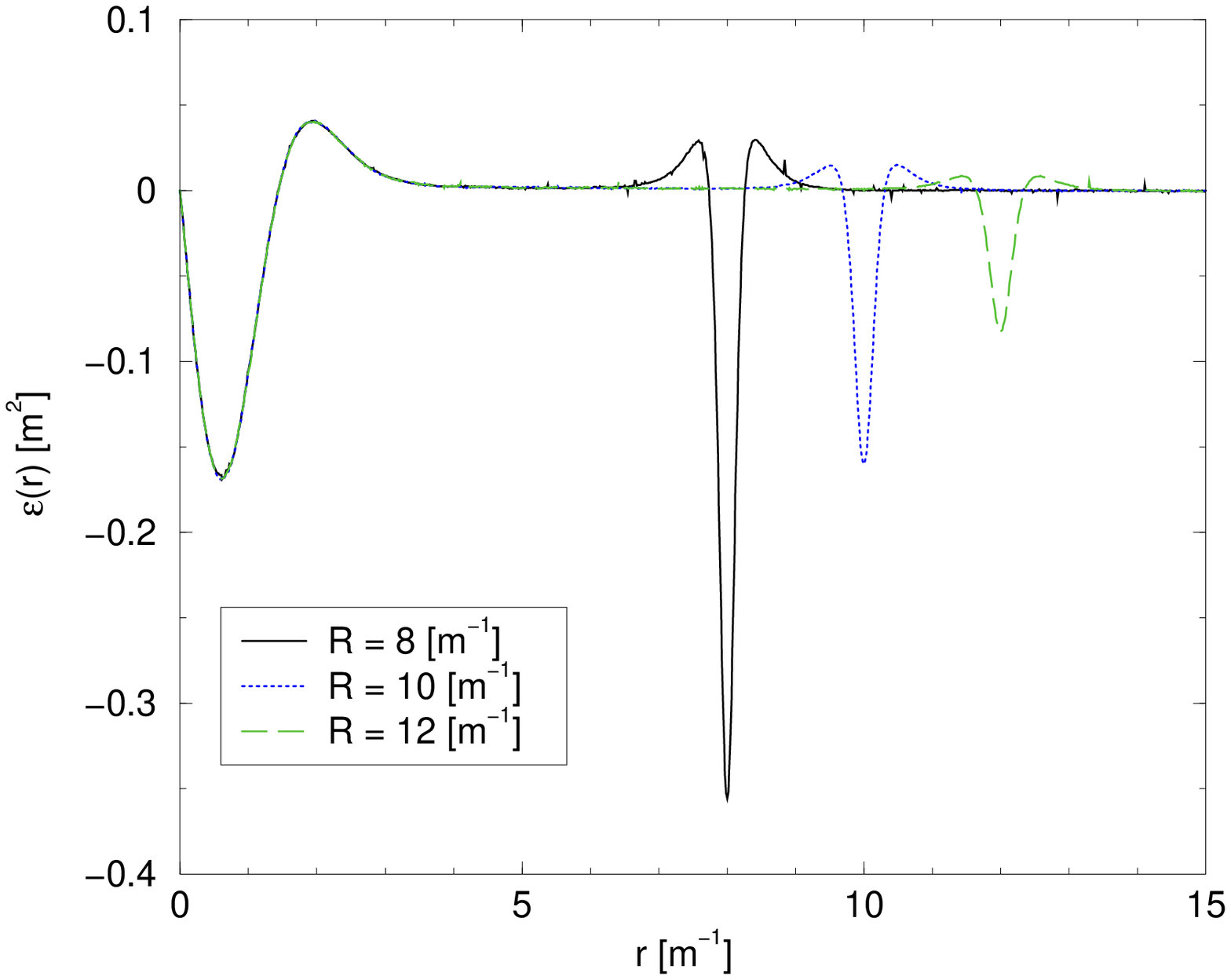}
\hskip1cm
\includegraphics[width=7cm]{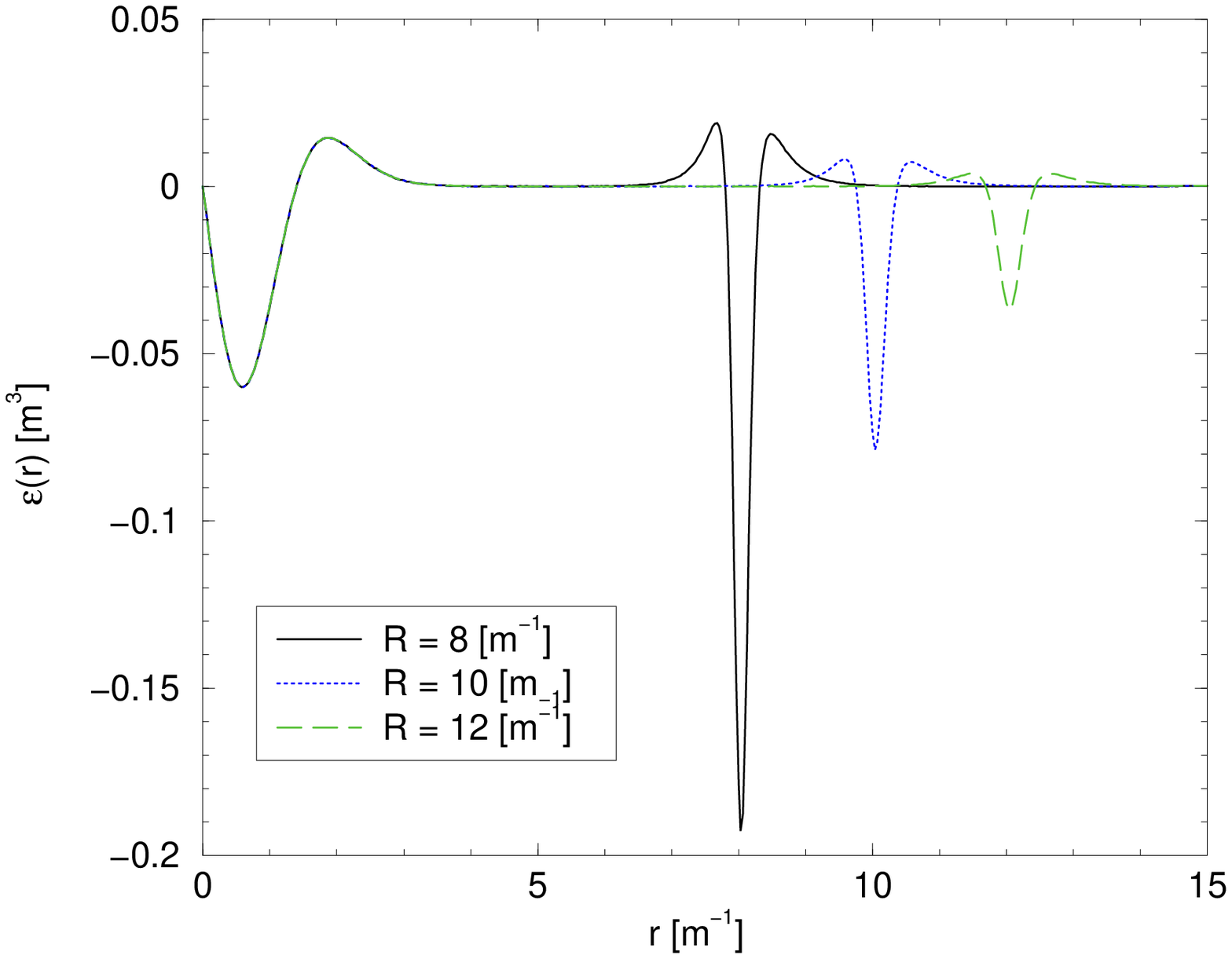}}
\caption{\label{fig_6.1}\sl The vacuum polarization energy densities
of configurations with zero net flux as functions of the separation
$R=8,10,12/m$ between $B_G$ and $B_R$. We choose $eB(0)=4m^2$ and
$w=1.5/m$. The left and right panels show $D=2+1$ and $D=3+1$,
respectively.}
\end{figure}

We define the radial energy density in terms of the renormalized
vacuum expectation value of the energy-momentum tensor $\hat T_{\mu
\nu}(x)$ by $\epsilon(r)= 2 \pi r \langle \hat{T}_{00}(x) \rangle$.  In 
Appendix~\ref{Feynmanappendix} we discuss $\hat{T}_{00}(x)$ in more detail,
with particular attention to total derivative terms.
The radial energy density $\epsilon(r)$ has been normalized so that
the total energy is $E^{(D)}_{\rm vac}=\int_0^\infty dr\, \epsilon(r)$.
In figure~\ref{fig_6.1} we show the energy density for the
configuration defined in eq.~(\ref{eq:zeroNF}).  As the separation
$R$ between $B_G$ and $B_R$ increases, the contribution from the
return flux region decreases, going to zero as
$R\to\infty$.  In addition, we observe that the energy
density localized around the flux tube at $r=0$ remains unchanged
when we increase the separation.  That part of the energy
density does not depend on whether the return flux
configuration is included once $B_G$ and $B_R$ are well separated.
Finally, the intermediate region in which the density $\epsilon(r)$
vanishes becomes more pronounced. 
Once the region with $\epsilon(r)\approx0$ is large enough we can clearly
distinguish between the energy densities due to $B_G$ and $B_R$.

\begin{figure}
\centerline{
\includegraphics[width=6cm]{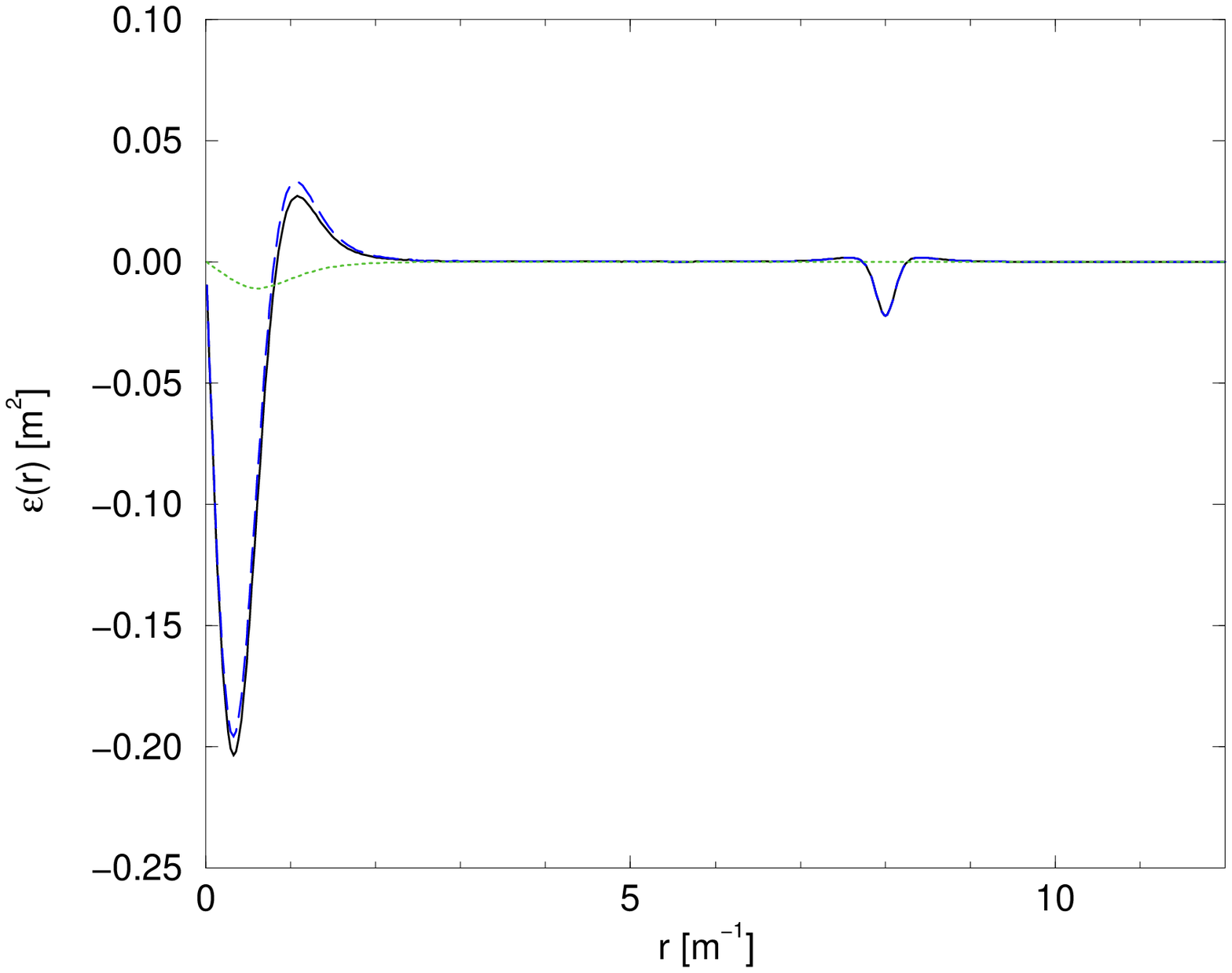}
\hskip1cm
\includegraphics[width=6cm]{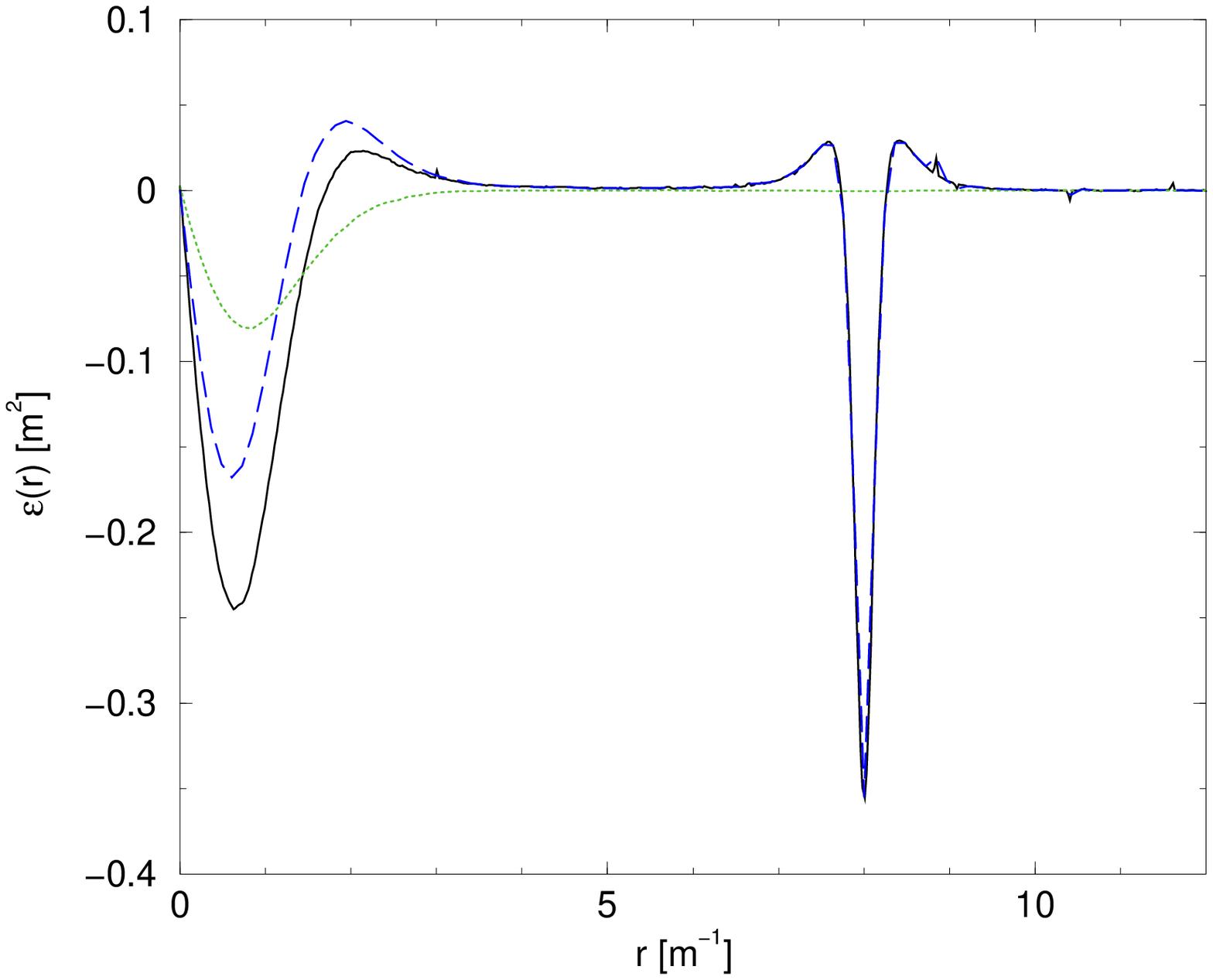}}
\vskip0.1cm
\centerline{
\includegraphics[width=6cm]{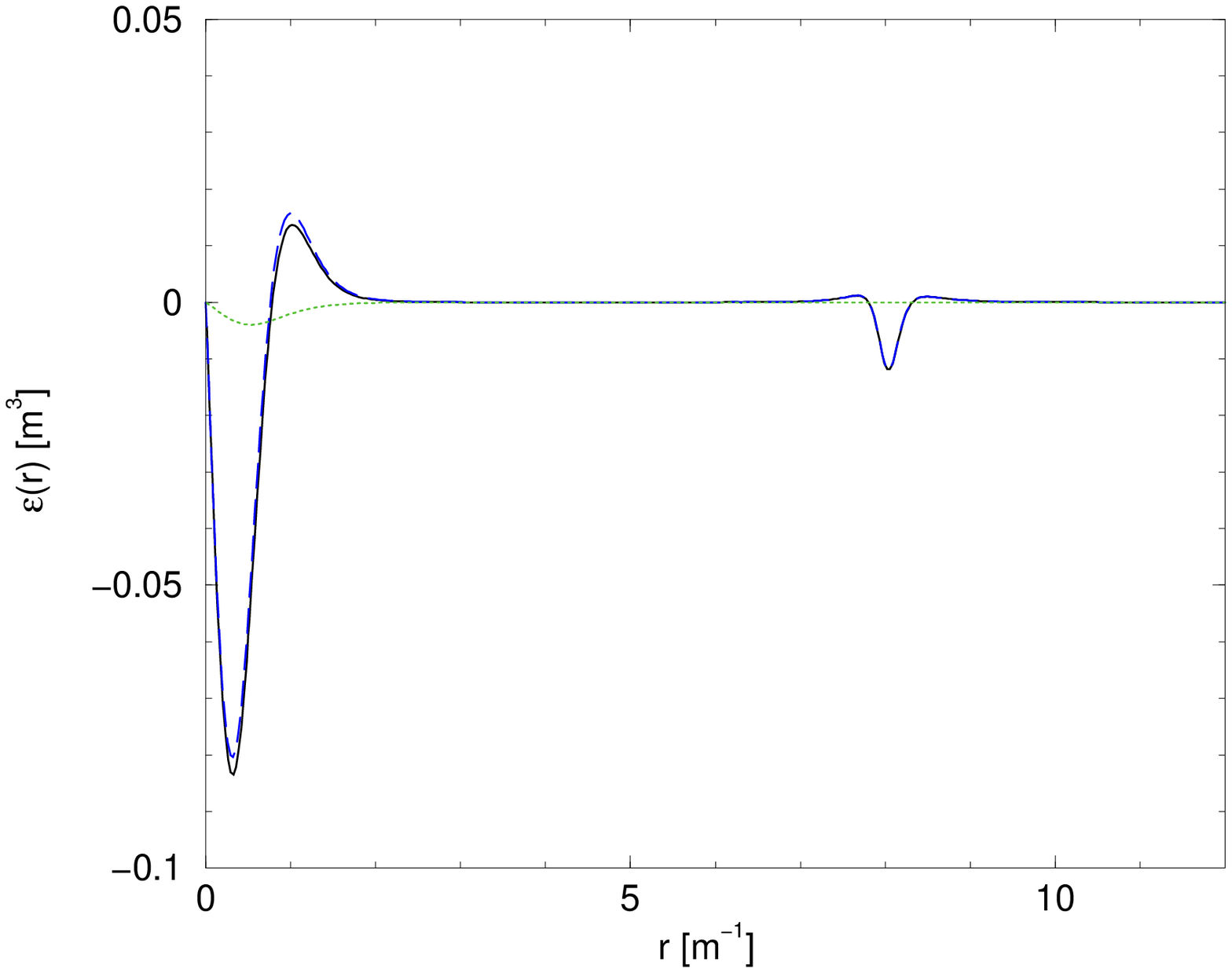}
\hskip1cm
\includegraphics[width=6cm]{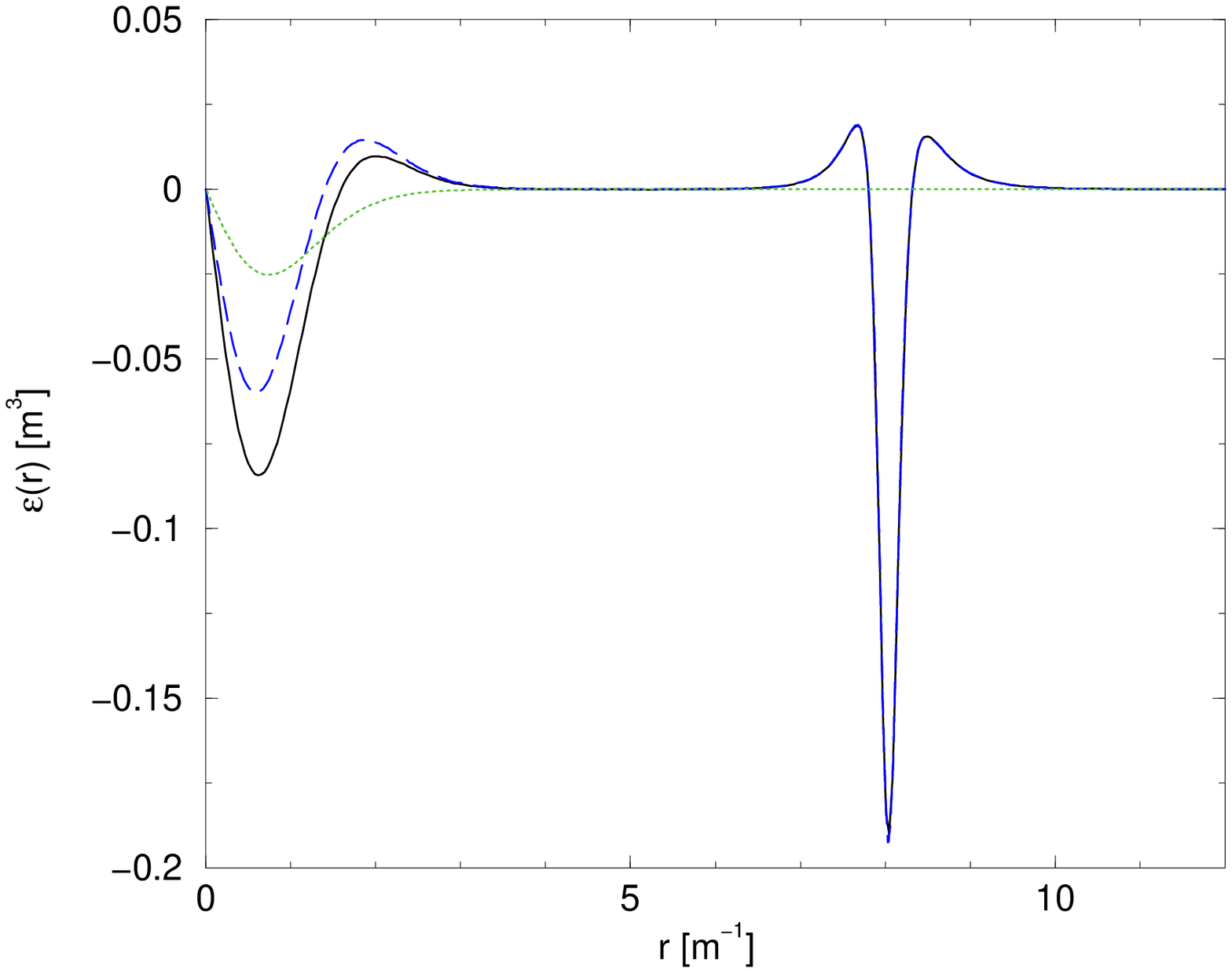}}
\caption{\label{fig_6.2}\sl Feynman diagram (dashed) and scattering
data (dotted) contributions to the energy density for configurations with
zero net flux. Their sum (full) is the full vacuum polarization
energy density.
Two cases are considered, $w=0.75/m$ (left panel) and
$w=1.5/m$ (right panel), both with $eB(0)=4m^2$ and $R=8/m$,
for $D=2+1$ (top) and $D=3+1$ (bottom).}
\end{figure}

In figure~\ref{fig_6.2} we show the renormalized Feynman 
diagram and scattering data contributions to the energy density. 
We observe that for small widths of the central Gau{\ss}ian 
flux tube, the scattering data contributions are 
negligible, while for larger widths they contribute as much as 30\%
to the total energy.   Note that $B(0)$ is kept fixed.
We also see that the scattering data contribution to the energy
density essentially vanishes in the return flux region.
This numerical result reflects a cancellation when summing over many orbital 
angular momentum channels; the individual channels themselves contain
sizable contributions to the energy density in the vicinity of
$r=R$.  This result is to be expected:  Since the return flux region
extends to large radii, large angular momentum channels contribute to
the energy density there.  We find that several hundred channels must
be summed for convergence.

\begin{figure}
\centerline{
\includegraphics[width=7cm,height=6cm]{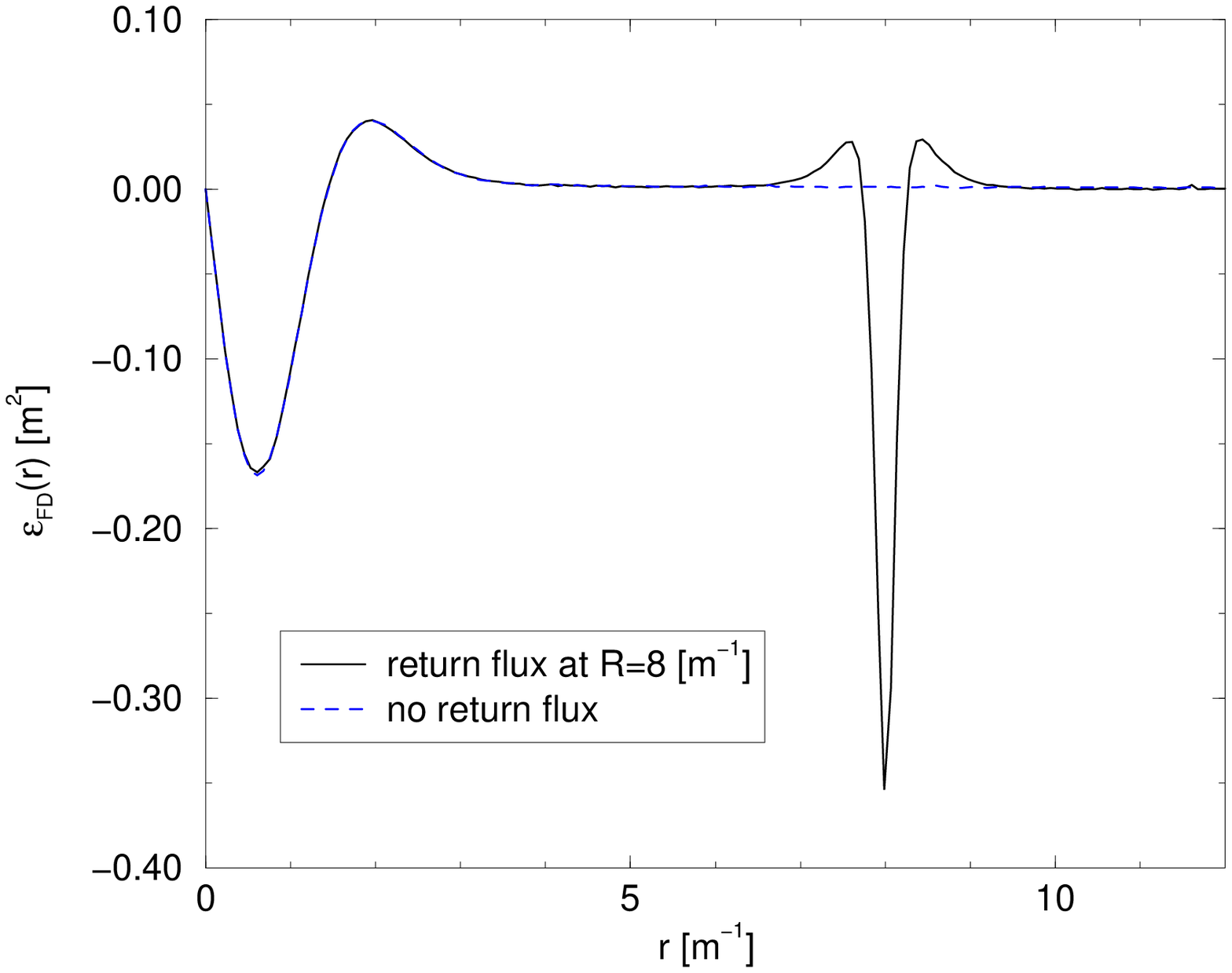}
\hskip1cm
\includegraphics[width=7cm,height=6cm]{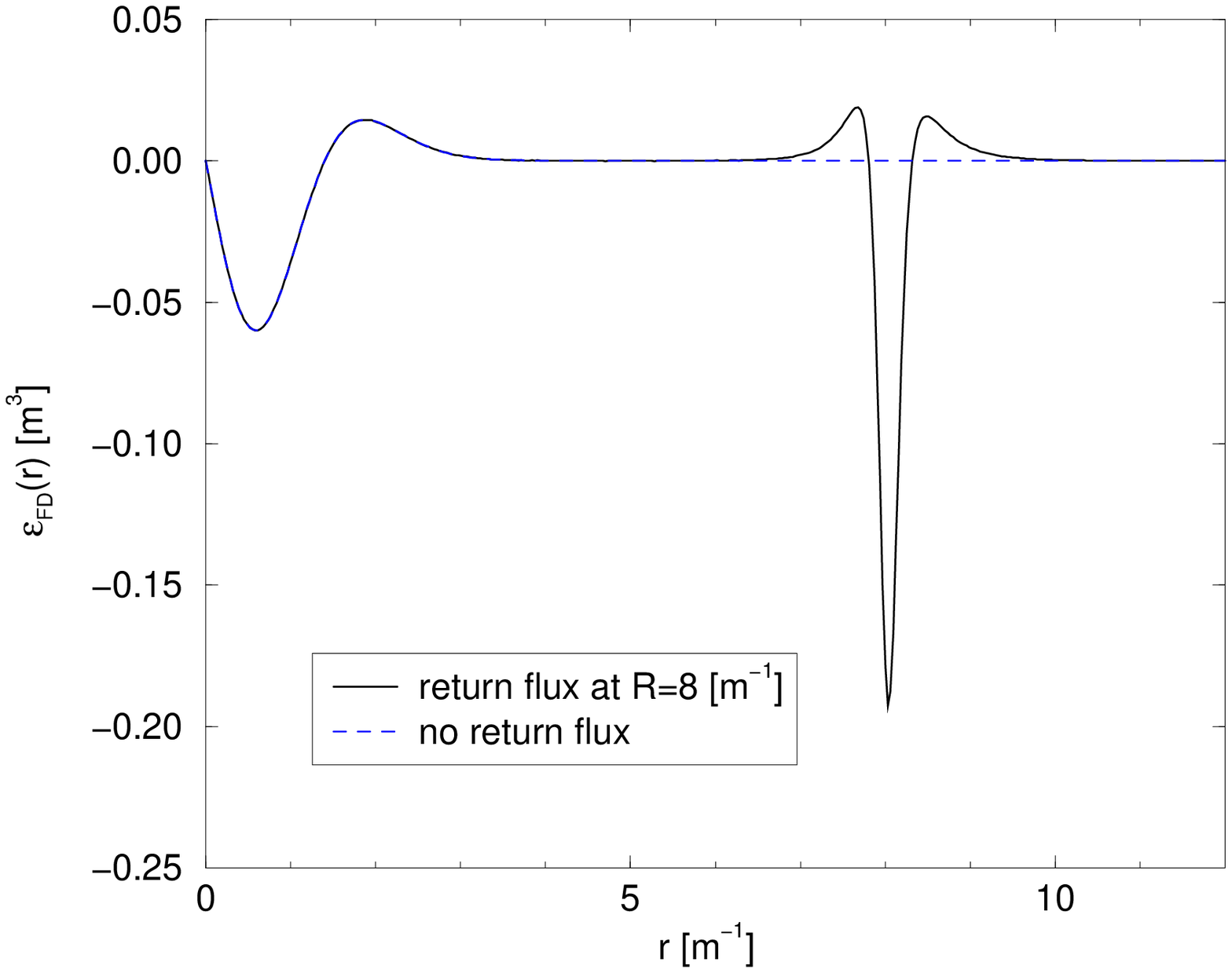}}
\caption{\label{fig_6.3}\sl Renormalized Feynman diagram contributions 
to the energy density for configurations with and without return
flux for $w=1.5/m$, $eB(0)=4m^2$ and $R=8/m$, in the case of 
zero net flux. The left and right panels show $D=2+1$ and $D=3+1$
respectively.}
\end{figure}

The computation of the Feynman diagram contribution to the
energy density does not rely on any conditions for the background field. 
Since the energy densities are well approximated by
their Feynman diagram contributions for small widths, we can
compare energy densities for configurations with and without $B_R$
in cases with small $w$. This comparison is shown in figure~\ref{fig_6.3}.

Integrating the energy density over the region of return 
flux, we find that even at moderate separation it only 
contributes a small amount to the vacuum polarization energy.
However, as figure~\ref{fig_6.3} illustrates,
this result arises from the cancellation of significant positive and negative
densities in this region.  It is not surprising that the energy
density of the return flux is nontrivial:  We keep $B(0)$ fixed, so as
the width of the flux tube increases, its flux does too, according to 
eq.~(\ref{eq:gaussf}).  To have zero net flux, the amplitude of the
return flux piece must increase when $w$ gets larger and $R$ remains
unchanged, as can be seen from eq.~(\ref{eq:returnB}).

To compare the results for the exact vacuum polarization
energy density with the derivative expansion,
we must identify the energy density in that approximation.  The derivative
expansion for the vacuum polarization energy,
eqs.~(\ref{eq:derexpD3explicit},\ref{eq:derexpD4explicit}),
originates from the expansion for the action.  Hence
simply omitting the radial integrals in those expressions
can only be expected to yield an approximation to the action density.
Formally we may write
\begin{equation}
E_{{\rm DE},n}^{(D)}=\int_0^\infty dr r
\left(\frac{d B(r)}{dr}\right)^n f^{(D)}_n(B(r))\,,
\label{eq:formalEDE}
\end{equation}
with $f_n^{(D)}(B(r))$ to be read off
eqs.~(\ref{eq:derexpD3explicit},\ref{eq:derexpD4explicit}) for 
$n=0,2$.  As discussed in Appendix~\ref{Feynmanappendix} the 
action and energy densities differ by total derivatives, which 
vanish when integrated over space.
These total derivates only affect the orders $n\ge2$ because for 
constant magnetic fields both the energy and action density should 
be constant.  Thus only $E_{\rm DE,2}^{(D)}$ is affected at 
next-to-leading order. The direct determination of this 
total derivate term involves all orders in perturbation theory
and seems difficult to accomplish.  We therefore introduce the 
parameter $\xi$ to define the energy density at next-to-leading 
order in the derivative expansion,
\begin{equation}
\epsilon_{\rm DE}^{(D)}(r)=r\left[f^{(D)}_0(B(r))+
\left(\frac{d B(r)}{dr}\right)^2 f^{(D)}_2(B(r))\right]
+\xi\frac{d}{dr}\left[r\frac{d B(r)}{dr}f^{(D)}_2(B(r))\right]\, ,
\label{eq:defepsDE}
\end{equation}
where the case $\xi=0$ corresponds to the the derivative expansion for the
action density.  Then we can fit for $\xi$ by comparing to the exact
result $\epsilon(r)$.

\begin{figure}
\centerline{
\includegraphics[width=7cm,height=5cm]{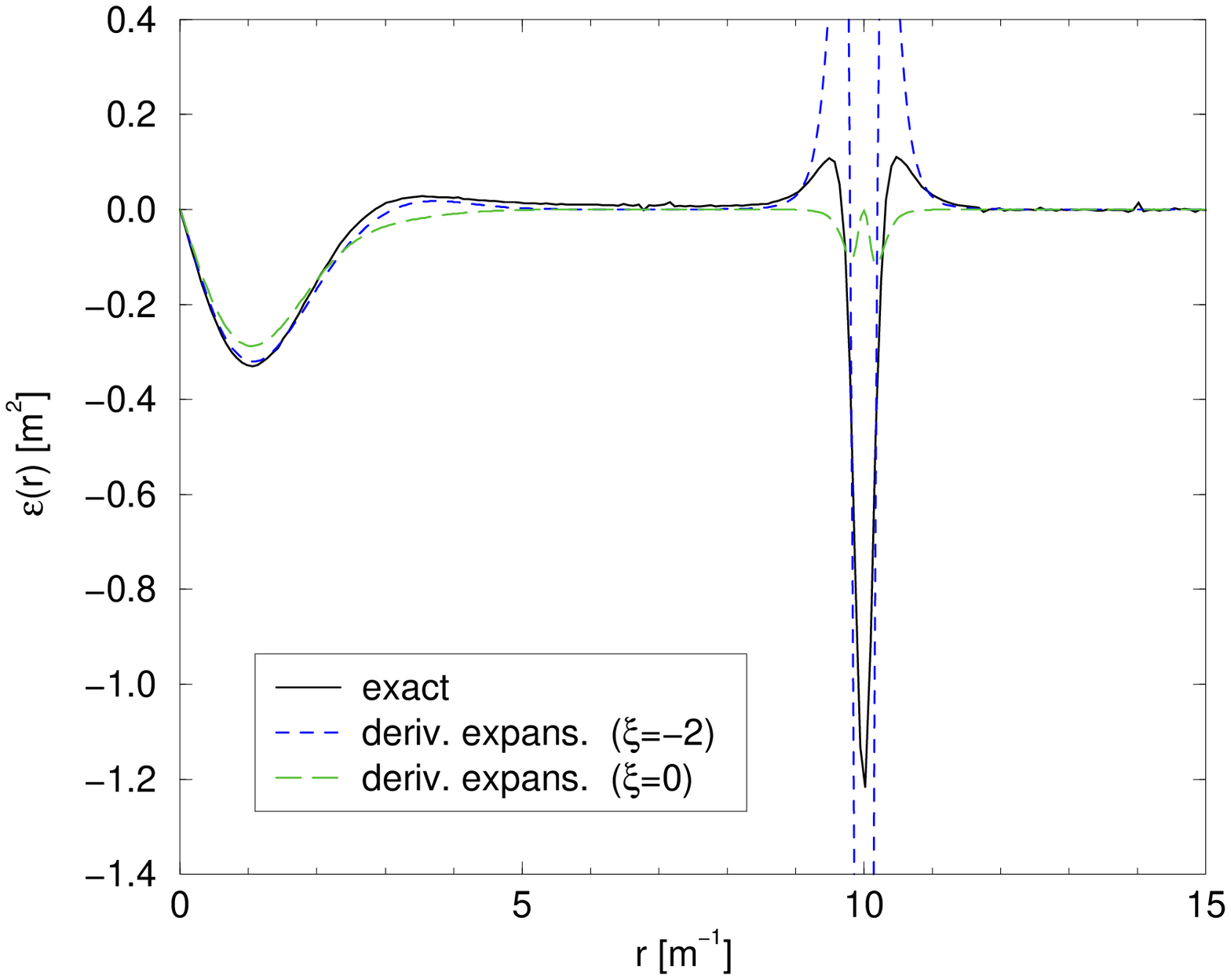}
\hskip1cm
\includegraphics[width=7cm,height=5cm]{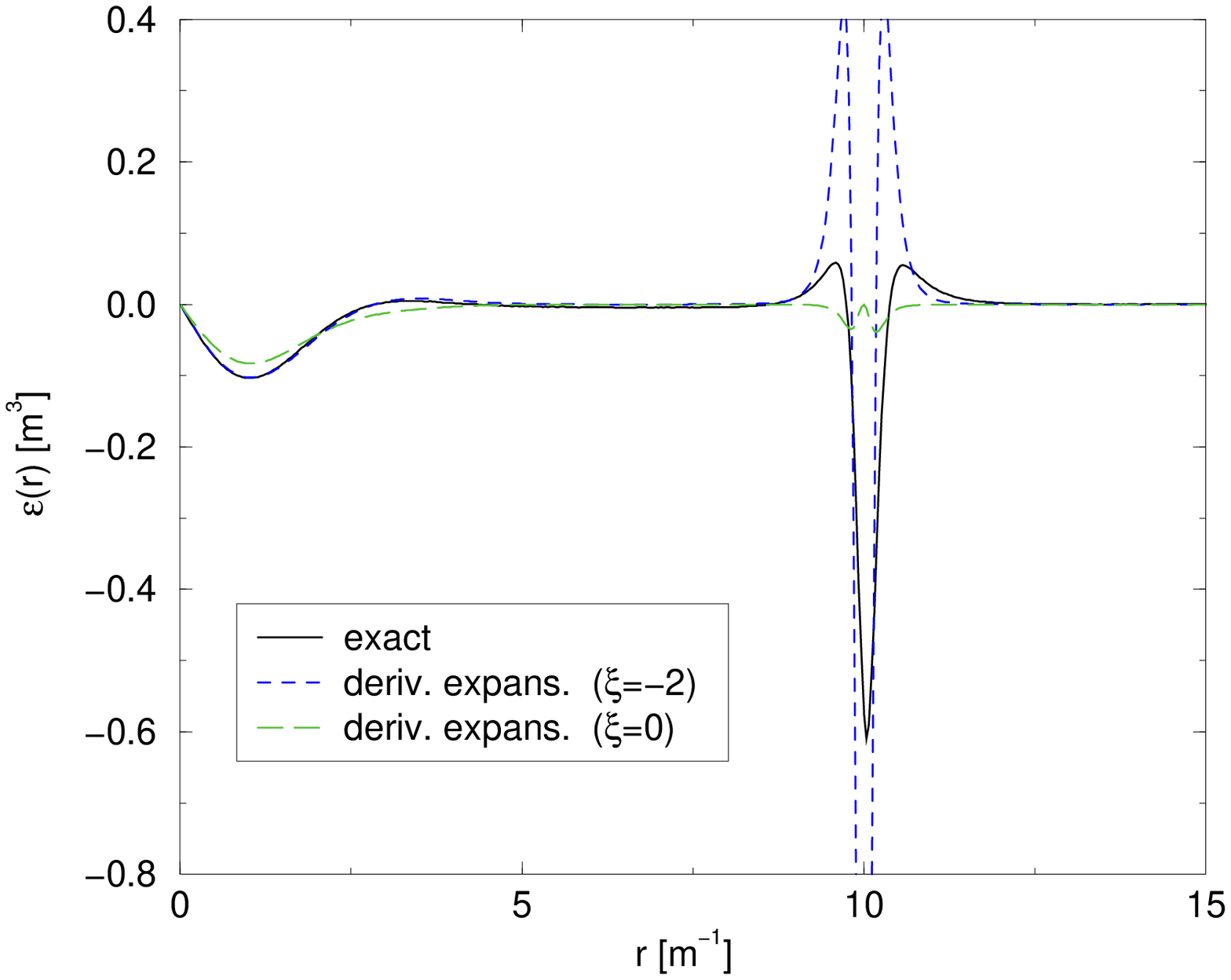}}
\caption{\label{fig_6.4}\sl Comparison between the computed 
energy density for configurations with zero net flux and the
derivative expansion approximation. We take $w=2.5/m$, $eB(0)=4m^2$ 
and $R=10/m$ and consider the case without ($\xi=0$) and
with ($\xi=-2$) total derivative term,~(\ref{eq:defepsDE}). 
The left and right panels show $D=2+1$ and $D=3+1$, respectively.}
\end{figure}

For the derivative expansion to be applicable we need to consider 
increasing widths~$w$ with $B(0)$ unchanged.  As discussed above, for
such configurations we may not omit the scattering data contribution
and therefore we consider configurations with zero net flux. The
comparison of the energy density is displayed in figure~\ref{fig_6.4}. 
We find reasonable agreement 
with the derivative expansion for sizable $w$ in the region of the 
central flux tube for both $D=2+1$ and $D=3+1$, where the magnetic
field is slowly varying.  Since we keep the position
and the width of the return flux unchanged as we increase the width
of the flux tube, we do not expect the derivative expansion
to match the exact result in the vicinity of the return flux.
From figure~\ref{fig_6.4} we observe that $\xi=-2$ appears optimal for
both $D=2+1$ and $D=3+1$.
\begin{figure}
\centerline{
\includegraphics[width=7cm,height=5cm]{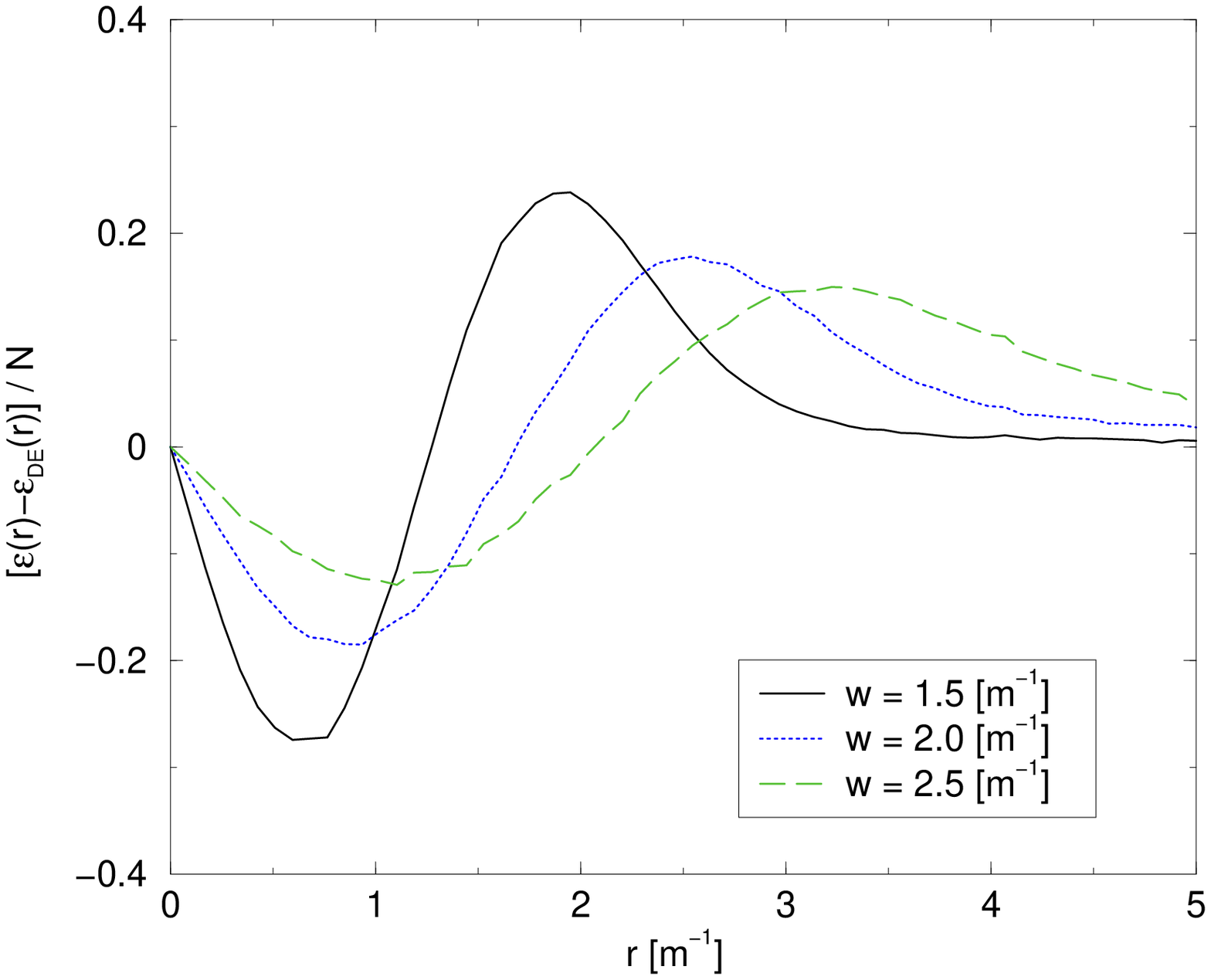}
\hskip1cm
\includegraphics[width=7cm,height=5cm]{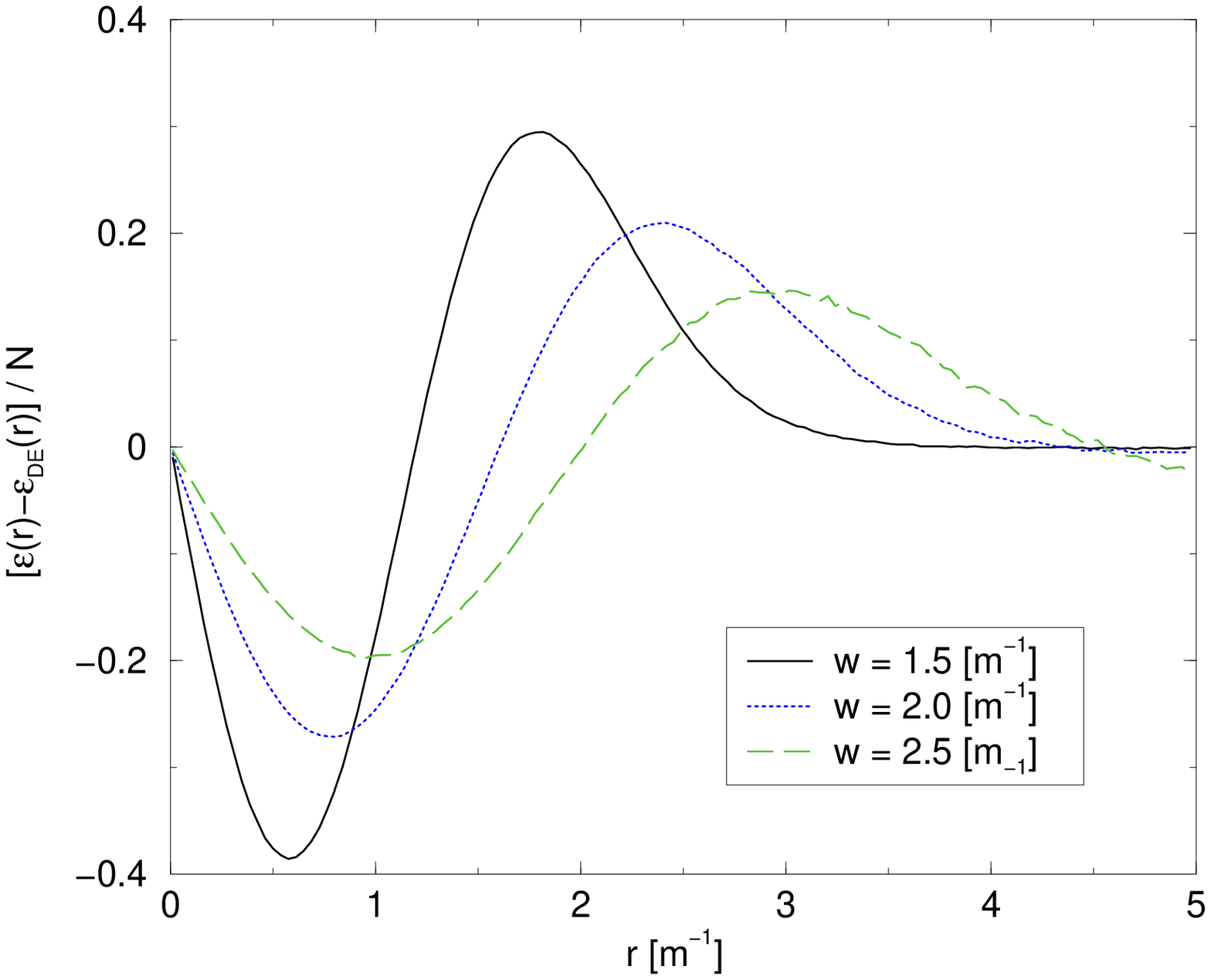}}
\centerline{
\includegraphics[width=7cm,height=5cm]{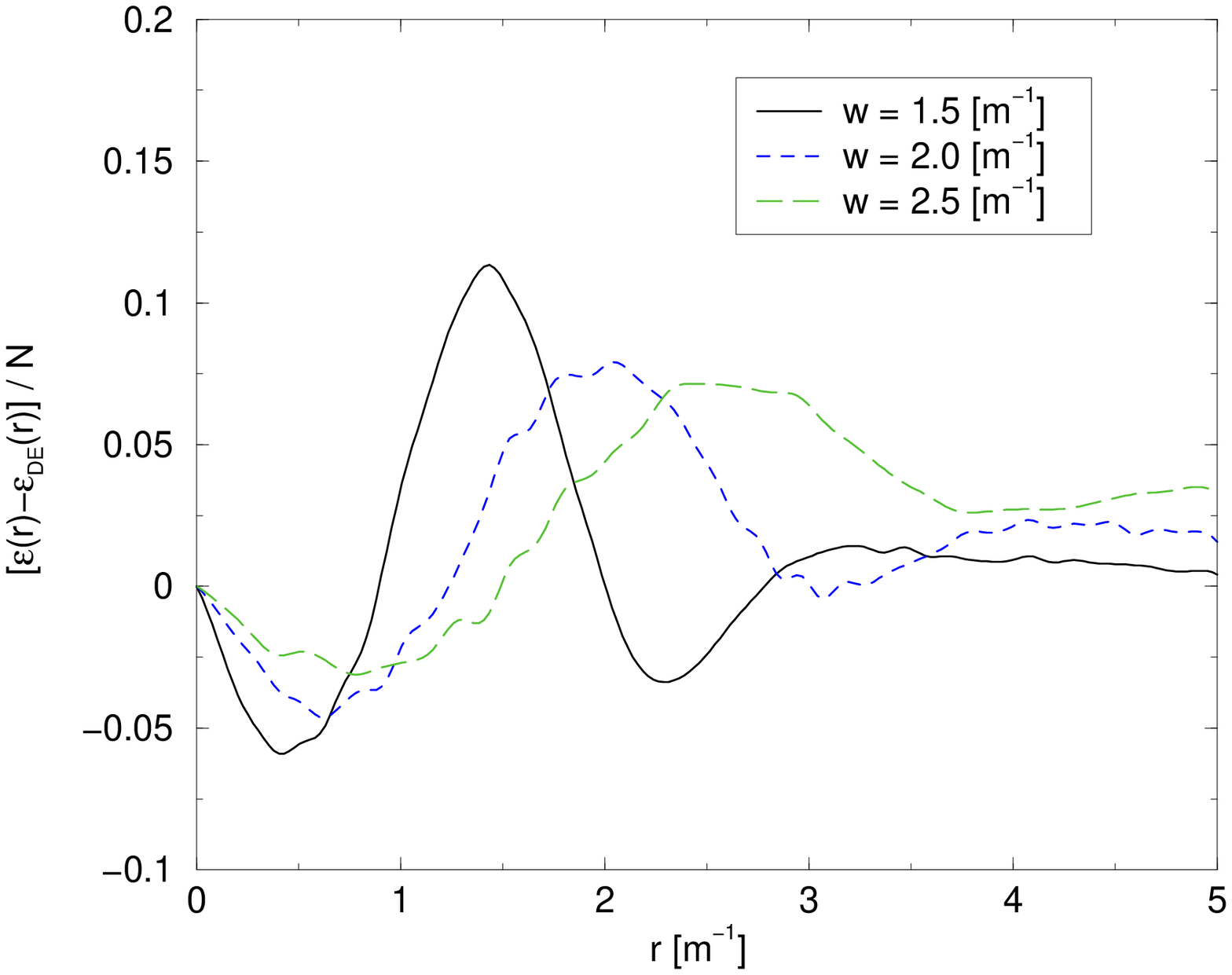}
\hskip1cm
\includegraphics[width=7cm,height=5cm]{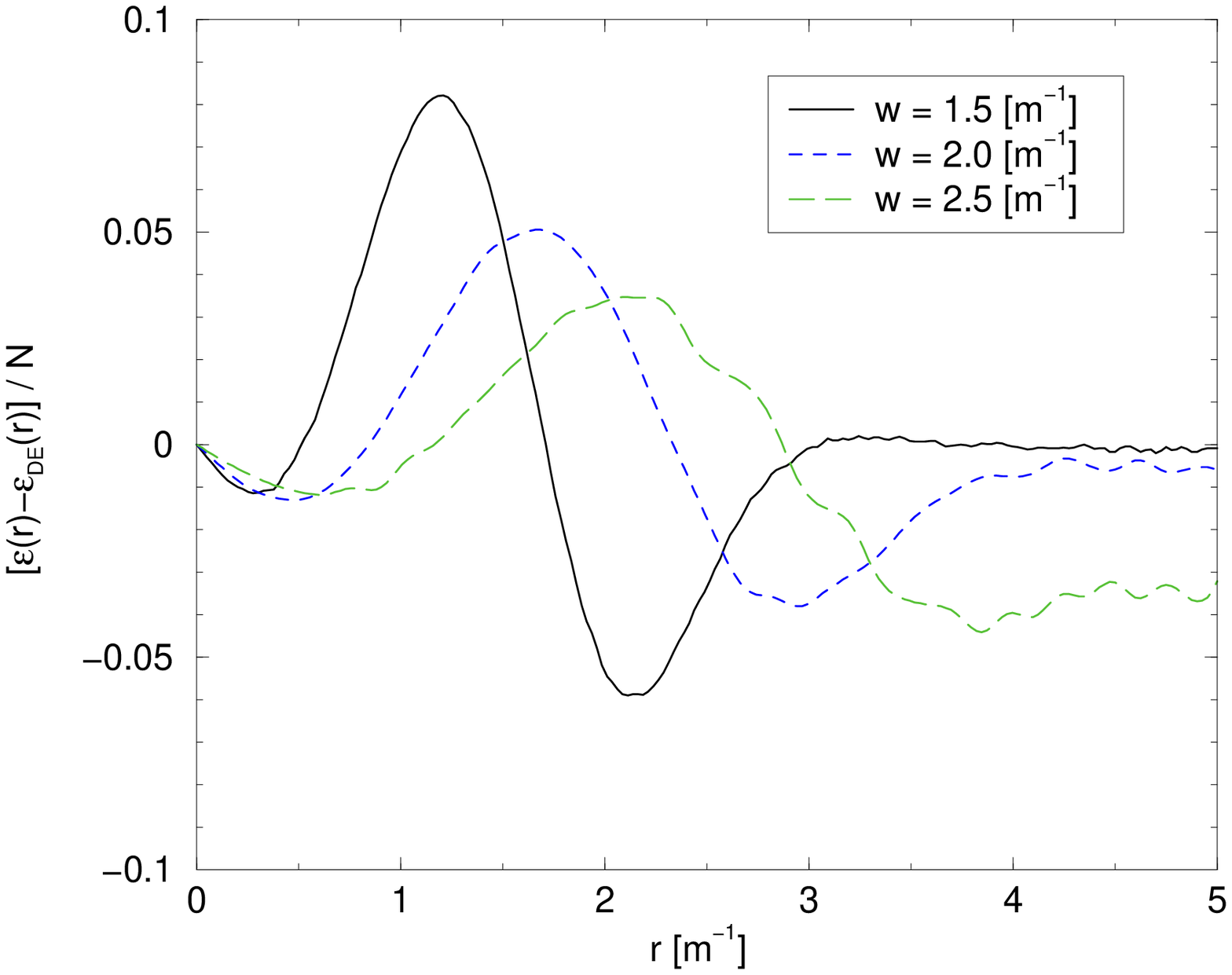}}
\caption{\label{fig_6.4b}\sl The difference between the exact
result and derivative expansion for the
energy density in the vicinity of the central flux tube 
as a function of its width. These differences are always
normalized to maximal absolute value of the exact energy
density in the flux tube region.  The left and
right panels show $D=2+1$ and $D=3+1$, respectively.
Also we consider $\xi=0$ (top) and $\xi=-2$ (bottom). Note
the different scale of the latter.}
\end{figure}
This result confirms the assertion of Appendix~\ref{Feynmanappendix}
that the energy and action densities differ by total derivatives.

In fig.~\ref{fig_6.4b} we display the normalized difference 
$\left[\epsilon(r)-\epsilon_{\rm DE}(r)\right]/N$ with the
normalization factor
$N={\rm max}\{|\epsilon(r)|,r\le 2w\}$ for various widths $w$. 
 Of course, we have to ensure that there
is no overlap with the return flux when we consider large
values of $w$. For $w=2.5/m$ (the largest value we consider) this
requirement is well satisfied for $R=10/m$.  As expected, this normalized
difference decreases as $w$ increases.   A non-zero
difference extends to larger $r$ as $w$ increases, reflecting
the growing extension of the central flux tube. Again, we observe
that the inclusion of the total derivative term improves
the agreement with derivative expansion approximation.
As can be clearly seen from fig.~\ref{fig_6.4b}, 
even though the difference between the derivative expansion and the
exact result for the density is quite sizable at small widths, it is
close to a total derivative, which explains the good convergence of
the derivative expansion for the \emph{total} energy, even at moderate
widths $w$.

To summarize, we find reasonable
agreement for the energy density between our exact results and 
the derivative expansion approximation at next--to--leading order 
for both $D=2+1$ and $D=3+1$.  Our result for the $D=3+1$ 
case contradicts the findings of ref.~\cite{Langfeld:2002vy}.
That work uses the world-line formalism as a non-perturbative
computation of the action density and finds a discrepency with 
the derivative expansion approximation in $D=3+1$.  We note that 
this discrepency cannot originate from the total derivative 
terms in the derivative expansion, since that work also 
considered the action density.  Also, we have seen that 
the total derivative terms affect $D=2+1$ and $D=3+1$ similarly.

\begin{figure}
\centerline{
\includegraphics[width=9cm,height=5cm]{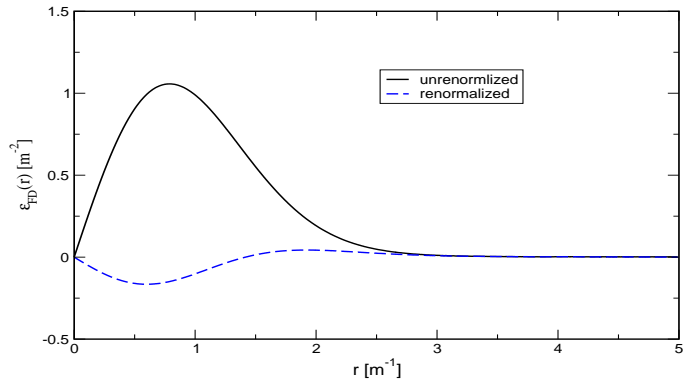}}
\caption{\label{fig_6.5}\sl Comparison of renormalized and
unrenormalized $2^{\rm nd}$ order Feynman diagram contribution
to the energy density for $D=2+1$. For simplicity we set
$B_R\equiv0$ and take $eB(0)=4m^2$ and $w=1.5/m$.}
\end{figure}

On the other hand, in ref.~\cite{Langfeld:2002vy}, a fixed flux
sequence of magnetic fields has been employed to test the
derivative expansion. We have already argued that a fixed
peak magnetic field configuration should be used instead. 
Thus one might wonder why ref.~\cite{Langfeld:2002vy} finds a good
agreement for a fixed flux configuration in the $D=2+1$ case.
We ascribe this result to the choice of renormalization scheme, in
which the finite counterterm has been omitted.  For both the energy
density and the total  energy, we find that the inclusion of  the
counterterm leads to considerable changes in the case of $D=2+1$.
Since the counterterm only affects the second-order contribution to
the energy and energy density, it is sufficient to consider the
corresponding Feynman diagrams to study the influence of
renormalization for $D=2+1$. This comparison is shown in figure~\ref{fig_6.5}. 
The renormalized energy density is about an order of 
magnitude smaller and of opposite sign than the unrenormalized one. 
Thus if one does not add a counterterm in $D=2+1$, the leading order
term $E_{\rm DE,0}$ contains a large, though unphysical, contribution
to the energy and energy density.  But then the comparision with the
derivative expansion merely tests the agreement in this term, which
masks the underlying differences between the exact and approximate results.
We thus conjecture that the authors of ref.~\cite{Langfeld:2002vy}
would also have found disagreement between their results and the
derivative expansion approximation in $D=2+1$ if they had imposed 
the on--shell renormalization condition for their fixed flux 
background configuration. In substance, the (dis)agreements
between the derivative expansion and the world line formalism
observed in ref.~\cite{Langfeld:2002vy} do not reflect short-comings
of either of these approaches but merely the inappropriateness
of the fixed flux configuration to check these approaches 
against each other.

\subsection{Charge Density}

\begin{figure}
\centerline{\parbox[t]{2.5cm}{~}
\includegraphics[width=6cm,height=8.5cm,angle=270]{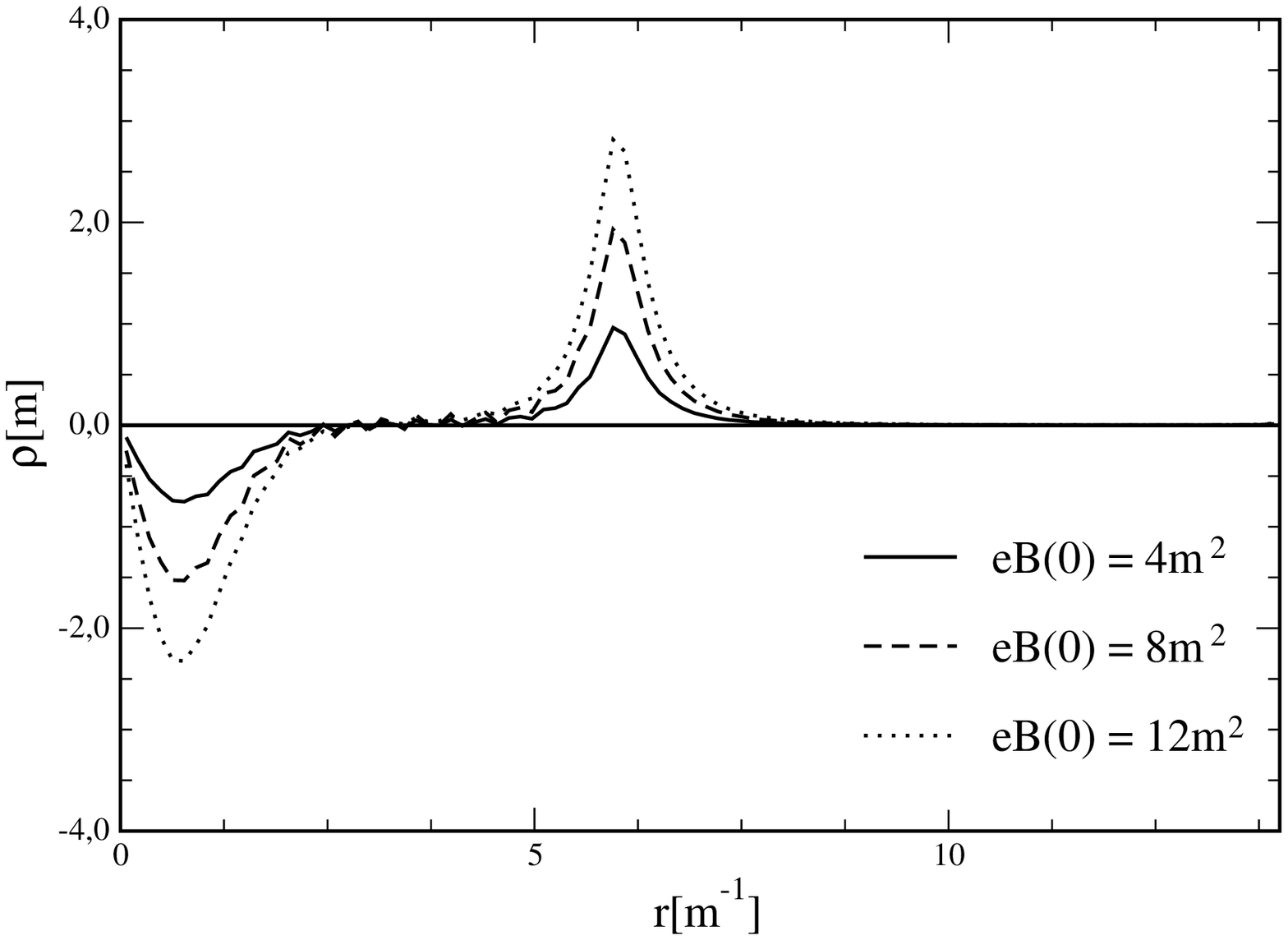}
\includegraphics[width=6cm,height=8.5cm,angle=270]{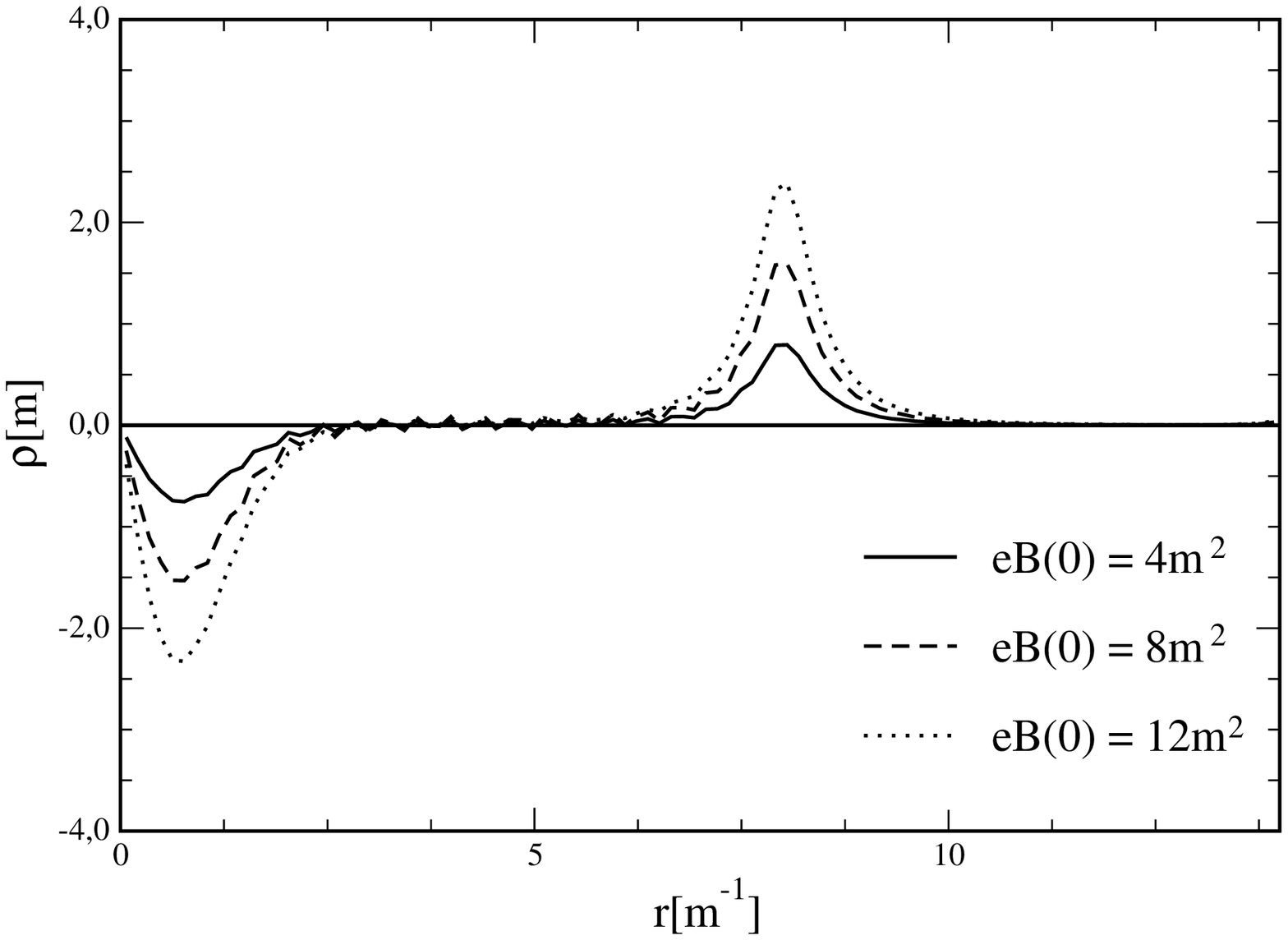}}
\caption{\label{fig_charge}\sl The charge density computed 
from eq.~(\ref{defcharge}) for various locations of the return
flux and strength of the magnetic field.
The return flux is at $6/m$ in the left panel and at
$8/m$ in the right panel.}
\end{figure}

\begin{figure}
\centerline{\hskip0.5cm
\includegraphics[width=6cm,height=7cm,angle=270]{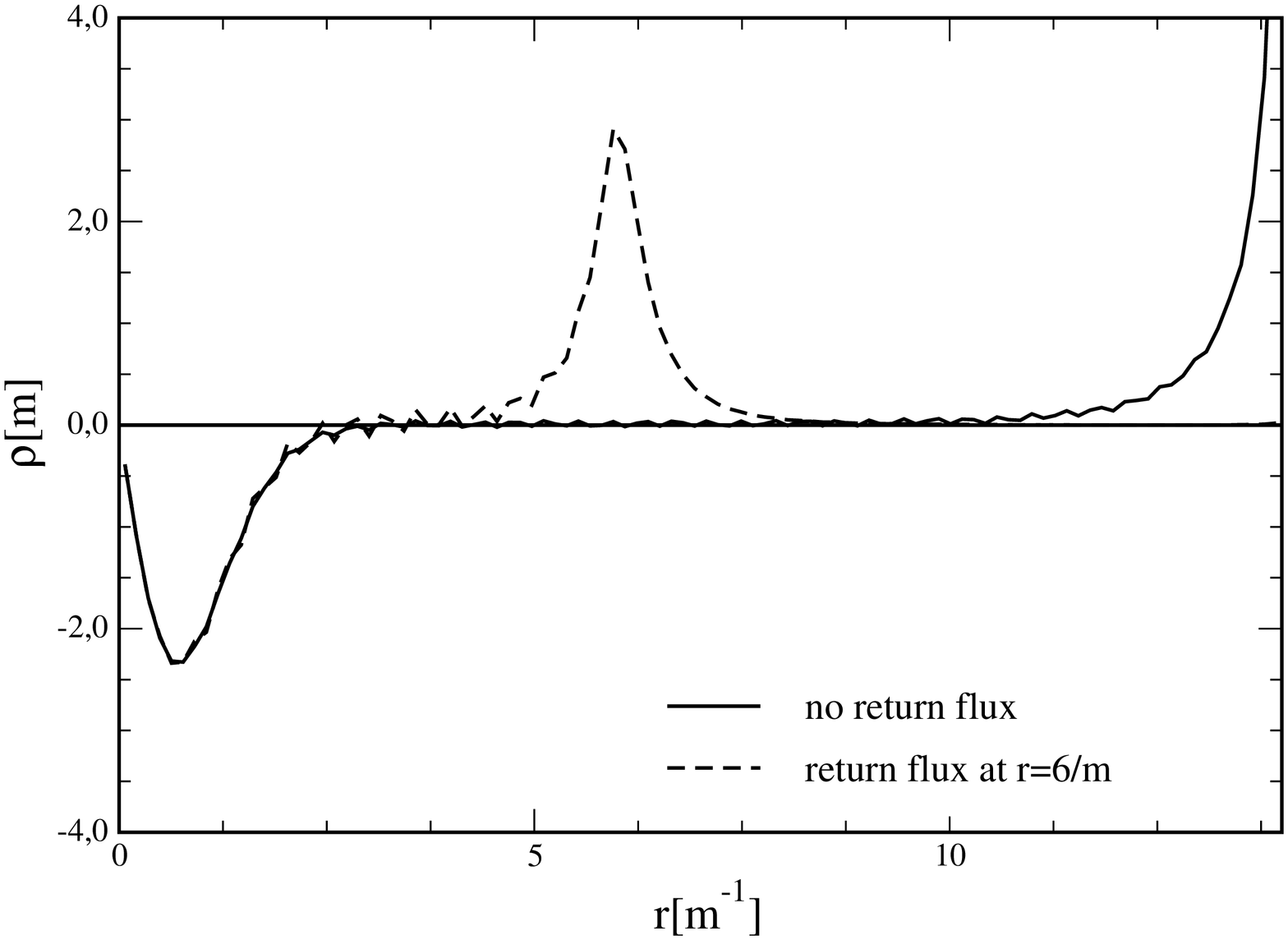}~~
\includegraphics[width=6cm,height=7cm,angle=270]{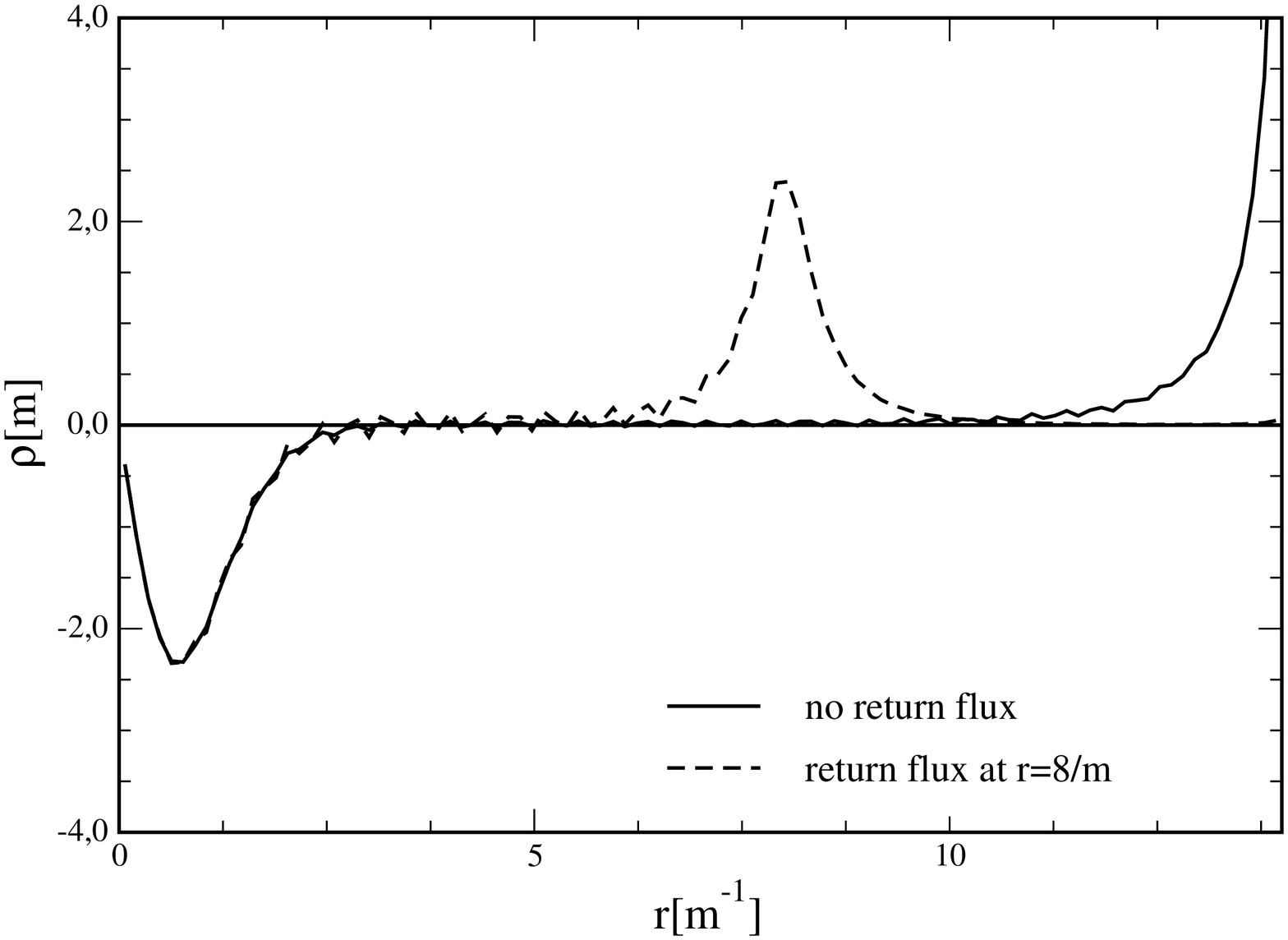}}
\caption{\label{fig_chargenr}\sl The charge density computed
from eq.~(\ref{defcharge}) with and without return flux.  In both
cases $eB(0) = 12m^2$.}
\end{figure}

In this subsection, we concentrate on ${\rm QED}_{2+1}$ with a single 
two--component fermion since for a four--component fermion the charge
density vanishes identically.\footnote{For two--component spinors we have 
${\rm tr}_2\left(\gamma_\mu\gamma_\rho\gamma_\sigma\right)=\pm
2i\epsilon_{\mu\rho\sigma}$ where the sign depends on the chosen
representation. The four--component spinors may be understood as the
sum of these two possible representation and hence obey
${\rm tr}_4\left(\gamma_\mu\gamma_\rho\gamma_\sigma\right)=0$.}
Since the charge density is a component of a conserved current, it is
not renormalized by quantum effects.  Therefore, we can diagonalize the
Dirac Hamiltonian within a Hilbert space that contains a large 
but finite number of states. 
We then simply compute
\begin{equation}
\rho (r) =  r\, \sum_{j} {\rm sign}(\epsilon_j)\,
\int_0^{2\pi}d\varphi\,
\Psi_j^\dagger(\vec{x\,}) 
\Psi_j(\vec{x\,})
\label{defcharge}
\end{equation}
where $\epsilon_j$ and $\Psi_j(\vec{x\,})$ are the
eigenvalues and eigenspinors respectively. 
The technical details of this calculation are summarized in 
Appendix~C.  We have verified the stability of the 
sum~(\ref{defcharge}) with respect to variation of the cut-off
that restricts the Hilbert space. In figure~\ref{fig_charge} 
we display the resulting densities for the background field defined 
in eq.~(33).
Clearly the charge density is localized at the regions
of the flux tube and return flux.  As these regions are separated,
so are the peaks of the charge density. While the total charge vanishes,
the integral up to some intermediate point $r_M\approx R/2$, 
with $\rho(r_M)\approx0$ gives
\begin{equation}
\int_0^{r_M}dr\,\rho(r)\approx - \frac{\res{\flux}}{2}\,
\label{halfcharge}
\end{equation}
exactly as expected from a single vortex located at $r=0$
\cite{Kiskis97,Blankenbecler:1986ft}.
In figure~\ref{fig_chargenr}, we display the charge density
obtained from the same calculation for background fields without
return flux.\footnote{The calculation as outlined in Appendix~A
is a unitary transformation on states describing a CP--invariant
spectrum. Thus the computed total charge will always vanish. 
However, in the case with non--zero net flux, the compensating
contribution arises from a peak at a position that in the
numerical treatment corresponds to spatial infinity, as shown in 
figure~\ref{fig_chargenr}, and is thus considered unphysical.}
We observe that in the vicinity of the central vortex, the
resulting densities with and without return flux are identical,
again verifying that we may consider the vortex as
part of a configuration with vanishing total flux. 

\begin{figure}
\centerline{\hskip0.5cm
\includegraphics[width=6cm,height=9cm,angle=270]{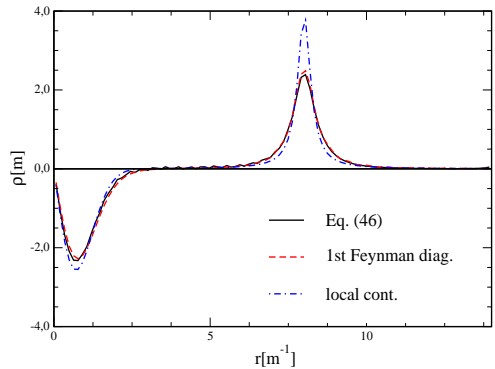}}
\caption{\label{fig_rhofd}\sl The charge density computed
from the Feynman diagram~(\ref{rhofd}) compared to the 
calculation from eq.~(\ref{defcharge}) and the local contribution
$\frac{{\cal F}}{2} \frac{d f(r)}{dr}$.  Here $eB(0) = 12m^2$ and the return
flux is at $8/m$.}
\end{figure}

It is illuminating to study the charge density using the
Green's function formalism as we did for the energy density
in the previous Section, using the results of Appendix A.  For background
configurations with vanishing total flux, the solutions to Dirac
equation come in pairs with opposite signs of the energy eigenvalues.
From equations~(\ref{poseng}) and~(\ref{negeng}) in Appendix~A 
we then find that the scattering states obey
\begin{equation}
\sum_j{\rm sign}(\epsilon_j)\,\frac{1}{2}
\left[\Theta(|\epsilon_j|-\delta)-\Theta(|\epsilon_j|+\delta)\right]
\int_0^{2\pi}d\varphi\, \Psi_j^\dagger(\vec{x\,}) \Psi_j(\vec{x\,})=0\,.
\label{scatteringzero}
\end{equation}
for arbitrarily small but non--vanishing $\delta$.
Hence they do not contribute to the charge density. Of course,
this argument applies to any order of the Born expansion as well.
However, to apply our Green's function formalism we have to subtract
enough Born terms that integrations in the complex momentum plane
become well defined.  For the charge density in $D=2+1$ only the
first-order subtraction is required. The corresponding Feynman diagram
gives
\begin{equation}
\rho_{\rm FD}(r)=r {\cal F}
\int_0^\infty dp \left[\int_0^\infty dr^\prime \,
\frac{d f(r^\prime)}{dr^\prime}\, J_0(pr^\prime)\right]{\rm arctan}
\left(\frac{p}{2m}\right) J_0(pr)\,.
\label{rhofd}
\end{equation}
Since we argued that the scattering Green's function does 
not contribute to the charge density we expect the charge to be
given exactly by this Feynman diagram. This result is confirmed
by numerical calculations, shown
in figure~\ref{fig_rhofd}.  We also compare to the local 
contribution, $\frac{{\cal F}}{2} \frac{d f(r)}{dr}$, which is the leading 
term of the derivative expansion to $\rho_{\rm FD}(r)$.
We see that the charge density contains sizable non--localities,
mostly in the vicinity of the return flux where the background is not
slowly varying.

\section{Summary}
We have studied exact one-loop quantum energies and energy densities
of static electromagnetic flux tubes in three and four spacetime
dimensions.  To this order only fermion fluctuations contribute to the
vacuum polarization energy and energy density, which we compute
exactly from scattering data.  In general, this quantum contribution
contains ultraviolet divergences and an important feature of our
approach is that it allows us to impose the standard
renormalization conditions of perturbative quantum 
electrodynamics.  Even though the calculation in three spacetime
dimensions does not suffer from such divergences, a meaningful
comparison between three and four dimensions
can only be made when identical renormalization conditions are
imposed.  Thus we must include a finite renormalization of the
three-dimensional result.  Furthermore, the use of 
scattering data to compute the vacuum polarization energy of
an individual flux tube leads to subtleties arising from the
long-range potential induced by the flux tube background,
which does not satisfy the standard conditions of scattering
theory.  Consequently, the scattering data do not necessarily have the
standard analytic properties and the phase shifts are discontinuous at
small momenta.  We have circumvented these problems by considering field 
configurations in which the flux tube is embedded with 
a well-separated return flux so that the total flux vanishes.
We have constructed a limiting procedure in which this return flux
does not contribute to the energy, enabling us to compute the energy
of an isolated flux tube.  While the return flux can give
a nontrivial contribution to the charge and energy densities (which 
integrates to zero in the case of the energy density), such
contributions are well separated from the flux tube and thus easily
identified.  Thus we have an unambiguous definition of the energy and
charge of the isolated flux tube even in the embedded configuration.
This embedding is analogous to considering a kink-antikink rather than
an isolated kink to avoid boundary effects in one dimensional $\phi^4$
theory \cite{Graham:1999pp}.

We do not see qualitative differences between 
three and four dimensions for either the energy or energy density,
once identical renormalization conditions have been imposed.  However,
we stress that renormalization in the case of three dimensions proved
essential to this result because the (finite) counterterm
contribution turned out to be large, thus causing sizable
cancellations in the final result.

We have tested various approximation schemes to the
vacuum polarization energy of the flux.  Though we have 
observed convergence of the perturbative expansion 
for small fluxes, we have concentrated on the derivative
expansion.  While we find good agreement between our exact 
calculation and the derivative expansion approximation for the
total energy, there are discrepancies for the energy density, which
arise because the derivative expansion result is an approximation to the
effective \emph{action}, which differs from
the appropriate matrix element of the energy momentum tensor by a
total derivative.

This study gives an initial step toward understanding flux tubes and
vortices in more complicated  theories.  The next step would be to
consider the  complete $U(1)$ model, including a Higgs field with
spontaneous symmetry breaking, which would describe flux tubes in a Type II
superconductor.  It could then be generalized to non-Abelian gauge bosons
for the study of $Z$-strings in the Standard Model, which are flux
tubes in the $Z$ gauge boson field.  The methods developed here could
then be used to determine whether quantum corrections stabilize the
classically unstable $Z$-string.

\subsection*{Acknowledgments}
We gratefully acknowledge helpful discussions with 
G.~V.~Dunne, E. Farhi, H.~Gies and R. L. Jaffe.
N. G. was supported in part by the Vermont Experimental Program to
Stimulate Competitive Research (VT-EPSCoR).
V. K. is supported in part by the U.S. Department of 
Energy (D.O.E.) under cooperative research agreements DF-FC02-94ER40818 
and DE-FG02-92ER40716.
O.S. was supported by the \textit{Deutsche Forschungsgemeinschaft}
under grant DFG Schr 749/1-1.  This work is also supported in part
by funds provided by the U.S. Department of Energy (D.O.E.) under
cooperative research agreement DF-FC02-94ER40818.

\appendix

\section{Energy Density from Green's Function}

\label{energyappendix}

In this Appendix we describe the construction of the exact
fermion Green's function in the flux tube background in terms of
scattering data.  We then use it to compute the quantum energy density
due to the fermion fluctuations.

We will begin by considering four--component Dirac spinors $\Psi$ 
in $D=3+1$ dimensions.  The calculation for two--component Dirac
spinors in $D=2+1$ dimensions is then a simple manipulation. For the
electromagnetic field $A^\mu=(0,A_\varphi\,\hat{e}_\varphi)^\mu$, the
$4\times4$ Dirac equation decomposes into $2\times2$ blocks
\begin{equation}
H\Psi=
\begin{pmatrix}
m & H_2 \cr  H_2 & -m
\end{pmatrix}\Psi =\omega\Psi\,,
\label{Dirac4}
\end{equation}
where we now use the Bjorken--Drell representation for the Dirac matrices.
Since the system is translationally invariant in the $z$ direction, we can  
choose $\Psi$ to be an eigenstate of $z$ momentum with eigenvalue $p$.
Then the $2\times2$ block becomes
\begin{eqnarray}
H_2&=&\begin{pmatrix}
\ p \ & \ L^\dagger \ \cr L &-p
\end{pmatrix}, 
\cr\cr
L&=&-i{\rm e}^{i\varphi}\left[\partial_r+eA_\varphi(r)
+\frac{i}{r}\partial_\varphi\right], \quad
L^\dagger=-i{\rm e}^{-i\varphi}\left[\partial_r-eA_\varphi(r)
-\frac{i}{r}\partial_\varphi\right]\,.
\label{Dirac2}
\end{eqnarray}
where $r=\sqrt{x^2+y^2}$.
Accordingly we decompose the four-spinor into two two-spinors
$\Psi=\begin{pmatrix}\xi ~~ \eta\end{pmatrix}$, which obey the
second-order equations
\begin{equation}
H_2^2\eta=\left(\omega-m\right)H_2\xi=\left(\omega^2-m^2\right)\eta
\hbox{\quad with \quad}
H_2^2=\begin{pmatrix}p^2+L^\dagger L & 0 \cr 0& LL^\dagger+p^2
\end{pmatrix}
\label{diagh22}
\end{equation}
and similarly for $\xi$.  Thus we only have to diagonalize $L^\dagger L$
and $LL^\dagger$ by solving
\begin{eqnarray}
\label{LL}
L^\dagger L \left[g_\ell^{(1)}(r){\rm e}^{i\ell \varphi}\right]&=&
{\rm e}^{i\ell \varphi}\left\{-\partial_r^2-\frac{1}{r}\partial_r
+\frac{\left(l-h(r)\right)^2}{r^2} -
\frac{h^\prime(r)}{r}\right\}g_\ell^{(1)}(r)\cr
&=&k^2 \left[g_\ell^{(1)}(r){\rm e}^{i\ell \varphi}\right] \\ 
LL^\dagger \left[g_\ell^{(2)}(r){\rm e}^{i(\ell+1)\varphi}\right]&=&
{\rm e}^{i(\ell+1)\varphi}\left\{-\partial_r^2-\frac{1}{r}\partial_r
+\frac{\left(l+1-h(r)\right)^2}{r^2} 
+ \frac{h^\prime(r)}{r}\right\}g_\ell^{(2)}(r)\cr
&=&k^2\left[g_\ell^{(2)}(r){\rm e}^{i(\ell+1)\varphi}\right]\,,
\nonumber
\end{eqnarray}
where $h(r)=erA_\varphi(r)=\res{F}f(r)$, as in eqs.~(\ref{eq:vortex})
and~(\ref{flux1}). Note that $\ell$ takes positive and  
negative integer values.  

The equations in~(\ref{LL}) represent problems in ordinary scattering theory,
with potentials
\begin{eqnarray}
V_1^{(\ell)}(r)&=&\frac{h(r)^2}{r^2} - 
2 \ell\frac{h(r)}{r^2}+\frac{h^\prime(r)}{r} \cr
V_2^{(\ell)}(r)&=&\frac{h(r)^2}{r^2}-2(\ell+1)\frac{h(r)}{r^2} -
\frac{h^\prime(r)}{r}
\label{background}
\end{eqnarray}
and their solutions possess the standard analytic properties in the
complex $k$ plane if these potentials obey the usual
conditions\cite{Newton:1982qc}.  This
is the case for the embedded flux tube, where we can use the techniques of
ref.~\cite{Graham:2002xq} to construct solutions for~$g_\ell^{(i)}$
that obey the Jost and scattering boundary conditions.  Since the
rotation to the imaginary $k$ axis removes the explicit contribution
of any bound states to the energy density\cite{Bordag:1998tg},  it is
sufficient to concentrate on the scattering solutions.

We construct the scattering eigenspinors of eq.~(\ref{Dirac4})
in analogy to the plane wave solution to a free Dirac field.
The eigenvalues of the full Dirac problem are
$\omega=\pm\sqrt{k^2+p^2+m^2}$, and we define the positive quantity
$E=\sqrt{k^2+p^2+m^2}$.  Spinors with eigenvalue $\omega=+E$ are
\begin{eqnarray}
\Psi_\ell^{(1)}(k,p;r,z)&=&\sqrt{\frac{E+m}{4\pi E}} {\rm e}^{ipz}
\begin{pmatrix}
g_\ell^{(1)}(k;r){\rm e}^{i\ell \varphi} \cr 0 \cr
\frac{p}{E+m}g_\ell^{(1)}(k;r){\rm e}^{i\ell \varphi}\cr
\frac{L}{E+m}g_\ell^{(1)}(k;r){\rm e}^{i\ell\varphi}
\end{pmatrix}\\
\label{poseng} \cr
&=&\sqrt{\frac{E+m}{4\pi E}}\,{\rm e}^{i\ell \varphi} {\rm e}^{ipz}
\begin{pmatrix}
g_\ell^{(1)}(k;r)\cr 0 \cr \frac{p}{E+m}g_\ell^{(1)}(k;r)\cr
\frac{-i{\rm e}^{i\varphi}}{E+m}L_r^{(-)}g_\ell^{(1)}(k;r)\cr
\end{pmatrix}\cr\cr
\Psi_\ell^{(2)}(k,p;r,z)&=&\sqrt{\frac{E+m}{4\pi E}}\,
{\rm e}^{i(\ell+1)\varphi} {\rm e}^{ipz}
\begin{pmatrix}
0\cr -g_\ell^{(2)}(k;r) \cr 
\frac{i{\rm e}^{-i\varphi}}{E+m}L_r^{(+)}g_\ell^{(2)}(k;r)\cr
\frac{p}{E+m}g_\ell^{(2)}(k;r)
\end{pmatrix}\,,
\nonumber
\end{eqnarray}
where we have introduced the linear differential operators
\begin{equation}
L_r^{(+)}=\partial_r+\frac{\ell+1-h(r)}{r}\qquad
{\rm and} \qquad
L_r^{(-)}=\partial_r-\frac{\ell-h(r)}{r}\,.
\label{raddiff}
\end{equation}
These equations exhibit the symmetry
\begin{equation}
g_{-\ell-1}^{(1)}(k;r)[h]=g_\ell^{(2)}(k;r)[-h]\,.
\label{symm}
\end{equation}
Similarly, the spinors for the anti--fermions with  eigenvalues $\omega=-E$ are
\begin{eqnarray}
\Psi_\ell^{(3)}(k,p;r,z)&=&\sqrt{\frac{E+m}{4\pi E}}
{\rm e}^{i\ell\varphi}  {\rm e}^{ipz}
\begin{pmatrix}
\frac{-p}{E+m}g_\ell^{(1)}(k;r)\cr
\frac{i{\rm e}^{i\varphi}}{E+m}L_r^{(-)}g_\ell^{(1)}(k;r)\cr
g_\ell^{(1)}(k;r)\cr 0
\end{pmatrix}\cr\cr
\Psi_\ell^{(4)}(k,p;r,z)&=&\sqrt{\frac{E+m}{4\pi E}}\,
{\rm e}^{i(\ell+1)\varphi}  {\rm e}^{ipz}
\begin{pmatrix}
\frac{i{\rm e}^{-i\varphi}}{E+m}L_r^{(+)}g_\ell^{(2)}(k;r)\cr
\frac{p}{E+m}g_\ell^{(2)}(k;r)\cr
0 \cr g_\ell^{(2)}(k;r)
\end{pmatrix}\,.
\hspace{0.2cm}
\label{negeng}
\end{eqnarray}
Assuming the normalization condition for boson wavefunctions
as in ref.~\cite{Graham:2002xq}
\begin{equation}
\int_0^\infty r dr \,
g_\ell^{(i)*}(k^\prime;r)g_\ell^{(i)}(k,r)=\frac{\pi}{2}\delta(k^\prime-k)
\qquad i=1,2
\label{bosnorm}
\end{equation}
the spinors obey
\begin{equation}
\int d^2r \,
\Psi_{\ell^\prime}^{(\alpha^\prime)\dagger}(k^\prime,p;r)
\Psi_\ell^{(\alpha)}(k;r)=\delta_{\alpha^\prime\alpha}
\delta_{\ell^\prime\ell}\frac{\pi}{2}\delta(k^\prime-k)\,.
\label{spinornorm}
\end{equation}

The solutions in the two-dimensional case are obtained by setting $p=0$.
Then the spinors are also eigenstates of ${\rm diag}(1,-1,-1,1)$,
reflecting the decomposition of the four-spinor into two decoupled 
two-spinors.

We would like to compute the renormalized radial energy density,
\begin{equation}
\epsilon(r)=2 \pi r
\langle \Omega |\hat{T}_{00}(x)|\Omega \rangle_{\rm ren}\,.
\label{engdens0}
\end{equation}
Here $|\Omega\rangle$ denotes the vacuum of fluctuating fermions that
are polarized by the background electromagnetic field, $A_\mu(x)$ and
\begin{equation}
\hat{T}_{00}(x)=\bar{\Psi}(x)\left(-i\gamma^i\partial_i
+e\Aslash+m\right)\Psi(x)
\label{t00}
\end{equation}
is the energy field operator.  The polarized vacuum is annihilated
by the operators $a_\ell^{(s)}(k,p)$ and $b_\ell^{(s)}(k,p)$ that appear
in the decomposition of the field operator
\begin{eqnarray}
\Psi(x)&=&\int\frac{dp}{\sqrt{2\pi}}
{\rm e}^{ipz}\,\int_0^\infty \frac{dk}{\sqrt{\pi E}}
\sum_{\ell=-\infty}^\infty \sum_{s=1}^2
\Big[\Psi_\ell^{(s)}(k,p;r)\,{\rm e}^{-iEt}\,a_\ell^{(s)}(k,p)
\cr &&\hspace{7cm}
+\Psi^{(s+2)}_\ell(k,p;r)\,{\rm e}^{iEt}\,b_\ell^{(s)\dagger}(k,p)\Big]\,.
\label{fieldoperator}
\end{eqnarray}
Again, we have omitted potential bound state contributions.

The ultraviolet divergences in the matrix element~(\ref{engdens0})
are canceled by counterterms that we determine by suitable
renormalization conditions.  These divergences are at most
quadratic in the background field, $A_\varphi$. 
The Taylor expansion in $A_\varphi$ for $\left[\epsilon(r)\right]_2$,
the radial energy density with the first two Born approximations
subtracted, starts at cubic order and is free of ultraviolet 
divergences.  We will  therefore first compute
$\left[\epsilon(r)\right]_2$ and in the next Appendix add back the
subtracted pieces in form of Feynman diagrams combined with the counterterms.
We obtain $\left[\epsilon(r)\right]_2$  using the standard anti-commutation
relations for $a_\ell^{(s)}(k,p)$ and $b_\ell^{(s)}(k,p)$. Since the
spinors $\Psi_\ell^{(s)}(k,p;r)$ obey the Dirac equation this yields
\begin{equation}
\left[\epsilon(r)\right]_2=-r\int_0^{2\pi}d\varphi 
\int\frac{dp}{2\pi}\int_0^\infty\frac{dk}{\pi}
\sum_{\ell=-\infty}^{+\infty}\sum_{\alpha=1}^4
\sqrt{k^2+p^2+m^2}\,
\left[\Psi_\ell^{(\alpha)\dagger}(k,p;r)
\Psi_\ell^{(\alpha)}(k,p;r)\right]_2\,.
\label{engdens}
\end{equation}
We note that
$L_r^{(-)}g_\ell^{(1)}(k;r)$ and $g_\ell^{(2)}(k;r)$ satisfy the same 
linear second order differential equation, as do 
$L_r^{(+)}g_\ell^{(2)}(k;r)$ and $g_\ell^{(1)}(k;r)$. 
To relate one solution to the other, we need to consider the boundary
conditions.  The radial functions that enter the Green's
functions~(\ref{Greenfct}) obey physical  scattering
boundary conditions~\cite{Graham:2002xq}. Thus we identify
\begin{equation}
L_r^{(-)}g_\ell^{(1)}(k;r)=-ikg_\ell^{(2)}(k;r) \qquad
{\rm and}\qquad
L_r^{(+)}g_\ell^{(2)}(k;r)=ikg_\ell^{(1)}(k;r)\,,
\label{identification}
\end{equation}
which simplifies the sum in eq.~(\ref{engdens}) 
considerably.  Our main objective is to express the right hand
side of eq.~(\ref{engdens}) in terms of the Green's functions
for $g_\ell^{(1)}(k;r)$ and $g_\ell^{(2)}(k;r)$,
\begin{equation}
G^{(s)}_\ell(r,r^\prime,k)=-\frac{2r}{\pi}\int_0^\infty dq\,
\frac{g_\ell^{(s)*}(q;r)g_\ell^{(s)}(q;r^\prime)}
{(k+i\epsilon)^2-q^2}\,,\qquad s=1,2 \,.
\label{Greenfct}
\end{equation}
by noting that the imaginary part of the Green's 
function at coincident points may be related to
the wavefunctions,
$$
\mathsf{Im}G_\ell^{(s)}(r,r,k)=\frac{r}{k}\,|g_\ell^{(s)}(k;r)|^2\,.
$$
Collecting pieces finally yields the energy density 
\begin{equation}
\left[\epsilon(r)\right]_2=-2\int_{-\infty}^\infty\frac{dp}{2\pi}
\int_{-\infty}^\infty\frac{dk}{2\pi i}
\sum_{\ell=-\infty}^{+\infty}\sum_{s=1}^2
E \left[k\,G_\ell^{(s)}(r,r,k)\right]_2\,,
\label{engdens2}
\end{equation}
where we have also used that $\mathsf{Im}G_\ell^{(s)}(r,r,k)$ 
is odd in the momentum~$k$.  We have succeeded in finding the energy
density from fermion fluctuations in terms of bosonic Green's functions
$G_\ell^{(s)}(r,r^\prime,k)$. For further details on the decomposition
of these bosonic Green's functions into Jost
and regular  solutions we refer to ref.~\cite{Graham:2002xq}.  In that
paper also the computation of the corresponding phase shifts
$\delta_\ell^{(s)}$ is described. In the present case,
$\delta_\ell^{(1)}(k)$ and $\delta_\ell^{(2)}(k)$ actually refer to
the same Dirac spinor. They must be equal, which is
a consequence of the identification~(\ref{identification}).

It is also straightforward to obtain the expansion in the background
field, $A_\varphi$.  We iterate the differential equations
starting from free solutions obeying outgoing wave boundary conditions,
$g^{(s)}_\ell(k;r)\Big|_0=H^{(1)}_{\ell+s-1}(kr)$  \cite{Graham:2002xq}.

Once we have removed the ultraviolet divergences by subtracting
the leading contributions in $A_\varphi$, the integral in the complex 
$k$ plane over the semi-circle at infinity vanishes and we can compute 
the $k$ integral by contour integration. Then only branch cuts from 
multi-valued functions (such as $\sqrt{k^2+p^2+m^2}$) and poles due 
to bound states contribute.  In our calculations, any contributions
from bound state poles will be canceled by the explicit contributions of
the bound states themselves, so we need not consider them.  Thus if 
$f(p,k)$ is an analytic function in $k$ and goes to zero fast enough at
infinity that it does not contribute to the integral over the semicircle,
the replacement $E\to E-f(p,k)$ 
does not alter the value of the integral~(\ref{engdens2}).  In particular, 
we may choose $f(p,k)=\sqrt{p^2+m^2}+k^2/2\sqrt{p^2+m^2}$.
This modification allows us to carry out the $p$ integration first,
\begin{equation}
\left[\epsilon(r)\right]_2=\frac{1}{\pi}
\int_{-\infty}^\infty\frac{dk}{2\pi i}
\sum_{\ell=-\infty}^{+\infty}\sum_{s=1}^2
\omega_k^2\,{\rm ln}\left(\frac{\omega_k}{m}\right) 
\left[kG_\ell^{(s)}(r,r,k)\right]_2\,,
\label{engdens3}
\end{equation}
where $\omega_k=\sqrt{k^2+m^2}$ and
we have dropped pieces analytic in $k$ because they vanish by the
above argument as well.
In the upper half plane, ${\rm ln}(\omega)$ has a branch 
cut along the imaginary axis, starting at $k=im$, where it jumps 
by $i\pi$.  Therefore the energy density is given as
an integral along the imaginary $k$-axis,
\begin{equation}
[\epsilon(r)]_2=
\int_m^\infty\frac{dt}{2\pi}
\sum_{\ell=-\infty}^{+\infty}\sum_{s=1}^2
\left(t^2-m^2\right)\left[tG_\ell^{(s)}(r,r,it)\right]_2\,,
\label{engdens5}
\end{equation}
where $k=it$.

The integral of the Green's function is related to the Jost function
$F_\ell^{(s)}(k)$ by~\cite{Graham:2002xq}
\begin{equation}
2k\,\int_0^\infty dr \left[G^{(s)}(r,r,k)\right]_2
=-\frac{d}{dk}\,\left[{\rm ln}\, F_\ell^{(s)}(k)\right]_2\,\qquad
{\rm for}\,\, \mathsf{Im}k>0\,,
\label{jost1}
\end{equation}
which implies that the twice-subtracted
total energy obtained by integrating $\left[\epsilon(r)\right]_2$ 
in the form of eq.~(\ref{engdens3}) is given by the phase shift 
formula
\begin{eqnarray}
\left[E\right]_2=\int_0^\infty dr \left[\epsilon(r)\right]_2&=&
\int_0^\infty\frac{dk}{2\pi^2} \sum_{\ell=\infty}^\infty \sum_{s=1}^2 \,
\omega_k^2\,{\rm ln}\frac{\omega_k}{m}\,
\frac{d}{dk}\left[\delta_\ell^{(s)}(k)\right]_2\cr
&=&-\int_0^\infty\frac{dk}{2\pi^2} \sum_{\ell=-\infty}^\infty \sum_{s=1}^2\,
k\,{\rm ln}\frac{\omega_k^2}{m^2}\left[\delta_\ell^{(s)}(k)\right]_2\,
\label{engtotal}
\end{eqnarray}
where again terms that do not cause branch cuts in the 
complex plane have been omitted. Taking into account that 
the two phase shifts are identical we may define
$\bar{\delta}_{\ell+\half}(k)=4\left[\delta_\ell^{(1)}(k)\right]_2
=4\left[\delta_\ell^{(2)}(k)\right]_2$ to write
$$
\left[E\right]_2=-\int_0^\infty\frac{dk}{2\pi^2} 
k\,{\rm ln}\frac{\omega_k}{m}
\sum_{\ell=-\infty}^\infty\bar{\delta}_{\ell+\half}(k)\,,
$$
which is eq.~(\ref{eq:phaseshiftD4}) in the absence of
bound states.  (On the real axis, the bound state contribution does
not cancel, so to find it requires that we restore
the contribution from the analytic terms in the momentum integrands.)
The fourfold degeneracy for each momentum $k$ represents the 
two possible energies $\pm\omega_k$ and the two decoupled spin
channels $s=1,2$.

Finally let us consider the $D=2+1$ case. In that case 
we drop the $p$ integral in eq.~(\ref{engdens2}),
yielding
\begin{equation}
\left[\epsilon_{2D}(r)\right]_2=
-\int_{-\infty}^\infty\frac{dk}{\pi i}
\sum_{\ell=-\infty}^{+\infty}\sum_{s=1}^2
\sqrt{k^2+m^2}\,\left[k G_\ell^{(s)}(r,r,k)\right]_2\,.
\label{engdens2d1}
\end{equation}
Integrating along the branch cut we find
\begin{equation}
\left[\epsilon_{2D}(r)\right]_2=2\int_m^\infty\frac{dt}{\pi}
\sum_{\ell=-\infty}^{+\infty}\sum_{s=1}^2
\sqrt{t^2-m^2}\, [tG_\ell^{(s)}(r,r,it)]_2\,.
\label{engdens2d2}
\end{equation}
We recover the expression for the total energy, eq.~(\ref{eq:phaseshiftD3}),
similarly to the $D=3+1$ case using the spatial integral in
eq.~(\ref{jost1}).

\subsection{Phase shifts for non-zero flux}
\label{subsec:nonzeroNF}

When there is no net flux, the potential, eq.~(\ref{background})
approaches zero as $r\to\infty$ and it is convenient to parameterize
the radial fermion wavefunctions as
\begin{equation}
g_\ell^{(s)}(k;r)={\rm e}^{i\beta_\ell^{(s)}(k;r)}
H^{(1)}_{\ell+s-1}(kr)\,.
\label{netflux0}
\end{equation}
These {\it ans\"atze} induce non-linear
second order differential equations for the complex
radial functions $\beta_\ell^{(s)}(k;r)$.  Imposing the boundary 
condition $\lim_{r\to\infty}\beta_\ell^{(s)}(k;r)=0$, the
solutions to these differential equations yield
the phase shifts~\cite{Graham:2002fi}
\begin{equation}
\delta^{(s)}_\ell(k)=-\lim_{r\to0}\mathsf{Re}
\left(\beta_\ell^{(s)}(k;r)\right)\,.
\label{phase1}
\end{equation}
As usual, the Born expansion for the phase shifts is obtained
by expanding $\beta_\ell^{(s)}(k;r)$ in powers of the backgroud field.
The solution to the iterated differential equations 
for $\beta_\ell^{(n,s)}(k;r)$ then yields the $n$-th Born approximation
for the phase shifts $\delta^{(n,s)}_\ell(k)=-\lim_{r\to0}\mathsf{Re}
\left(\beta_\ell^{(n,s)}(k;r)\right)$ \cite{Graham:2002fi}.

In the case of non--zero flux, {\it i.e.} 
$\lim_{r\to\infty}h(r)={\mathcal F}\ne0$, the situation is
more complicated. The {\it ans\"atze}~(\ref{netflux0}) are 
changed to 
\begin{equation}
g_\ell^{(s)}(k;r)={\rm e}^{i\beta_\ell^{(s)}(k;r)}
H^{(1)}_{\ell+s-1+{\mathcal F}}(kr)\,,
\label{netflux1}
\end{equation}
to accommodate the modified asymptotic behavior.  The phase shifts
are again defined as in eq.~(\ref{phase1}), but the Born
expansion needs to be modified.  It is essentially a series in 
powers of ${\mathcal F}$, but in the product {\it ansatz}~(\ref{netflux1}),
both factors depend on ${\mathcal F}$.  For the computation of the
Born series it is therefore more useful to decompose as in
eq.~(\ref{netflux0}),
\begin{equation}
g_\ell^{(s)}(k;r)={\rm e}^{i\beta_{\ell \mathcal F}^{(s)}(k;r)}
H^{(1)}_{\ell+s-1}(kr)\,,
\label{netflux2}
\end{equation}
and expand the exponent according to
$\beta_{\ell \mathcal F}^{(s)}(k;r)=\sum_{n}
{\mathcal F}^n\beta_{\ell \mathcal F}^{(n,s)}(k;r)$ where $n$
labels the order in the Born series. Thus the $n$-th Born approximation
for the phase shift is formally the same
with and without net flux. However, for ${\mathcal F}\ne0$
the background potential approaches zero only slowly as $r\to\infty$.
Then the numerical treatment requires special care because we 
start integrating from large but finite $r_\infty$, where it is no
longer accurate enough to adopt the na{\"\i}ve boundary
conditions $\beta_{\ell \mathcal F}^{(n,s)}(k;r_\infty)=0$. Rather we
take \cite{Pasipoularides:2000gg}
\begin{eqnarray}
\beta_{\ell \mathcal F}^{(1,s)}(k;r_\infty)&=&
-\frac{1}{k}\frac{\ell+s-1}{r_\infty}\,, \qquad 
\frac{d}{dr}\beta_{\ell \mathcal F}^{(1,s)}(k;r)\Big|_{r=r_\infty}
=\frac{1}{k}\frac{\ell+s-1}{r^2_\infty}\,, \cr
\beta_{\ell \mathcal F}^{(2,s)}(k;r_\infty)&=&\frac{1}{2kr_\infty}
\,, \qquad
\frac{d}{dr}\beta_{\ell \mathcal F}^{(2,s)}(k;r)\Big|_{r=r_\infty}
=-\frac{1}{2kr_\infty^2}\,.
\label{bcflux}
\end{eqnarray}
to integrate the corresponding differential equations from 
$r=r_\infty$ towards $r=0$.

\subsection{Known properties of threshold states}
\label{subsec:threshold}
In this Appendix we review results for 
threshold states in magnetic backgrounds \cite{Aharonov:1979,
Jackiw:1984ji}. We concentrate on a single two-component spinor in
$2+1$ dimensions, obeying the Dirac equation in
eq.~(\ref{Dirac2}).  We parameterize the two-component spinor as
$\Psi=\left({\rm e}^{i\ell\varphi}g_\ell^{(1)}(r),
{\rm e}^{i(\ell+1)\varphi}g_\ell^{(2)}(r)\right)^{\rm T}$.  
At the positive energy threshold, $\omega=+m$, we find the solutions 
\be 
g_\ell^{(1)}(r) = r^{\ell}
\exp{\left(-\Phi(r)\right)}\quad {\rm and}\quad  g_\ell^{(2)}(r)\equiv 0 \,,
\ee 
where
\be
\Phi(r) = \int_0^{r}  d r' \frac{h(r')}{r'}=
\res{\flux} \int_0^{r}  d r' \frac{f(r')}{r'}\,.
\ee
At the negative energy threshold, $\omega=-m$, 
we find
\be 
g_\ell^{(2)}(r) = r^{-\ell-1} \exp{\left(\Phi(r)\right)}
\quad {\rm and}\quad g_\ell^{(1)}(r)\equiv 0\,.
\ee
In a system with net flux $\res{\flux}\ne0$, $f(r')$ approaches
unity as $r' \to \infty$, and thus $\Phi(r) \to \res{\flux}
\ln{r}$ as $r \to \infty$. For $r \to 0$ we demand
$f(r') \approx r'{}^2$, and thus $\lim_{r \to 0}
\Phi(r) \to 0$. Normalizability of the wave function thus
requires:
\bea
\omega= +m: & & \res{\flux} > \ell +1 > 0 \,, \\
\omega= -m: & & 0 > \ell >  \res{\flux}\,.
\label{eq:fluxcond}
\eea
Hence a positive (negative) flux is required to generate
threshold states at positive (negative) threshold. 
The same analysis can be performed for 3+1 dimensions, but there we
have two two-component spinors of opposite chirality.  As a result,
for a given flux we find equal numbers of states at the two thresholds in
that case.

The threshold states have a very close relation to the famous Landau
levels in a constant magnetic field $B_0$, where
the total number of threshold states is infinite.  For $B_0>0$, we
find positive energy threshold states
\be
g_\ell^{(1)}(r)=r^{\ell}{\rm e}^{-\frac{eB_0}{4} r^2}
\quad {\rm and}\quad g_\ell^{(2)}(r)\equiv0\,,
\ee
Analogously for $B_0<0$ we find at negative energy threshold states
\be
g_\ell^{(2)}(r)=r^{-\ell-1}{\rm e}^{\frac{eB_0}{4} r^2}
\quad {\rm and}\quad g_\ell^{(1)}(r)\equiv0\,.
\ee
Infinitely many threshold states emerge because these wavefunctions
are normalizable for any value $\ell$ that allows a regular solution
at $r=0$.   These solutions lead to the standard Landau density of
states, appearing at the appropriate threshold.

\section{Feynman Diagrams}
\label{Feynmanappendix}

We begin this Appendix by showing formally that
for any static electromagnetic background field,
the total energy obtained from the spatial integral 
of the vacuum matrix element of the energy density operator
equals minus the action per unit time, as one would expect.  We show
that this identification does not extend to the energy density, however.
Finally, we use these results to compute the first two terms in the
Feynman series for the flux tube background.

\subsection{Feynman series for the total energy}

We would like to develop a Feynman series for the total energy in a
static background.  It will contain divergent terms, so we must
carry out our calculation using a gauge-invariant regulator, such as
dimensional regularization.  Since we are considering only static
background fields, the time coordinate is special.  Thus we can use
dimensional regularization in one time dimension and $D-1$ space
dimensions, where only the latter may be fractional.

The energy density is obtained as the vacuum expectation value of the 
operator defined in eq.~(\ref{t00}). In order to generate 
the Feynman series we adopt the functional language,
and for simplicity we rescale the vector field $A_\mu$ to contain
the coupling constant $e$.  We have
\begin{eqnarray}
\langle\hat{T}_{00}(x)\rangle&=&
-i\,{\rm Tr}\left\{\left(i\Dslash-m\right)^{-1}
\delta(x-\hat{x})\left[-i\gamma^i\partial_i+\Aslash+m\right]\right\}
\cr
&=&-i\,{\rm Tr}\left\{\left(i\Dslash-m\right)^{-1}
\delta(x-\hat{x})\left[i\gamma^0\partial_0-i\Dslash+m\right]\right\}
\cr
&=&-i\,{\rm Tr}\left\{\left(i\Dslash-m\right)^{-1}
\delta(x-\hat{x})i\gamma^0\partial_0\right\}
\cr
&=&-i\,{\rm Tr}\left\{i\gamma^0\partial_0
\left(i\Dslash-m\right)^{-1}\delta(x-\hat{x})\right\}\,,
\label{ftrace}
\end{eqnarray}
where we have omitted $A_\mu$ independent terms because they are
canceled by subtracting the zeroth order term of the Born series,
the cosmological constant. The rest of the Feynman diagrams are 
obtained by expanding
\begin{eqnarray}
\langle\hat{T}_{00}(x)\rangle&=&
-i\,{\rm Tr}\left\{i\gamma^0\partial_0\left(1-S\Aslash\right)^{-1}
S\delta(x-\hat{x})\right\}
\label{FD1}
\end{eqnarray}
in powers of $A_\mu$. Here $S=(i\gamma^\mu\partial_\mu-m)^{-1}$ is
the free Dirac propagator. The $n^{\rm th}$ term in the expansion is
\begin{eqnarray}
\langle\hat{T}_{00}(x)\rangle_n&=&
-i\,{\rm Tr}\left\{i\gamma^0\partial_0
\left(S\Aslash\right)^nS\delta(x-\hat{x})\right\}\,.
\label{FD2}
\end{eqnarray}
Since the background field $A_\mu$ is static, it is useful to 
introduce frequency states $|\omega\rangle$ with
$\langle\omega |A_\mu |\omega^\prime\rangle=
A_\mu\delta(\omega-\omega^\prime)$.
Then the expectation value becomes
\begin{eqnarray}
\langle\hat{T}_{00}(x)\rangle_n&=&-i
\int\frac{d\omega}{2\pi}{\rm Tr^\prime}
\left\{\gamma^0\omega\left[S(\omega)\Aslash\right]^n
S(\omega)\delta(\vec{x}-\hat{\vec{x}})\right\}\,.
\label{FD3}
\end{eqnarray}
where 
$S(\omega)=(\gamma^0\omega-i\vec{\gamma}\cdot\vec{\partial}-m)^{-1}$
and ${\rm Tr^\prime}$ is the trace over all remaining
degrees of freedom, spatial and discrete.
The total energy at order $n$ is 
\begin{eqnarray}
E_n=\int d^{D-1}x\langle\hat{T}_{00}(x)\rangle_n
&=&-i\int\frac{d\omega}{2\pi}{\rm Tr^\prime}
\left\{S(\omega)\gamma^0\omega\left[S(\omega)\Aslash\right]^n\right\}\cr
&=&-i\int\frac{d\omega}{2\pi}{\rm Tr^\prime}
\left\{S(\omega)\gamma^0\omega 
S(\omega)\Aslash S(\omega)\Aslash S(\omega)\Aslash \ldots \right\}.
\label{Etot1}
\end{eqnarray}
We use the Ward identity
$\frac{\partial}{\partial\omega}S(\omega)=
-S(\omega)\gamma^0S(\omega)$ to write
\begin{eqnarray}
E_n&=&i\int\frac{d\omega}{2\pi}{\rm Tr^\prime}
\left\{\omega\left[\frac{\partial}{\partial\omega}S(\omega)\right]
\Aslash S(\omega)\Aslash S(\omega)\Aslash\ldots\right\}\cr
&=&-i\int\frac{d\omega}{2\pi}{\rm Tr^\prime}
\Bigg\{S(\omega)\Aslash S(\omega)\Aslash S(\omega)\Aslash \ldots
+\omega \Aslash \left[\frac{\partial}{\partial\omega}S(\omega)\right]
\Aslash S(\omega)\Aslash\ldots
\nonumber \\ &&\hspace{2cm}
+\omega \Aslash S(\omega)\Aslash 
\left[\frac{\partial}{\partial\omega}S(\omega)\right]\Aslash\ldots
\,\,+\,\,\ldots\Bigg\}
\cr
&=&-i\int\frac{d\omega}{2\pi}{\rm Tr^\prime}
\Bigg\{S(\omega)\Aslash S(\omega)\Aslash S(\omega)\Aslash \ldots
\nonumber \\ &&\hspace{2cm}
+(n-1)\omega \Aslash \left[\frac{\partial}{\partial\omega}S(\omega)\right]
\Aslash S(\omega)\Aslash\ldots\Bigg\}\,,
\label{partint}
\end{eqnarray}
where we have integrated by parts and used the cyclic properties
of the trace. We identify the first and the third terms of the above
equations to eliminate the derivative terms,
\begin{eqnarray}
E_n&=&-\frac{i}{n}\int\frac{d\omega}{2\pi}{\rm Tr^\prime}
\left\{\left[S(\omega)\Aslash\right]^n\right\}
=-\frac{i}{n}\frac{1}{T}{\rm Tr}
\left\{\left[S\Aslash\right]^n\right\} \,.
\label{Etot2}
\end{eqnarray}
The time interval, $T$, appears because we have re-established
the full functional trace. This expansion can then be re-summed
\begin{eqnarray}
E=\sum_n E_n=\frac{i}{T}{\rm Tr}\,{\rm ln}\left[1-S\Aslash\right]
=\frac{i}{T}{\rm Tr}\,{\rm ln}\left[i\Dslash-m\right]\,,
\label{effaction}
\end{eqnarray}
where again $A_\mu$ independent terms have been omitted.
We see that order by order the total energy can be obtained
from the effective action. 

However, this is not the case for the density. The action
density would be 
\begin{eqnarray}
\langle {\mathcal A}(x)\rangle&=&-i{\rm Tr}
\left\{{\rm ln}\left[1-S\Aslash\right]\delta(x-\hat{x})\right\}
=\sum_n\langle {\mathcal A}(x)\rangle_n\cr
\langle {\mathcal A}(x)\rangle_n&=&\frac{i}{n}{\rm Tr}
\left\{\left[S\Aslash\right]^n\delta(x-\hat{x})\right\}\,.
\label{expt00}
\end{eqnarray}
The terms in this expansion are simpler than those in 
eq.~(\ref{FD2}) because they only have $n$ Dirac propagators at 
order $n$, instead of the $n+1$ in eq.~(\ref{FD2}).

Upon inserting $1=\frac{\partial}{\partial\omega}\omega$
and integrating by parts, eq.~(\ref{expt00}) turns into
\begin{equation}
\langle {\mathcal A}(x)\rangle_n=\frac{i}{n}
\sum_{k=0}^{n-1}\int\frac{d\omega}{2\pi}{\rm Tr^\prime}
\Bigg\{S(\omega)\gamma^0\omega 
\left[S(\omega)\Aslash\right]^{n-k}
\delta(\vec{x}-\hat{\vec{x}})\left[S(\omega)\Aslash\right]^k\Bigg\}\,.
\label{puzzle1}
\end{equation}
These are the same operators as in eq.~(\ref{FD3}), but in
different orders. The differences are characterized by the 
commutator $[\delta(x-\hat{x}),S]$ which has the matrix
elements
\begin{equation}
\langle k^\prime | [\delta(x-\hat{x}),S] |k \rangle =
{\rm e}^{i(k-k^\prime)\cdot x}
\left[\frac{1}{k\hskip-0.5em/-m}
-\frac{1}{k\hskip-0.5em/^\prime-m}\right] \,.
\label{comm}
\end{equation}
These matrix elements have expansions in $k-k^\prime$, starting at
linear order. That is, in coordinate space they will give total
derivative contributions to the energy density (thus leading to the 
same total energy).  When expanding the energy density we deal with
three operators, $S,\Aslash$ and $\delta(x-\hat{x})$, while the
total energy contains only the combination $S-\Aslash$. Thus
the total only involves two operators, which that can 
always be brought under the trace in the desired order.

The expansion~(\ref{expt00}) will always be of the form
$$
{\rm tr}\Bigg\{\Aslash(x)\int d^4q_1 \ldots d^4q_{n-1}
\tilde{\Aslash}(q_1) \ldots \tilde{\Aslash}(q_{n-1})
{\rm e}^{ix\cdot(q_1 ...)}\Pi(q_1,\ldots, q_{n-1})\Bigg\}
$$
with some $(n-1)$--point function $\Pi(q_1,\ldots, q_{n-1})$.
Thus it is {\it semi-local} --- it vanishes everywhere where $A_\mu$
vanishes.

\subsection{Feynman diagrams for energy density}

In Appendix~\ref{energyappendix} we have computed the energy density with
the first two terms of the expansion in $A_\mu$ subtracted.  This
subtracted quantity was then amenable to contour integration in the
complex momentum plane.  Now we have to add back in
the terms we have subtracted, which we will do in terms of Feynman
diagrams, treating renormalization the standard way.

We start from the formal expansion in eq.~(\ref{FD3}). 
The first-order contribution is given by
\begin{eqnarray}
\langle\hat{T}_{00}(x)\rangle_1&=&-i\int \dbar k_1 \dbar k_2\, {\rm tr}\,
\left\{\gamma_0 k_1^0(\kslash_1-m)^{-1}\tilde{\Aslash}(k_1-k_2)
(\kslash_2-m)^{-1}{\rm e}^{i(k_2-k_1)\cdot x}\right\}\cr
&=&-i\int \dbar l \dbar q \tilde{A}_\mu(q){\rm e}^{-iq\cdot x}
\int_0^1d\xi \left[l^2-m^2+\xi(1-\xi)q^2\right]^{-2}\cr
&&\hspace{2cm}\times {\rm tr}\,
\left\{\gamma_0l^0[\lslash+\xi\qslash+m]\gamma^\mu
[\lslash+(\xi-1)\qslash+m]\right\}\,.
\label{ord11}
\end{eqnarray}
where $\dbar k=\frac{d^Dk}{(2\pi)^D}$ for momenta and
$\dbar x= d^D x$ for coordinates in $D$ dimensions.
Here we have already used that the background field is static,
so that $\tilde{A}_\mu(q)\propto \delta(q_0)$, which allows us 
to set $q^0=0$ inside the integrand and therefore the shift 
$k_1=l+\xi q$ does not affect the time component. We observe that
\begin{equation}
l^0\left[l^2-m^2+\xi(1-\xi)q^2\right]^{-2}=
-\frac{1}{2}\frac{\partial}{\partial l_0}
\left[l^2-m^2+\xi(1-\xi)q^2\right]^{-1}
\label{partder}
\end{equation}
and integrate by parts in $l_0$,
\begin{eqnarray}
\langle\hat{T}_{00}(x)\rangle_1&=&-
\frac{i}{2}\int \dbar l \dbar q \tilde{A}_\mu(q){\rm e}^{-iq\cdot x}
\int_0^1d\xi\, {\rm tr}\,
\frac{\gamma^\mu(\lslash+(2\xi-1)\qslash+m)}
{l^2-m^2+\xi(1-\xi)q^2}\cr
&=&-2i\int \dbar q q^\mu \tilde{A}_\mu(q){\rm e}^{-iq\cdot x}
\int_0^1d\xi \int \dbar l\,
\frac{2\xi-1}{l^2-m^2+\xi(1-\xi)q^2}
\cr &=&0 
\label{ord12}
\end{eqnarray}
because the $\xi$--integral vanishes.  As expected from Furry's
theorem, the first-order contribution vanishes.

The second-order contribution is
\begin{eqnarray}
\langle\hat{T}_{00}(x)\rangle_2&=&-i{\rm Tr}\left\{
i\gamma_0\partial_0 S\Aslash S \Aslash S \delta(x-\hat{x})\right\}\cr
&=&-2i\int \dbar q \dbar p\, \tilde{A}_\mu(p)\tilde{A}_\nu(q-p)
{\rm e}^{-iq\cdot x}\int_0^1d \xi \int_0^{1-\xi}d\eta 
\label{ord21} \\
&&\hspace{0.2cm}\times \int \dbar l\, {\rm tr}\,
\frac{l^0\gamma_0(\lslash+\kslash+m)\gamma^\mu
(\lslash+\kslash-\pslash+m)\gamma^\nu
(\lslash+\kslash-\qslash+m)}{\left[l^2-m^2+\Delta\right]^3} \,,
\nonumber
\end{eqnarray}
where we have introduced the abbreviations
\begin{equation}
k_\mu=\xi p_\mu + \eta q_\mu\, ,\quad 
\Delta=\xi(1-\xi)p^2+\eta(1-\eta)q^2-2\xi\eta p\cdot q=
\xi p^2 +\eta q^2 -k^2\,,
\label{abbr}
\end{equation}
and again used that the background field is static to set
$p^0=q^0=0$ inside the integrand. We integrate by parts in $l^0$
to obtain
\begin{equation}
\langle\hat{T}_{00}(x)\rangle_2=-\frac{i}{2}
\int \dbar q \dbar p\, \tilde{A}_\mu(p)\tilde{A}_\nu(q-p)
{\rm e}^{-iq\cdot x}\,\Pi^{\mu\nu}(p,q)\,,
\label{ord22}
\end{equation}
where 
\begin{eqnarray}
\Pi^{\mu\nu}(p,q)\hspace{-0.2cm}&=&\hspace{-0.2cm}
\int_0^1\hspace{-0.1cm} d\xi 
\int_0^{1-\xi}\hspace{-0.3cm}d\eta
\int \frac{\dbar l}{\left[l^2-m^2+\Delta\right]^2}\,
{\rm tr}\, \Bigg\{
2\gamma^\mu(\lslash+\kslash-\pslash+m)\gamma^\nu
(\lslash+\kslash-\frac{\qslash}{2}+m)\hspace{1cm}\cr
&&\hspace{4.0cm}
+\gamma^0(\lslash+\kslash+m)\gamma^\mu\gamma^0\gamma^\nu
(\lslash+\kslash-\qslash+m)\Bigg\}\,.
\label{defPi}
\end{eqnarray}
Obviously, $\Pi^{\mu\nu}(p,q)$ contains ultraviolet divergences as 
$D\to4$.  However, $\Pi^{\mu\nu}(p,q)$ should become finite by merely
adding the conventional counterterm proportional to $F_{\mu\nu}^2$;
no other counterterms are available in this theory.
To see how this result emerges from the equations, we evaluate the trace in
eq.~(\ref{defPi}) and keep  only terms that do not vanish under the
$l$--integral 
\begin{eqnarray}
{\rm tr}\Big\{\ldots\Big\}&=&
{\rm tr}\Bigg\{2\left(\frac{2-D}{D}l^2+m^2-\Delta\right)
\gamma^\mu\gamma^\nu+2\Delta\gamma^\mu\gamma^\nu\cr
&&\hspace{1.0cm}
+\left(\frac{2-D}{D}l^2+m^2-\Delta\right)
\gamma^0\gamma^\mu\gamma^0\gamma^\nu
+\Delta\gamma^0\gamma^\mu\gamma^0\gamma^\nu\cr
&&\hspace{1.0cm}
+2\gamma^\mu(\kslash-\pslash)\gamma^\nu(\kslash-\frac{\qslash}{2})
+\gamma^0\kslash\gamma^\mu\gamma^0\gamma^\nu
(\kslash-\qslash)\Bigg\}\,.
\label{Pi1}
\end{eqnarray}
and the quadratic divergences drop out, giving
\begin{equation}
\Pi^{\mu\nu}(p,q)=\int_0^1d \xi \int_0^{1-\xi}d\eta
\int \dbar l \frac{N^{\mu\nu}(p,q;\xi,\eta)}
{\left[l^2-m^2+\Delta\right]^2}
\label{Pi2}\
\end{equation}
with
\begin{equation}
N^{\mu\nu}(p,q;\xi,\eta)={\rm tr}\, \Big\{
2\Delta\gamma^\mu\gamma^\nu+
2\gamma^\mu(\kslash-\pslash)\gamma^\nu(\kslash-\frac{\qslash}{2})
+\Delta\gamma^0\gamma^\mu\gamma^0\gamma^\nu
+\gamma^0\kslash\gamma^\mu\gamma^0\gamma^\nu
(\kslash-\qslash)\Big\}\,.
\label{defN}
\end{equation}
To deal with the logarithmic divergence, we first consider
$\Pi^{\mu\nu}(p,0)$. In that case the $\eta$--integral 
becomes trivial and yields a factor $1-\xi$. 
The $\xi$--integral simplifies due to the identity
\small
\begin{equation}
\int_0^1 d\xi \xi^2(1-\xi)f\big(\xi(1-\xi)\big)=
\int_0^1 d\xi \xi(1-\xi)^2f\big(\xi(1-\xi)\big)=
\frac{1}{2}\int_0^1 d\xi \xi(1-\xi)f\big(\xi(1-\xi)\big) \,,
\label{xiident}
\end{equation}
\normalsize
which leads to
\begin{equation}
\Pi^{\mu\nu}(p,0)=8\left(p^2g^{\mu\nu}-p^\mu p^\nu\right)
\int_0^1 d\xi \xi (1-\xi)
\int \frac{\dbar l}{\left[l^2-m^2+\xi(1-\xi)p^2\right]^2}\,.
\label{Pi3}
\end{equation}
This has the form of the unrenormalized second order contribution to 
the total energy, {\it cf.} eq.~(\ref{defPi}).  
Therefore one might assume that in the decomposition 
\begin{equation}
\langle\hat{T}_{00}(x)\rangle_2=-\frac{i}{2}
\int \dbar q \dbar p\, \tilde{A}_\mu(p)\tilde{A}_\nu(q-p)
{\rm e}^{-iq\cdot x} \left\{\Pi^{\mu\nu}(p,0)
+\left[\Pi^{\mu\nu}(p,q)-\Pi^{\mu\nu}(p,0)\right]
\right\}\,,
\label{ord23}
\end{equation}
the divergence from the first term in curly brackets 
would be canceled by the usual counterterm and 
the difference $\Pi^{\mu\nu}(p,q)-\Pi^{\mu\nu}(p,0)$ would 
give a finite result on its own. The latter conjecture can easily be
proven incorrect, since the traces in eq.~(\ref{Pi2}) are
different for $q\ne0$ and $q=0$, hence the integrands in the
$\Pi^{\mu\nu}(p,q)-\Pi^{\mu\nu}(p,0)$ do \emph{not}
cancel at large $l$. Nevertheless, let us continue and consider the 
first term in eq.~(\ref{ord23}),
\begin{eqnarray}
T_1(x) &=&-\frac{i}{2}
\int \dbar q \dbar p\, \tilde{A}_\mu(p)\tilde{A}_\nu(q-p)
{\rm e}^{-iq\cdot x} \Pi^{\mu\nu}(p,0)
\label{t11} \\
&=&-4i\int \dbar q \dbar p\, \tilde{A}_\mu(p)\tilde{A}_\nu(q-p)
{\rm e}^{-iq\cdot x}
\left(p^2g^{\mu\nu}-p^\mu p^\nu\right)\Pi(p^2)\,,
\nonumber
\end{eqnarray}
where, as usual
\begin{equation}
\Pi(p^2)=\int_0^1 d\xi \xi(1-\xi)\int
\frac{\dbar l}{\left[l^2-m^2+\xi(1-\xi)p^2\right]^2}\,.
\label{polten}
\end{equation}
Note that the piece $T_1(x)$ is semi-local ---
it vanishes everywhere where $A_\nu(x)$ is zero because
$\int \dbar q \tilde{A}_\nu(q-p){\rm e}^{-iq\cdot x}=
{\rm e}^{-ip\cdot x}A_\nu(x)$. Furthermore, eq.~(\ref{t11}) is the
second-order contribution, ${\mathcal A}_2(x)$, in eq.~(\ref{expt00}). 

Next we add the usual local counterterm determined at the
renormalization scale $M^2$.  (For the calculations reported
in the main text we used on-shell renormalization conditions
corresponding to $M^2=0$.)  This term,
\begin{eqnarray}
{\cal L}_{\rm ct}&=&-2i\Pi(M^2)F_{\mu\nu}(x)F^{\mu\nu}(x)\\
\label{lct1}
&=&-4i\Pi(M^2)\left\{\partial_\mu
\left[A_\nu(x)\partial^\mu A^\nu(x)
-A_\nu(x)\partial^\nu A^\mu(x)\right]
-A_\nu(x)\left[g^{\mu\nu}\partial^2
-\partial^\mu\partial^\nu\right]A_\mu(x)\right\} \,,
\nonumber
\end{eqnarray}
does not make the piece $T_1(x)$ finite because of the surface
term.\footnote{This result also indicates that the
expansion~(\ref{expt00}) cannot be used together with
the (standard) counterterm Lagrangian, eq.~(\ref{lct1}).}
Rather we find
\begin{eqnarray}
T_1(x)+\epsilon_{\rm ct}(x)&=&-4iA_\nu(x)\int \dbar p
\tilde{A}_\mu(p){\rm e}^{-ip\cdot x}
\left(p^2g^{\mu\nu}-p^\mu p^\nu\right)\Pi_{\rm R}(p^2,M^2)\cr
&&+4i\Pi(M^2)\int \dbar p \dbar q
\tilde{A}_\mu(p)\tilde{A}_\nu(q-p){\rm e}^{-iq\cdot x}
\left(q^\mu p^\nu-g^{\mu\nu}p\cdot q\right) \,,
\label{t12}
\end{eqnarray}
where the renormalized polarization tensor is
\begin{equation}
\Pi_{\rm R}(p^2,M^2)=\Pi(p^2)-\Pi(M^2)\,.
\label{renpolten}
\end{equation}
Since the theory is renormalizable, the divergence in eq.~(\ref{t12}) must 
cancel the divergence in the difference $\Pi^{\mu\nu}(p,q)-\Pi^{\mu\nu}(p,0)$.
To evaluate that difference we have to compute the numerator 
trace~(\ref{defN}). Because the background is static and its time component 
vanishes, we may replace
$\gamma^0\gamma^\mu\gamma^0\gamma^\nu\to -\gamma^\mu\gamma^\nu$
and $\gamma^0\kslash\gamma^\mu\gamma^0\gamma^\nu
(\kslash-\qslash)\to \kslash\gamma^\mu\gamma^\nu(\kslash-\qslash)$.
We obtain
\begin{eqnarray}
N^{\mu\nu}(p,q;\xi,\eta)&=&4\left[\xi(3-2\xi)p^2
+(2\eta-4\xi\eta-1)p\cdot q +\eta(1-2\eta)q^2\right]g^{\mu\nu}\cr
&&\hspace{0.2cm}
+16\xi(\xi-1)p^\mu p^\nu +8\eta(2\eta-1)q^\mu q^\nu
+8\xi q^\mu p^\nu\cr
&&\hspace{0.2cm}
+4\left[(\xi-1)(2\eta-1)+2\xi\eta-\xi\right]
\left(p^\mu q^\nu+q^\mu p^\nu\right)\,.
\label{Nfinal}
\end{eqnarray}
Using
\begin{equation}
\int_0^1d\xi \int_0^{1-\xi}d\eta\left[
N^{\mu\nu}(p,q;\xi,\eta)-N^{\mu\nu}(p,0;\xi,\eta)\right]
=\frac{4}{3}\left(q^\mu p^\nu- p\cdot q g^{\mu\nu}\right)\,,
\label{Nint}
\end{equation}
we have
\begin{equation}
\frac{i}{2}\int_0^1d\xi \int_0^{1-\xi}d\eta\int \dbar l\,
\frac{N^{\mu\nu}(p,q;\xi,\eta)-N^{\mu\nu}(p,0;\xi,\eta)}
{\left[l^2-m^2\right]^2}
=4i\Pi(0)\left(q^\mu p^\nu- p\cdot q g^{\mu\nu}\right)\,,
\label{Nintren}
\end{equation}
which indeed cancels the UV divergence in eq.~(\ref{t12}).

Putting everything together, we have
\begin{eqnarray}
\langle\hat{T}_{00}(x)\rangle_2+\epsilon_{\rm ct}(x)&=&
4i\Pi_{\rm R}(M^2,0)\partial_\mu
\left[A_\nu(x)\partial^\mu A^\nu(x)
-A_\nu(x)\partial^\nu A^\mu(x)\right]\cr
&&\hspace{-3.0cm}
-4iA_\nu(x)\int \dbar p
\tilde{A}_\mu(p){\rm e}^{-ip\cdot x}
\left(p^2g^{\mu\nu}-p^\mu p^\nu\right)\Pi_{\rm R}(p^2,M^2)\cr
&&\hspace{-3.0cm}
-\frac{1}{2(4\pi)^2}\int \dbar q \dbar p 
\tilde{A}_\mu(p)\tilde{A}_\nu(q-p){\rm e}^{-iq\cdot x}
\int_0^1d \xi \int_0^{1-\xi} d\eta\cr
&&\hspace{-1.2cm}\times
\Bigg\{N^{\mu\nu}(p,q;\xi,\eta)
{\rm ln}\frac{m^2-\Delta}{m^2}-N^{\mu\nu}(p,0;\xi,\eta)
{\rm ln}\frac{m^2-\xi(1-\xi)p^2}{m^2}\Bigg\}\cr
&& \hspace{-3.0cm}
=2i\Pi_{\rm R}(M^2,0)F_{\mu\nu}F^{\mu\nu}
-4iA_\nu(x)\int \dbar p
\tilde{A}_\mu(p){\rm e}^{-ip\cdot x}
\left(p^2g^{\mu\nu}-p^\mu p^\nu\right)\Pi_{\rm R}(p^2,0)\cr
&&\hspace{-3.0cm}
-\frac{1}{2(4\pi)^2}\int \dbar q \dbar p
\tilde{A}_\mu(p)\tilde{A}_\nu(q-p){\rm e}^{-iq\cdot x}
\int_0^1d \xi \int_0^{1-\xi} d\eta\cr
&&\hspace{-1.2cm}\times
\Bigg\{N^{\mu\nu}(p,q;\xi,\eta)
{\rm ln}\frac{m^2-\Delta}{m^2}-N^{\mu\nu}(p,0;\xi,\eta)
{\rm ln}\frac{m^2-\xi(1-\xi)p^2}{m^2}\Bigg\}\,.
\label{result}
\end{eqnarray}

We define
\begin{eqnarray}
8\pi i\Pi_{\rm R}(M^2,0)&=&F_1(M)=\frac{-1}{2\pi}
\left[\frac{5}{18}+\frac{2m^2}{3M^2}
+\frac{M^4-2m^2M^2-8m^4}{3M^3\sqrt{4m^2-M^2}}
\,{\rm arcsin}\left(\frac{M}{2m}\right)\right]\,,\cr
8\pi i\Pi_{\rm R}(-\vec{p\,}^2,0)&=&F_2(p)=\frac{-1}{2\pi}
\left[\frac{5}{18}-\frac{2m^2}{3p^2}
-\frac{p^4+2m^2p^2-8m^4}{6p^3\sqrt{4m^2+p^2}}
\,{\rm ln}\frac{\sqrt{4m^2+p^2}+p}{\sqrt{4m^2+p^2}-p}\right]\,,
\hspace{0.6cm}
\label{defF12}
\end{eqnarray}
for $M\le 2m$ with $p=|\vec{p\,}|$ and
\begin{equation}
\tilde{h}(p)=-\frac{2\pi}{p}\int dr J_0(pr) h^\prime(r) \,,
\label{FT1}
\end{equation}
so that
\begin{equation}
\tilde{\vec{A\,}}(\vec{p\,})=i\tilde{h}(p) \hat{z}\times\hat{p}\,,
\label{FT2}
\end{equation}
where $\vec{p\,}$ is in the $x-y$ plane.  We finally find
\begin{eqnarray}
\langle\hat{T}_{00}(x)\rangle_2+\epsilon_{\rm ct}(x)&=&
\frac{1}{\pi}F_1(M)\left(\frac{h^\prime(r)}{r}\right)^2
+\frac{h(r)}{4\pi^2r}\int_0^\infty p^3 dp 
F_2(p) \tilde{h}(p) J_1(pr)
\nonumber \\*
&& -\frac{1}{64\pi^5}\int q dq J_0(qr) \int dp\, \tilde{h}(p)
\int_0^{2\pi} d\phi\, 
\frac{\tilde{h}(\sqrt{p^2+q^2-2pq\,{\rm cos}\phi}\,)}
{\sqrt{p^2+q^2-2pq\,{\rm cos}\phi}}
\nonumber \\*
&&\hspace{0.5cm}\times\left[
\left(p^2-pq\,{\rm cos}\phi\right)Z_1(p,q,{\rm cos}\phi)
+(pq\,{\rm sin}\phi)^2 Z_2(p,q,{\rm cos}\phi)\right]\,.\hspace{1cm}
\label{final1}
\end{eqnarray}
We have used the Lorentz gauge condition, which in momentum space
llows us to substitute
$p_\nu\tilde{A}^\nu(q-p)\to q_\nu\tilde{A}^\nu(q-p)$ and
$p_\mu\tilde{A}^\mu(p)\to0$, so that the only tensor structures are
$\delta_{mn}$ and $q_m q_n$. They multiply Feynman parameter 
integrals
\begin{eqnarray}
Z_1(p,q,z)&=&\int_0^1d\xi
\Bigg\{\xi(3-2\xi)p^2I_0
+\left[(2-4\xi)I_1-I_0\right]pqz+(I_1-2I_2)q^2 \,, \cr
&&\hspace{1.5cm}-2\xi(1-\xi)p^2{\rm ln}\left[1+
\xi(1-\xi)\frac{p^2}{m^2}\right]\Bigg\}\cr
Z_2(p,q,z)&=&\int_0^1d\xi \left[I_0+4(\xi-1)I_1+4I_2\right]\,.
\label{z1z2}
\end{eqnarray}
The $\eta$ integrals 
\begin{equation}
I_n(p,q,z;\xi)=\int_0^{1-\xi} d\eta\, \eta^n \,
{\rm ln}\left[1+\xi(1-\xi)\frac{p^2}{m^2}
+\eta(1-\eta)\frac{q^2}{m^2}-2\xi\eta 
\frac{pq}{m^2}z\right]
\label{intIn}
\end{equation}
can be performed analytically. We define
$\eta_0=q-2\xi pz$ and  
$s=\sqrt{4m^2+4\xi(1-\xi)p^2+\eta_0^2}$ and find
\begin{eqnarray}
I_0&=&(1-\xi)\left\{{\rm ln}
\frac{m^2+\xi(1-\xi)(p^2+q^2-2pqz)}{m^2}-2\right\}\cr
&&\hspace{0.2cm}
-\frac{\eta_0}{2q}\,{\rm ln}\,\frac{m^2+\xi(1-\xi)(p^2+q^2-2pqz)}
{m^2+\xi(1-\xi)p^2}
+\frac{s}{2q}\,{\rm ln}\Bigg|
\frac{\left[2q(1-\xi)-\eta_0+s\right](\eta_0+s)}
{\left[2q(1-\xi)-\eta_0-s\right](\eta_0-s)}\Bigg|\cr
I_1&=&\frac{1}{2}(1-\xi)^2\left\{
{\rm ln}\frac{m^2+\xi(1-\xi)(p^2+q^2-2pqz)}{m^2}-1\right\}
-\frac{1}{2}(1-\xi)\frac{\eta_0}{q}\cr
&&\hspace{0.2cm}
-\frac{\eta_0^2+s^2}{8q^2}\,
{\rm ln}\frac{m^2+\xi(1-\xi)(p^2+q^2-2pqz)}
{m^2+\xi(1-\xi)p^2}
+\frac{\eta_0s}{4q^2}\,{\rm ln}\Bigg|
\frac{\left[2q(1-\xi)-\eta_0+s\right](\eta_0+s)}
{\left[2q(1-\xi)-\eta_0-s\right](\eta_0-s)}\Bigg|\cr
I_2&=&\frac{1}{3}(1-\xi)^3\left\{
{\rm ln}\frac{m^2+\xi(1-\xi)(p^2+q^2-2pqz)}{m^2}
-\frac{2}{3}\right\}\cr
&&\hspace{0.2cm}
-\frac{1}{6}(1-\xi)^2\frac{\eta_0}{q}
-\frac{1}{6}(1-\xi)\frac{\eta_0^2+s^2}{q^2}\cr
&&\hspace{0.2cm}
-\frac{\eta_0}{8q^3}\left(\frac{\eta_0^2}{3}+s^2\right)\,
{\rm ln}\frac{m^2+\xi(1-\xi)(p^2+q^2-2pqz)}
{m^2+\xi(1-\xi)p^2}\,,\cr
&&\hspace{0.2cm}
+\frac{s}{8q^3}\left(\frac{s^2}{3}+\eta_0^2\right)\,
{\rm ln}\Bigg|
\frac{\left[2q(1-\xi)-\eta_0+s\right](\eta_0+s)}
{\left[2q(1-\xi)-\eta_0-s\right](\eta_0-s)}\Bigg| \,.
\label{I0I1I2}
\end{eqnarray}
Although these formulas look awkward, they can straightforwardly be
included in a numerical program.  The separation of 
$\Pi^{\mu\nu}(p,0)$ was advantageous to integrate over the loop 
momentum. Once that integral was computed, the $\eta$--integral in 
eq.~(\ref{result}) could be done. It then turned out that this term 
exactly canceled the semi-local contribution. That is,
\begin{eqnarray}
\langle\hat{T}_{00}(x)\rangle_2+\epsilon_{\rm ct}(x)&=&
2i\Pi_{\rm R}(M^2,0)F_{\mu\nu}F^{\mu\nu}\cr
&&\hspace{-3.0cm}
-\frac{1}{2(4\pi)^2}\int \dbar q \dbar p
\tilde{A}_\mu(p)\tilde{A}_\nu(q-p){\rm e}^{-iq\cdot x}
\int_0^1d \xi \int_0^{1-\xi} d\eta
N^{\mu\nu}(p,q;\xi,\eta){\rm ln}\frac{m^2-\Delta}{m^2}\,.
\label{result3}
\end{eqnarray}
This turns into
\begin{eqnarray}
\langle\hat{T}_{00}(x)\rangle_2+\epsilon_{\rm ct}(x)&=&
\frac{1}{\pi}F_1(M)\left(\frac{h^\prime(r)}{r}\right)^2\cr
&&-\frac{1}{64\pi^5}\int q dq J_0(qr) \int dp\, \tilde{h}(p)
\int_0^{2\pi} d\phi\,
\frac{\tilde{h}(\sqrt{p^2+q^2-2pq\,{\rm cos}\phi})}
{\sqrt{p^2+q^2-2pq\,{\rm cos}\phi}}
\nonumber \\*
&&\hspace{0.5cm}\times\left[
\left(p^2-pq\,{\rm cos}\phi\right)\bar{Z}_1(p,q,{\rm cos}\phi)
+(pq\,{\rm sin}\phi)^2 Z_2(p,q,{\rm cos}\phi)\right]\,.\hspace{1cm}
\label{final3}
\end{eqnarray}
Here
\begin{equation}
\bar{Z}_1(p,q,z)=\int_0^1d\xi
\left\{\xi(3-2\xi)p^2I_0
+\left[(2-4\xi)I_1-I_0\right]pqz+(I_1-2I_2)q^2\right\}
\label{z1z23}
\end{equation}
while $Z_2$ is the same as in eq.~(\ref{z1z2}). Though eq.~(\ref{final3})
looks simpler than eq.~(\ref{final1}) it is not obvious whether it
is easier to handle numerically because in eq.~(\ref{final1}) 
the semi-local and the total derivative terms have been
separated.  The latter vanishes under spatial integration, and tends
to be smaller because in eq.~(\ref{z1z2}) the leading
$p^2 \ln p$ terms cancel for $Z_1$, but they do not for $\bar{Z_1}$. 
The numerical computations presented in the main text are based on  
eq.~(\ref{final1}).

The case $D=2+1$ is similarly given by eqs.~(\ref{ord22})
and~(\ref{Pi2}). In particular the numerator trace~(\ref{Nfinal})
is identical, since we chose to work with four--component spinors.
However, now the loop integral~(\ref{Pi2}) is finite and
there is no need to split up $\Pi^{\mu\nu}(p,q)$ into
$\Pi^{\mu\nu}(p,0)$ and $\Pi^{\mu\nu}(p,q)-\Pi^{\mu\nu}(p,0)$.
Otherwise, the result is very similar to eq.~(\ref{final1}),
\begin{eqnarray}
\langle\hat{T}_{00}(x)\rangle_2&=&
\frac{1}{32\pi^4}\int q dq J_0(qr) \int dp\, \tilde{h}(p)
\int_0^{2\pi} d\phi\,
\frac{\tilde{h}(\sqrt{p^2+q^2-2pq\,{\rm cos}\phi})}
{\sqrt{p^2+q^2-2pq\,{\rm cos}\phi}}
\nonumber \\*
&&\hspace{1cm}\times\left[
\left(p^2-pq\,{\rm cos}\phi\right)Z_1(p,q,{\rm cos}\phi)
+(pq\,{\rm sin}\phi)^2 Z_2(p,q,{\rm cos}\phi)\right]\,.\hspace{1cm}
\label{final2}
\end{eqnarray}
While the integral $Z_1$ is different from the $D=3+1$ case,
splitting up $\Pi^{\mu\nu}(p,q)$ does not affect $Z_2$,
\begin{eqnarray}
Z_1(p,q,z)&=&\int_0^1d\xi
\left\{\xi(3-2\xi)p^2I_0
+\left[(2-4\xi)I_1-I_0\right]pqz+(I_1-2I_2)q^2\right\}\cr
Z_2(p,q,z)&=&\int_0^1d\xi \left[I_0+4(\xi-1)I_1+4I_2\right]\,.
\label{z1z22}
\end{eqnarray}
Of course, the $\eta$ integrals reflect the reduced dimension
\begin{equation}
I_n(p,q,z;\xi)=\int_0^{1-\xi} d\eta\, 
\frac{\eta^n} {\left[m^2+\xi(1-\xi)p^2
+\eta(1-\eta)q^2-2\xi\eta
pqz\right]^{1/2}}\,.
\label{intIn2}
\end{equation}
These integrals can be computed analytically,
\begin{eqnarray}
I_0&=&\frac{-1}{q}\left\{
{\rm arcsin}\left(\frac{\eta_0-2q+2\xi q}{s}\right)
-{\rm arcsin}\left(\frac{\eta_0}{s}\right)\right\} \,, \cr
I_1&=&\frac{-1}{q^2}\left\{\sqrt{m^2+\xi(1-\xi)(p^2+q^2-2pqz)}
-\sqrt{m^2+\xi(1-\xi)p^2}\right\}+\frac{\eta_0}{2q}I_0\,,\cr
I_2&=&\frac{-3\eta_0}{q^3}\left\{\sqrt{m^2+\xi(1-\xi)(p^2+q^2-2pqz)}
-\sqrt{m^2+\xi(1-\xi)p^2}\right\}\cr
&&\hspace{0.2cm}
+\frac{\xi-1}{2q^2}\sqrt{m^2+\xi(1-\xi)(p^2+q^2-2pqz)}
+\frac{2\eta_0^2+s^2}{8q^2}I_0\,,
\label{I0I1I22}
\end{eqnarray}
with $\eta_0$ and $s$ already defined above eq.~(\ref{I0I1I2}).

Finally we allow for a finite counterterm as in eq.~(\ref{lct1}).
This adds 
\begin{eqnarray}
\epsilon_{\rm ct}(x)&=&-F_3(M)F_{\mu\nu}F^{\mu\nu}
=-2F_3(M)\left(\frac{h^\prime}{r}\right)^2\,,\cr
F_3(M)&=&\frac{1}{4\pi}\int_0^1d\xi
\frac{\xi(1-\xi)}{\sqrt{m^2-\xi(1-\xi)M^2}}\,,\cr
&=&\frac{-m^2}{32\pi M^3}\left\{4\frac{M}{m}
+\left(4+\frac{M^2}{m^2}\right){\rm ln}\,\frac{2m-M}{2m+M}\right\}
\label{ect2d}
\end{eqnarray}
to the energy density. This regulator has a finite 
limit as $M\to0$, specifically $\lim_{M\to0}F_3(M)=\frac{1}{24\pi}$.

\section{Charge Density}

In this Appendix we describe our computation of the charge
density matrix element
\begin{equation}
\rho=\langle \Omega | \Psi^\dagger(x)\Psi(x) | \Omega\rangle
\label{defdens}
\end{equation}
in the vacuum that is polarized by the flux tube background.
Here we consider only the case $D=2+1$.  The charge density is
finite without renormalization and is not affected by the finite
counterterm.  It can therefore be reliably computed in a simplified
approach in which we put the system in a large box and sum the
contributions from all modes up to a finite momentum cutoff $\Lambda$.  
Since no renormalization is required for this quantity, saturation of
the matrix element~(\ref{defdens}) is observed for large but finite
$\Lambda$.  We follow the method used to obtain the 
soliton in the regularized Nambu--Jona--Lasinio
model~\cite{Al96} (but with two-component spinors in this case).
We start with the spinors
\begin{equation}
\Psi^{(0)}_{\ell n}(r)=
\frac{N_{\ell n}}{\sqrt{2\pi}}\,{\rm e}^{i(\ell+1)\varphi}
\begin{pmatrix}
J_{l}(q_{\ell n}r)\,{\rm e}^{-i\varphi}\cr
{\rm sign}(E_{\ell n})\sqrt{\frac{E_{\ell n}-m}{E_{\ell n}+m}}\,
J_{l+1}(q_{\ell n}r)
\end{pmatrix}
=\frac{{\rm e}^{i\ell\varphi}}{\sqrt{2\pi}}
\begin{pmatrix}
u_{\ell,n}(r)\cr
l_{\ell,n}(r)\,{\rm e}^{i\varphi}
\end{pmatrix}
\,,
\label{Psi0}
\end{equation} 
where $J_\ell(qr)$ are Bessel functions. Discretization is achieved 
by the boundary condition $J_\ell(q_{\ell n}R)=0$ at a
large radius $R$.  This boundary condition ensures that no flux runs
through the boundary at $r=R$ and determines the allowed momenta
$q_{\ell n}$. Assume there are $N_\ell$ such 
momenta below the cutoff.  Then we label the energy eigenvalues as
\begin{eqnarray}
E_{\ell n}&=&-\sqrt{q_{\ell n}^2+m^2}\,,\qquad n=1,\ldots,N_\ell \,,\cr
E_{\ell n}&=&\sqrt{q_{\ell n-N_\ell}^2+m^2}\,,
\qquad n=N_\ell+1,\ldots,2N_\ell\,,
\label{energies}
\end{eqnarray}
where $n$ runs from $1,\ldots,2N_\ell$. The normalization
coefficients are easily verified to be
\begin{equation}
N_{\ell n}=\frac{1}{R|J_{\ell+1}(q_{\ell n}R)|}
\sqrt{\frac{E_{\ell n}+m}{E_{\ell n}}}\,,
\label{norm1}
\end{equation}
with the momenta $q_{\ell n}$ assigned as in eq.~(\ref{energies}).
Note that there is no problem with the square roots in 
eqs.~(\ref{Psi0}) and~(\ref{norm1}) because the definition
of $E_{\ell n}$ ensures that the arguments are positive for
any value of $n$. 

These states diagonalize the free Dirac operator in $D=2+1$
dimensions. The interaction associated with the flux tube background is
\begin{equation}
H_{\rm int}=
\frac{h(r)}{r}\begin{pmatrix}
0 & {\rm e}^{-i\varphi} \cr
{\rm e}^{i\varphi} & 0
\end{pmatrix}\,.
\label{interaction}
\end{equation}
We now evaluate the matrix elements 
\begin{equation}
\langle m \ell^\prime | H_0+H_{\rm int} | n\ell\rangle=
\delta_{\ell\ell^\prime}\left(E_{\ell n}\delta_{nm}
+\int_0^R dr h(r)\left[u_{\ell m}(r)l_{\ell n}(r)
+l_{\ell m}(r)u_{\ell n}(r)\right]\right)\,
\label{Hmatrix}
\end{equation}
which by construction are diagonal in the angular momentum $\ell$.
Diagonalization of the Hamiltonian matrix~(\ref{Hmatrix})
yields eigenvalues ${\mathcal E}_{\ell \mu}$ and eigenstates 
$\Psi_{\ell\mu}=\sum_{n=1}^{2N_\ell}C^{(\ell)}_{\mu n}\Psi^{(0)}_{\ell n}$.
The charge density is then simply given by 
\begin{eqnarray}
\rho(r)&=&2\pi\sum_{\ell=-\infty}^\infty\sum_{\mu=1}^{2N_\ell}
\frac{{\rm sgn}({\mathcal E}_{\ell \mu})}{2}
\Psi^\dagger_{\ell\mu}(r)\Psi_{\ell\mu}(r)\cr
&=&\sum_{\ell=-\infty}^\infty\sum_{\mu=1}^{2N_\ell}
\frac{{\rm sgn}({\mathcal E}_{\ell \mu})}{2}
\sum_{n,m=1}^{2N_\ell}C^{(\ell)}_{\mu n}C^{(\ell)}_{\mu m}
\left[u_{\ell m}(r)u_{\ell n}(r)
+l_{\ell m}(r)l_{\ell n}(r)\right]\,.
\label{charge}
\end{eqnarray}
In order to mitigate unphysical boundary effects\footnote{In the 
vicinity of $r,r^\prime\approx R$ the basis states~(\ref{Psi0}) do not
satisfy the completeness relation 
$\displaystyle \sum_{\ell\mu} \Psi^{(0)}_{\ell \mu}(r)
\Psi^{(0)\dagger}_{\ell \mu}(r^\prime) \sim
\mbox{{\sf 1}\hspace{-0.55mm}\rule{0.04em}{1.53ex}}\,\,
\delta(r-r^\prime)$.}
we subtract the free charge density in the same geometry,
$\rho^{(0)}(r)=\sum_{\ell n} {\rm sgn}(E_{\ell n})
\Psi^\dagger_{\ell n}(r) \Psi_{\ell n}(r)/2$. 
The difference $\rho(r)-\rho^{(0)}(r)$
smoothly approaches zero as $r\to R$. 

This discretization procedure is not unique.  However, a different
choice would only cause a change in the unphysical boundary effects as
$R\to\infty$. Hence, subtraction  of $\rho^{(0)}(r)$ also removes
ambiguities in the choice of the discretization condition.

\end{document}